# Parameter-Free Deterministic Global Search with Central Force Optimization

## Richard A. Formato[1]

***Abstract.*** This note describes a parameter-free implementation of Central Force Optimization for deterministic multidimensional search and optimization. The user supplies only one input: the objective function to be maximized, nothing more. The CFO equations of motion are simplified by assigning specific values to CFO's basic parameters, and this particular algorithmic implementation also includes hardwired internal parameters so that none is user-specified. The algorithm's performance is tested against a widely used suite of twenty three benchmark functions and compared to other state-of-the-art algorithms. CFO performs very well indeed.

# Parameter-Free Deterministic Global Search with Central Force Optimization

**Richard A. Formato[1]**

## 1. Introduction

Central Force Optimization (CFO) is a *deterministic* Nature-inspired metaheuristic for an evolutionary algorithm (EA) that performs multidimensional global search and optimization [1-13]. This note presents a parameter-free CFO implementation that requires only one user-specified input: the objective function to be maximized. It works quite well across a wide range of functions.

A major frustration with many EAs is the plethora of setup parameters. For example, the three Ant Colony Optimization (ACO) algorithms described in [14] (AS, $\mathcal{MAX}$-$\mathcal{MIN}$, ACS) require the user to specify parameters $m$ (number of ants), $\rho$ (pheromone evaporation rate), $Q$ (pheromone length constant), $\alpha/\beta$ (weighting exponents for pheromone/heuristic information), $\tau_{max}/\tau_{min}$ (upper/lower pheromone bounds), $\varphi$ (pheromone decay coefficient), $\tau_0$ (initial pheromone value), and $q_0$ (random variable crossover point). The generalized Particle Swarm Optimization (PSO) algorithm in [15] requires the user to specify a population size and six "suitable bounded coefficients" ($\chi$, $w^k$, $c_j$, $r_j$, $c_g$, $r_g$ with upper/lower bounds). The frustration of having to specify these numbers is compounded by the facts that (a) there usually is no methodology for picking "good" values; (b) the "right" parameters are often problem-specific; (c) the solutions often are sensitive to small changes; and (d) run upon run, exactly the same parameters never yield the same results because the algorithm is inherently stochastic.

CFO is quite different. It is based on Newton's laws of motion and gravity for real masses moving through the real Universe; and just as these laws are mathematically precise, so too is CFO. It is completely deterministic at every step, with successive runs employing the same setup parameters yielding precisely the same results. Its inherent determinism distinguishes CFO from all the stochastic EAs that fail completely if randomness is removed. Moreover, CFO's metaphor of gravitational kinematics appears to be more than a simple analogy. While many metaheuristics are inspired by natural processes, ant foraging or fish swarming, for example, the resulting algorithm is not (necessarily) an accurate mathematical model of the actual process. Rather, it truly is a metaphor. By contrast, CFO and real gravity appear to be much more closely related. Indeed, CFO's "metaphor" actually may be reality, because its "probes" often exhibit behavior that is strikingly similar to that of gravitationally trapped near earth objects (NEOs). If so, the panoply of mathematical tools and techniques available in celestial mechanics may be applicable to CFO directly or with modification or extension (see [7-9]). In addition, CFO may be interpreted in different ways depending upon the underlying model (vector force field, gradient of gravitational potential, kinetic or total energy) [13]. These observations, although peripheral to the theme of this note, provide additional context for the CFO metaheuristic.

## 2. CFO Algorithm

Pseudocode for the parameter-free CFO implementation appears in Fig. 1. Simplification of the algorithm is achieved by hardwiring all of CFO's basic parameters, which results in simplified equations of motion as described below. CFO searches for the global *maxima* of an *objective*

---

**Procedure** $CFO\,[\,f(\vec{x}), N_d, \Omega\,]$

**Set internal parameters:** $N_t$, $F_{rep}^{init}$, $\Delta F_{rep}$, $F_{rep}^{min}$, $\left(\dfrac{N_p}{N_d}\right)_{MAX}$, $\gamma_{start}$, $\gamma_{stop}$, $\Delta\gamma$.

**Initialize** $f_{max}^{global}(\vec{x}) =$ very large negative number, say, $-10^{+4200}$.

**For** $N_p/N_d = 2$ **to** $\left(\dfrac{N_p}{N_d}\right)_{MAX}$ **by** $2$:

(a.0)        Total number of probes: $N_p = N_d \cdot \left(\dfrac{N_p}{N_d}\right)$

**For** $\gamma = \gamma_{start}$ **to** $\gamma_{stop}$ **by** $\Delta\gamma$:

(a.1)        Re-initialize data structures for Position/Acceleration vectors & Fitness matrix.

(a.2)        Compute IPD [see text].

(a.3)        Compute initial fitness matrix, $M_0^p, 1 \le p \le N_p$.

(a.4)        Initialize $F_{rep} = F_{rep}^{init}$.

**For** $j = 0$ **to** $N_t$ [or earlier termination criterion – see text]:

(b)        Compute probe position vectors, $\vec{R}_j^p, 1 \le p \le N_p$ [eq.(2)].

(c)        Retrieve errant probes ($1 \le p \le N_p$):

        If $\vec{R}_j^p \cdot \hat{e}_i < x_i^{min} \therefore \vec{R}_j^p \cdot \hat{e}_i = \max\{x_i^{min} + F_{rep}(\vec{R}_{j-1}^p \cdot \hat{e}_i - x_i^{min}), x_i^{min}\}$.

        If $\vec{R}_j^p \cdot \hat{e}_i > x_i^{max} \therefore \vec{R}_j^p \cdot \hat{e}_i = \min\{x_i^{max} - F_{rep}(x_i^{max} - \vec{R}_{j-1}^p \cdot \hat{e}_i), x_i^{max}\}$.

(d)        Compute fitness matrix for current probe distribution, $M_j^p, 1 \le p \le N_p$.

(e)        Compute accelerations using current probe distribution and fitnesses [eq. (1)].

(f)        Increment $F_{rep}$: $F_{rep} = F_{rep} + \Delta F_{rep}$; If $F_{rep} > 1 \therefore F_{rep} = F_{rep}^{min}$.

(g)        If $j \ge 20$ and $j\ MOD\ 10 = 0 \therefore$

                (i) Shrink $\Omega$ around $\vec{R}_{best}$ [see text].

                (ii) Retrieve errant probes using procedure in Step (c).

**Next** $j$

(h)        Reset $\Omega$'s boundaries to their starting values before shrinking.

(i)        If $f_{max}(\vec{x}) \ge f_{max}^{global}(\vec{x}) \therefore f_{max}^{global}(\vec{x}) = f_{max}(\vec{x})$.

**Next** $\gamma$

**Next** $N_p/N_d$

---

Fig. 1. CFO Pseudocode.

function $f(x_1, x_2, ..., x_{N_d})$ defined on the $N_d$-dimensional decision space $\Omega$: $x_i^{\min} \leq x_i \leq x_i^{\max}$, $1 \leq i \leq N_d$. The $x_i$ are *decision variables*, and $i$ the coordinate number. The value of $f(\vec{x})$ at point $\vec{x}$ in $\Omega$ is its *fitness*. $f(\vec{x})$'s topology or "landscape" in the $N_d$-dimensional hyperspace is unknown, that is, there is no *a priori* information about the objective function's maxima. CFO searches $\Omega$ by flying "probes" through the space at discrete "time" steps (iterations). Each probe's location is specified by its position vector computed from two *equations of motion* that analogize their real-world counterparts for material objects moving through physical space under the influence of gravity without energy dissipation.

**Equations of Motion.** Probe $p$'s position vector at step $j$ is $\vec{R}_j^p = \sum_{k=1}^{N_d} x_k^{j,j} \hat{e}_k$, where the $x_k^{p,j}$ are its coordinates and $\hat{e}_k$ the unit vector along the $x_k$-axis. The indices $j$, $0 \leq j \leq N_t$, and $p$, $1 \leq p \leq N_p$, respectively, are the iteration number and probe number, where $N_t$ and $N_p$ are the corresponding *total* number of time steps and *total* number of probes.

In metaphorical "CFO space" each of the $N_p$ probes experiences an acceleration created by the "gravitational pull" of "masses" in $\Omega$. Probe $p$'s acceleration at step $j-1$ is given by

$$\vec{a}_{j-1}^p = \sum_{\substack{k=1 \\ k \neq p}}^{N_p} U(M_{j-1}^k - M_{j-1}^p) \cdot (M_{j-1}^k - M_{j-1}^p) \times \frac{(\vec{R}_{j-1}^k - \vec{R}_{j-1}^p)}{\left\| \vec{R}_{j-1}^k - \vec{R}_{j-1}^p \right\|}, \quad (1)$$

which is the first of CFO's two equations of motion. In (1), $M_{j-1}^p = f(x_1^{p,j-1}, x_2^{p,j-1}, ..., x_{N_d}^{p,j-1})$ is the objective function's fitness at probe $p$'s location at time step $j-1$. Each of the other probes at that step (iteration) has associated with it fitness $M_{j-1}^k, k = 1, ..., p-1, p+1, ..., N_p$. $U(\cdot)$ is the Unit Step function, $U(z) = \begin{Bmatrix} 1, & z \geq 0 \\ 0, & otherwise \end{Bmatrix}$. Note that $\vec{a}_{j-1}^p = 0$ for $\vec{R}_{j-1}^k = \vec{R}_{j-1}^p, k \neq p$, because probe $k$ then has coalesced with probe $p$ and cannot exert any gravitational force on $p$. In this case, equation (1) becomes indeterminate because $M_{j-1}^k = M_{j-1}^p$ [16]. The acceleration $\vec{a}_{j-1}^p$ causes probe $p$ to move from position $\vec{R}_{j-1}^p$ at step $j-1$ to position $\vec{R}_j^p$ at step $j$ according to the trajectory equation

$$\vec{R}_j^p = \vec{R}_{j-1}^p + \vec{a}_{j-1}^p, \quad j \geq 1, \quad (2)$$

which is CFO's second equation of motion. These simplified CFO equations result from hardwiring CFO's basic parameters to $G = 2$, $\Delta t = 1$, $\alpha = 1$, and $\beta = 1$ (see [1,12] for definitions and discussion of these parameters). Note that the "internal parameters" in Fig. 1 are *not* fundamental CFO parameters, but rather are specific to *this particular* CFO implementation.

**Initial Probe Distribution.** Every CFO run begins with an Initial Probe Distribution (IPD) defined by two variables: (a) the total number of probes used, $N_p$; and (b) where the probes are placed inside $\Omega$. The IPD used here is a an orthogonal array of $N_p/N_d$ probes per dimension deployed uniformly on "probe lines" parallel to the coordinate axes and intersecting at a point that slides along $\Omega$'s principal diagonal. Fig. 2 provides a two-dimensional (2D) example of this type of IPD (nine probes shown on each probe line, two overlapping, but any number may be used). The probe lines are parallel to the $x_1$ and $x_2$ axes intersecting at a point on $\Omega$'s principal diagonal marked by position vector $\vec{D} = \vec{X}_{min} + \gamma(\vec{X}_{max} - \vec{X}_{min})$, where $\vec{X}_{min} = \sum_{i=1}^{N_d} x_i^{min} \hat{e}_i$ and $\vec{X}_{max} = \sum_{i=1}^{N_d} x_i^{max} \hat{e}_i$ are the diagonal's endpoint vectors. The parameter $0 \le \gamma \le 1$ determines where along the diagonal the probe lines intersect.

Fig. 3 shows a typical 2D IPD for different values of $\gamma$, while Fig. 4 provides a 3D example for various $\gamma$ values. In Fig. 4 the principal diagonal is shown in red and the three orthogonal probe lines in blue. In this example each probe line contains six equally spaced probes. For certain values of $\gamma$ three probes overlap at the probe lines' intersection point on the diagonal. Of course, this IPD procedure is generalized to the $N_d$-dimensional decision space $\Omega$ to create $N_d$ probe lines parallel to the $N_d$ coordinate axes. While Fig. 2 shows equal numbers of probes on each probe line, a different number of probes per dimension can be used instead. For example, if equal probe spacing were desired in a decision space with unequal boundaries, or if overlapping probes were to be excluded in a symmetrical space, then unequal numbers could be used. Unequal numbers also might be appropriate if *a priori* knowledge of $\Omega$'s landscape, however obtained, suggests denser sampling in one region.

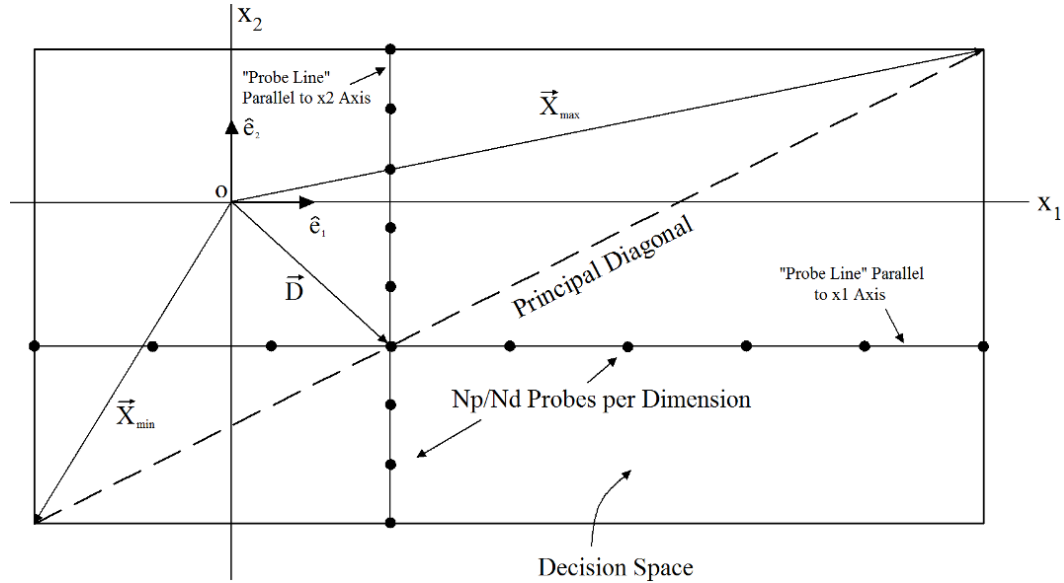

Fig. 2. Variable 2D Initial Probe Distribution.

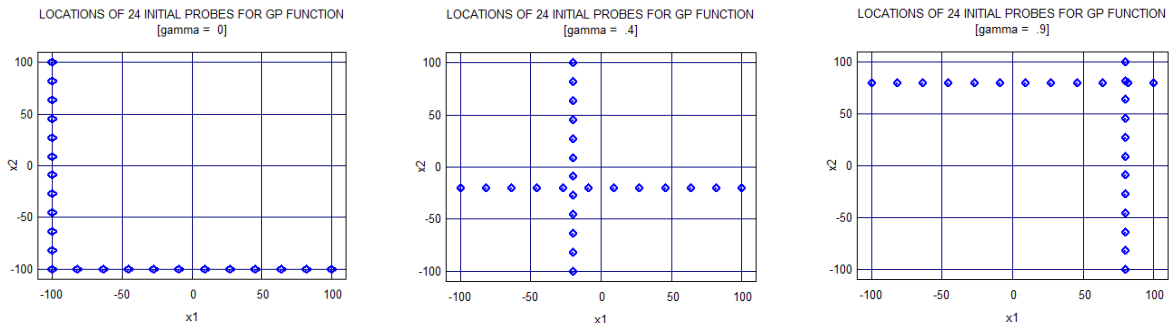

Fig. 3. Typical 2D IPDs for Different Values of $\gamma$.

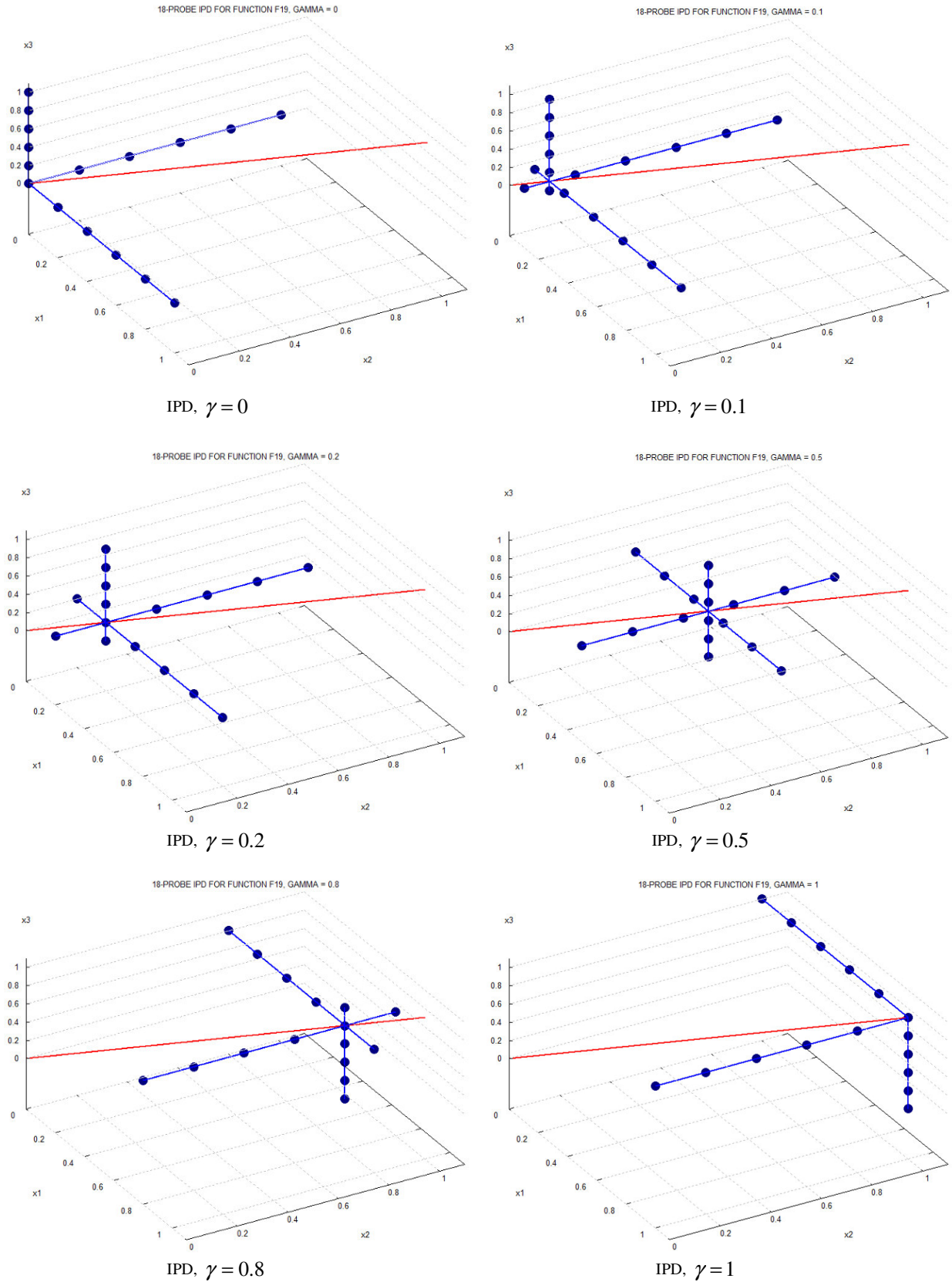

Fig. 4. Typical Variable 3D IPD (Initial Probe Distribution) for the GSO $f_{19}$ Function (probes as filled circles).

*Errant Probes.* It is possible that a probe's acceleration computed from equation (1) may be too great to keep it inside $\Omega$. If any coordinate $x_i < x_i^{\min}$ or $x_i > x_i^{\max}$, the probe enters a region of *unfeasible* solutions that are not valid for the problem at hand. The question (which arises in many algorithms) is what to do with an errant probe. While many schemes are possible, a simple, empirically determined one is used here. On a coordinate-by-coordinate basis, probes flying outside $\Omega$ are placed a fraction $\Delta F_{rep}^{\min} \leq F_{rep} \leq 1$ of the distance between the probe's starting coordinate and the corresponding boundary coordinate [see step (c) in Fig. 1 for details]. $F_{rep}$ is the "repositioning factor" introduced in [2]. $F_{rep}$ starts at an arbitrary initial value $F_{rep}^{init}$ which then is incremented at each iteration by an arbitrary amount $\Delta F_{rep}$ (subject to $F_{rep}^{\min} \leq F_{rep} \leq 1$). This procedure improves $\Omega$'s sampling by distributing probes throughout the space.

*Adaptation.* This particular CFO implementation includes adaptive decision space reconfiguration to improve convergence speed. Fig. 5 illustrates in 2D how $\Omega$'s size is adaptively reduced around $\vec{R}_{best}$, the location of the probe with best fitness throughout the run up to the current iteration. $\Omega$ shrinks every 10th step beginning at step 20. Its boundary coordinates are reduced by one-half the distance from the best probe's position to each boundary on a coordinate-by-coordinate basis, *viz*, $x_i'^{\min} = x_i^{\min} + \dfrac{\vec{R}_{best} \cdot \hat{e}_i - x_i^{\min}}{2}$ and $x_i'^{\max} = x_i^{\max} - \dfrac{x_i^{\max} - \vec{R}_{best} \cdot \hat{e}_i}{2}$, where the primed coordinate is $\Omega$'s new boundary, and the dot denotes vector inner product. For clarity in the diagram, Fig. 5 shows $\vec{R}_{best}$ as being fixed whereas generally it varies throughout a run. Changing $\Omega$'s boundary every ten steps instead of some other interval was chosen arbitrarily.

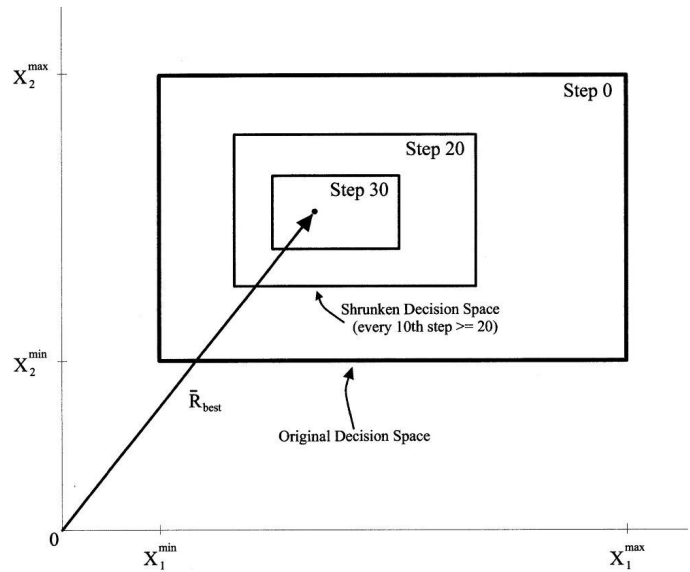

Fig. 5. Schematic 2D Decision Space Adaptation (constant $\vec{R}_{best}$ for illustration only)

    ***Internal Parameters.*** The hardwired internal parameter values used for the implementation reported here are: $N_t = 1000$, $F_{rep}^{init} = 0.5$, $\Delta F_{rep} = 0.1$, $F_{rep}^{\min} = 0.05$, $\left(\dfrac{N_p}{N_d}\right)_{MAX} = 14$ for $N_d \leq 6$, $\left(\dfrac{N_p}{N_d}\right)_{MAX} = 6$ for $21 \leq N_d \leq 30$, $\gamma_{start} = 0$, $\gamma_{stop} = 1$, $\Delta\gamma = 0.1$ The test for early termination is a difference between the average best fitness over 25 steps, including the current step, and the current best fitness of less than $10^{-6}$ (absolute value). This test is applied starting at iteration 35 (at least ten steps must be computed before testing for fitness saturation). The internal parameters, indeed all CFO-related parameters, have been chosen empirically. These particular values provide good results across a wide range of test functions. Beyond this empirical observation there currently is no methodology for choosing either CFO's basic parameters or the internal parameters for an implementation such as this one. There well may be (likely are) better parameter sets, which is an aspect of CFO development that merits further study.

## 3. Results

Group Search Optimizer (GSO) is a new, state-of-the-art stochastic metaheuristic that has gained some notoriety [17,18]. It mimics animal foraging behavior using the strategies of "producing" (searching for food) and "scrounging" (joining resources discovered by other group members). GSO was tested against a suite of twenty-three benchmark functions with dimensionalities ranging from 2 to 30, and also on a limited basis for scalability to 300 dimensions (see [17] for details). GSO thus provides a superior standard against which CFO can be evaluated. CFO was tested against the same twenty-three function 2-30D benchmark suite, and the results are reported here.

    In [17] GSO was compared to two other stochastic algorithms, "PSO" and "GA". PSO is a Particle Swarm implemented using "PSOt," a MATLAB-based toolbox that includes standard and variant PSO algorithms. GA is a Genetic Algorithm implemented using the GAOT toolbox (genetic algorithm optimization toolbox). Recommended default parameter values were used for PSOt and GA. Thus, while this note compares CFO and GSO directly, it indirectly compares CFO to PSO and GA as well.

    Table 1 summarizes CFO's results. $F$ is the test function using the same function numbering as [17]. $f_{\max}$ is the known global *maximum* (note that the negative of each benchmark in [17] is used here because, unlike the other algorithms, CFO locates maxima, not minima). $<\cdot>$ denotes average value. Because GSO, PSO and GA are inherently stochastic, their performance must be described statistically. The statistical data for those algorithms in Table 1 are reproduced from [17]. Of course, no statistical description is needed for CFO because it is deterministic. $N_{eval}$ is the total number of function evaluations made by CFO.

In the first group of high dimensionality unimodal functions ($f_1 - f_7$), CFO returned the best fitness on all seven functions, in one case by a wide margin ($f_5$). In the second set of six high dimensionality multimodal functions with many local maxima ($f_8 - f_{13}$), CFO performed best on three ($f_9 - f_{11}$) and essentially the same as GSO on $f_8$. In the last group of ten multimodal functions with few local maxima ($f_{14} - f_{23}$), CFO returned the best fitness on four ($f_{20} - f_{23}$, and by wide margins on $f_{21} - f_{23}$); equal fitnesses on four ($f_{14}, f_{17}, f_{18}, f_{19}$); and only very slightly lower fitness on two ($f_{15}, f_{16}$).

Of the twenty-three benchmark functions, CFO's performance was not quite as good as the other algorithms' on only two ($f_{12}, f_{13}$). However, CFO's performance can be improved considerably if a second run is made using a smaller decision space based on the results of the first run. As an example, for $f_{12}$ the thirty coordinates returned on the first run are in the range $[-1.03941458 \leq x_i \leq -0.99688049]$ (the known maximum is at the point $[-1]^{30}$). If based on this result $\Omega$ is shrunk from $[-50 \leq x_i \leq 50]$ to $[-5 \leq x_i \leq 5]$ and CFO run again, then the returned best fitness improves to $-7.39354 \times 10^{-35}$ with $N_{eval} = 273,780$, which is a good bit better than GSO's result.

Table 1. CFO Comparative Results for 23 Benchmark Functions

| $F^*$ | $N_d$ | $f_{max}^*$ | \<Best Fitness\>/ Other Algorithm | - - - CFO - - - | |
|---|---|---|---|---|---|
| | | | | Best Fitness | $N_{eval}$ |
| Unimodal Functions (other algorithms: average of 1000 runs) | | | | | |
| $f_1$ | 30 | 0 | -3.6927x10^-37 / PSO | 0 | 222,960 |
| $f_2$ | 30 | 0 | -2.9168x10^-24 / PSO | 0 | 237,540 |
| $f_3$ | 30 | 0 | -1.1979x10^-3 / PSO | -6.1861x10^-5 | 397,320 |
| $f_4$ | 30 | 0 | -0.1078 / GSO | 0 | 484,260 |
| $f_5$ | 30 | 0 | -37.3582 / PSO | -4.8623x10^-5 | 436,680 |
| $f_6$ | 30 | 0 | -1.6000x10^-2 / GSO | 0 | 176,580 |
| $f_7$ | 30 | 0 | -9.9024x10^-3 / PSO | -1.2919x10^-4 | 399960 |
| Multimodal Functions, Many Local Maxima (other algorithms: avg 1000 runs) | | | | | |
| $f_8$ | 30 | 12,569.5 | 12,569.4882 / GSO | 12,569.4865 | 415,500 |
| $f_9$ | 30 | 0 | -0.6509 / GA | 0 | 397,080 |
| $f_{10}$ | 30 | 0 | -2.6548x10^-5 / GSO | 4.7705x10^-18 | 518,820 |
| $f_{11}$ | 30 | 0 | -3.0792x10^-2 / GSO | -1.7075x10^-2 | 235,800 |
| $f_{12}$ | 30 | 0 | -2.7648x10^-11 / GSO | -2.1541x10^-5 | 292,080 |
| $f_{13}$ | 30 | 0 | -4.6948x10^-5 / GSO | -1.8293x10^-3 | 360,000 |
| Multimodal Functions, Few Local Maxima (other algorithms: avg 50 runs) | | | | | |
| $f_{14}$ | 2 | -1 | -0.9980 / GSO | -0.9980 | 78,176 |
| $f_{15}$ | 4 | -3.075x10^-4 | -3.7713x10^-4 / GSO | -5.6967x10^-4 | 143,152 |
| $f_{16}$ | 2 | 1.0316285 | 1.031628 / GSO | 1.03158 | 87,240 |
| $f_{17}$ | 2 | -0.398 | -0.3979 / GSO | -0.3979 | 82,096 |
| $f_{18}$ | 2 | -3 | -3 / GSO | -3 | 100,996 |
| $f_{19}$ | 3 | 3.86 | 3.8628 / GSO | 3.8628 | 160,338 |
| $f_{20}$ | 6 | 3.32 | 3.2697 / GSO | 3.3219 | 457,836 |
| $f_{21}$ | 4 | 10 | 7.5439 / PSO | 10.1532 | 251,648 |
| $f_{22}$ | 4 | 10 | 8.3553 / PSO | 10.4029 | 316,096 |
| $f_{23}$ | 4 | 10 | 8.9439 / PSO | 10.5364 | 304,312 |

* Negative of the functions in [17] are computed by CFO because CFO searches for maxima instead of minima.

## 4. Conclusion

Even though CFO is not nearly as highly developed as the other algorithms used for comparison, it performs very well against all of them while requiring nothing more than the objective function as a user input. CFO performed better than or essentially as well as GSO, GA and PSO on twenty one of twenty three test functions. By contrast, GSO compared to PSO and GA returned the best performance on only fifteen benchmarks. CFO's performance thus is considerably better than GSO's, which in turn performed better than PSO and GA.

CFO's further development lies along two lines: architecture and theory. While this note describes one particular CFO implementation that works well, there no doubt is room for improvement. It is clear that CFO's performance is a sensitive function of the IPD. Of course, there are limitless IPD possibilities. One potentially attractive approach applies precise geometric and combinatorial rules of construction to create function evaluation "centers" in a hypercube using extreme points, faces, and line segments [19-21]. Another improvement might be implementations that shrink $\Omega$ on successive independent runs as described above for $f_{12}$. On the theoretical front, gravitationally trapped Near Earth Objects may lead to a deeper understanding of the metaheuristic, as might alternative formulations involving the concepts of kinetic or total energy for masses moving in a gravitational field. These observations and the results reported here clearly show that CFO merits further study, and that there are many areas of potentially fruitful research. Hopefully this note will encourage talented researchers to become involved in CFO's further development.

The complete source code listing of the program used to generate the results reported here appears in the Appendix. The listing is available in electronic format upon request to the author (rf2@ieee.org).

*4 March 2010*

*Ver. 2, 20 March 2010* [A. Corrects typographical error in pseudocode in Fig. 1. "Very large" negative number should be $-10^{+4200}$ as shown in the source code listing page 17 instead of $-10^{-4200}$ as shown in error in the original version. B. Adds discussion after equation (1) for the case $\vec{R}_{j-1}^{k} = \vec{R}_{j-1}^{p}$ and modifies source code listing at the top of page 18 accordingly.]

[1] Richard A. Formato, JD, PhD
Registered Patent Attorney & Consulting Engineer
P.O. Box 1714, Harwich, MA 02645 USA
rf2@ieee.org

- - - - - - - - - -

# Appendix: CFO Source Code Listing

```
'Program 'CFO_02-27-2010(arXiv#5).BAS' compiled with
'Power Basic/Windows Compiler 9.03 (www.PowerBasic.com).

'THIS IS THE PROGRAM USED TO GENERATE DATA FOR arXiv PAPER #5
'"Parameter-Free Global Search Using Central Force Optimization"

'March 2010.

'LAST MOD 03-03-2010 -1728 HRS EST

'CHANGES MADE TO "CFO_11-26-09.BAS" TO CREATE THIS VERSION
'WHICH IS USED FOR arXiv CFO PSEUDORANDOMNESS PAPER VER #2:
'   (1).  OUTER LOOP Np/Nd = 2 to MaxProbesPerAxis by 2 ADDED
'   (2).  Np IGNORED & ARRAYS REDIMENSIONED AS REQUIRED BY (1).

'ADDITIONAL CHANGES MADE TO CFO_12-25-09.BAS TO CREATE THIS VERSION:
'   (1).  LOOPS ADDED TO VARY G AND ALPHA
'   (2).  SHRINK DS EVERY ?? STEPS INSTEAD OF 20
'   (3).  TERMINATE EARLY USING ?? STEP AVERAGE INSTEAD OF 50?

'IMPORTANT CHANGE MADE 01-15-10: MOVED REDIM STATEMENT INTO SUB CFO().
'THIS IS NECESSARY TO PROPERLY RE-INITIALIZE ALL MATRICES...

'SEPARATE ROSENBROCK & SPHERE FUNCTIONS ADDED 02-10-10.

'PROBE RETRIEVAL SCHEME MODIFIED 02-07-2010 PER MAHAMED 0MRAN's SUGGESTION.

'# STEPS FOR FITNESS SATURATION = 25, 02-07-2010.

'02-21-10: ADDED 3D IPD PLOTS FOR 3D FUNCTIONS.

'03-03-10: CHANGED VARIABLE NumProbesPerAxis% to NumProbesPerDimension%
'          AND  MaxProbesPerAxis% to MaxProbesPerDimension%.

'NOTE: ALL PBM FUNCTIONS HAVE WIRE RADIUS SET TO 0.00001LAMBDA
'=================================================================

'THIS PROGRAM IMPLEMENTS A SIMPLE VERSION OF "CENTRAL
'FORCE OPTIMIZATION."  IT IS DISTRIBUTED FREE OF CHARGE
'TO INCREASE AWARENESS OF CFO AND TO ENCOURAGE EXPERI-
'MENTATION WITH THE ALGORITHM.

'CFO IS A MULTIDIMENSIONAL SEARCH AND OPTIMIZATION
'ALGORITHM THAT LOCATES THE GLOBAL MAXIMA OF A FUNCTION.
'UNLIKE MOST OTHER ALGORITHMS, CFO IS COMPLETELY DETERMIN-
'ISTIC, SO THAT EVERY RUN WITH THE SAME SETUP PRODUCES
'THE SAME RESULTS.

'Please email questions, comments, and significant
'results to: CFO_questions@yahoo.com.  Thanks!

'(c) 2006-2010 Richard A. Formato

'ALL RIGHTS RESERVED WORLDWIDE

'THIS PROGRAM IS FREEWARE.  IT MAY BE COPIED AND
'DISTRIBUTED WITHOUT LIMITATION AS LONG AS THIS
'COPYRIGHT NOTICE AND THE GNUPLOT AND REFERENCE
'INFORMATION BELOW ARE INCLUDED WITHOUT MODIFICATION,
'AND AS LONG AS NO FEE OR COMPENSATION IS CHARGED,
'INCLUDING "TIE-IN" OR "BUNDLING" FEES CHARGED FOR
'OTHER PRODUCTS.

'-------------------------------------------------

'THIS PROGRAM REQUIRES wgnuplot.exe TO DISPLAY PLOTS.
'Gnuplot is a copyrighted freeware plotting program
'available at http://www.gnuplot.info/index.html.

'IT ALSO REQUIRES A VERSION OF THE Numerical Electromagnetics
'Code (NEC) in order to run the PBM benchmarks.  If this file
'is not present, a runtime error occurs.  Remove the code
'that checks for the NEC EXE is there is no interest in the
'PBM functions.

'-------------------------------------------------

'REFERENCES
'----------
'Suggestion:  Read references 1 and 2 first (both available online at no charge)...

'1. "Central Force Optimization: A New Metaheuristic with Applications in Applied Electromagnetics,"
'    Progress in Electromagnetics Research, PIER 77, 425-491, 2007 (online). DOI:10.2528/PIER07082403
'Abstract - Central Force Optimization (CFO) is a new deterministic multi-dimensional search
'metaheuristic based on the metaphor of gravitational kinematics.  It models "probes" that "fly"
'through the decision space by analogy to masses moving under the influence of gravity.  Equations
'are developed for the probes' positions and accelerations using the analogy of particle motion
'in a gravitational field.  In the physical universe, objects traveling through three-dimensional
'space become trapped in close orbits around highly gravitating masses, which is analogous to
'locating the maximum value of an objective function.  In the CFO metaphor, "mass" is a user-defined
'function of the value of the objective function to be maximized.  CFO is readily implemented in a
'compact computer program, and sample pseudocode is presented.  As tests of CFO's effectiveness,
'an equalizer is designed for the well-known Fano load, and a 32-element linear array is synthesized.
'CFO results are compared to several other optimization methods.

'2.  "Pseudorandomness in Central Force Optimization"
'     arXiv:1001.0317v1[cs.NE], online; www.arXiv.org
'Abstract - Central Force Optimization is a deterministic metaheuristic for an evolutionary algorithm
'that searches a decision space by flying probes whose trajectories are computed using a gravitational
'metaphor.  CFO benefits substantially from the inclusion of a pseudorandom component (a numerical
'sequence that is precisely known by specification or calculation but otherwise arbitrary).  The essential
'requirement is that the sequence is uncorrelated with the decision space topology, so that its
'effect is to pseudorandomly distribute probes throughout the landscape.  While this process may appear
'to be similar to the randomness in an inherently stochastic algorithm, it is in fact fundamentally
'different because CFO remains deterministic at every step.  Three pseudorandom methods are discussed
'(initial probe distribution, repositioning factor, and decision space adaptation).  A sample problem
'is presented in detail and summary data included for a 23-function benchmark suite.  CFO's performance
'is quite good compared to other highly developed, state-of-the-art algorithms.

'3.  "Central Force Optimization: A New Nature Inspired Computational Framework for Multidimensional
'     Search and Optimization," Studies in Computational Intelligence (SCI), vol. 129, 221-238 (2008),
'     www.springerlink.com, Springer-Verlag Berlin Heidelberg 2008.
'Abstract - This paper presents Central Force Optimization, a novel, nature inspired, deterministic
'search metaheuristic for constrained multi-dimensional optimization.  CFO is based on the metaphor
'of gravitational kinematics.  Equations are presented for the positions and accelerations experienced
'by "probes" that "fly" through the decision space by analogy to masses moving under the influence of
'gravity.  In the physical universe, probe satellites become trapped in close orbits around highly
'gravitating masses.  In the CFO analogy, "mass" corresponds to a user-defined function of the value
'of an objective function to be maximized.  CFO is a simple algorithm that is easily implemented in
'a compact computer program.  A typical CFO implementation is applied to several test functions.
'CFO exhibits very good performance, suggesting that it merits further study.
```

```
#COMPILE EXE

#DIM ALL

%USEMACROS = 1

#INCLUDE "Win32API.inc"

DEFEXT A-Z

'------ EQUATES -----
%IDC_FRAME1     = 101
%IDC_FRAME2     = 102

%IDC_Function_Number1  = 121
%IDC_Function_Number2  = 122
%IDC_Function_Number3  = 123
%IDC_Function_Number4  = 124
%IDC_Function_Number5  = 125
%IDC_Function_Number6  = 126
%IDC_Function_Number7  = 127
%IDC_Function_Number8  = 128
%IDC_Function_Number9  = 129
%IDC_Function_Number10 = 130
%IDC_Function_Number11 = 131
%IDC_Function_Number12 = 132
%IDC_Function_Number13 = 133
%IDC_Function_Number14 = 134
%IDC_Function_Number15 = 135
%IDC_Function_Number16 = 136
%IDC_Function_Number17 = 137
%IDC_Function_Number18 = 138
%IDC_Function_Number19 = 139
%IDC_Function_Number20 = 140
%IDC_Function_Number21 = 141
%IDC_Function_Number22 = 142
%IDC_Function_Number23 = 143
%IDC_Function_Number24 = 144
%IDC_Function_Number25 = 145
%IDC_Function_Number26 = 146
%IDC_Function_Number27 = 147
%IDC_Function_Number28 = 148
%IDC_Function_Number29 = 149
%IDC_Function_Number30 = 150
%IDC_Function_Number31 = 151
%IDC_Function_Number32 = 152
%IDC_Function_Number33 = 153
%IDC_Function_Number34 = 154
%IDC_Function_Number35 = 155
%IDC_Function_Number36 = 156
%IDC_Function_Number37 = 157
```

```
%IDC_Function_Number38 = 158
%IDC_Function_Number39 = 159
%IDC_Function_Number40 = 160
%IDC_Function_Number41 = 161
%IDC_Function_Number42 = 162
%IDC_Function_Number43 = 163
%IDC_Function_Number44 = 164
%IDC_Function_Number45 = 165
%IDC_Function_Number46 = 166
%IDC_Function_Number47 = 167
%IDC_Function_Number48 = 168
%IDC_Function_Number49 = 169
%IDC_Function_Number50 = 170

'---------------------------- GLOBAL CONSTANTS & SYMBOLS ----------------------------

GLOBAL XiMin(), XiMax(), DiagLength, StartingXiMin(), StartingXiMax() AS EXT 'decision space boundaries, length of diagonal

GLOBAL Aij() AS EXT 'array for Shekel's Foxholes function

GLOBAL EulerConst, Pi, Pi2, Pi4, TwoPi, FourPi, e, Root2 AS EXT 'mathematical constants

GLOBAL Alphabet$, Digits$, RunID$ 'upper/lower case alphabet, digits 0-9 & Run ID

GLOBAL Quote$, SpecialCharacters$ 'quotation mark & special symbols

GLOBAL Mu0, Eps0, c, eta0 AS EXT  'E&M constants

GLOBAL Rad2Deg, Deg2Rad, Feet2Meters, Meters2Feet, Inches2Meters, Meters2Inches AS EXT 'conversion factors

GLOBAL Miles2Meters, Meters2Miles, NautMi2Meters, Meters2NautMi AS EXT              'conversion factors

GLOBAL ScreenWidth&, ScreenHeight& 'screen width & height

GLOBAL xOffset&, yOffset&      'offsets for probe plot windows

GLOBAL FunctionNumber%

GLOBAL AddNoiseToPBM2$

'---------------------------- TEST FUNCTION DECLARATIONS ----------------------------

DECLARE FUNCTION F1(R(),Nd%,p%,j&)            'F1 (n-D)

DECLARE FUNCTION F2(R(),Nd%,p%,j&)            'F2(n-D)

DECLARE FUNCTION F3(R(),Nd%,p%,j&)            'F3 (n-D)

DECLARE FUNCTION F4(R(),Nd%,p%,j&)            'F4 (n-D)

DECLARE FUNCTION F5(R(),Nd%,p%,j&)            'F5 (n-D)

DECLARE FUNCTION F6(R(),Nd%,p%,j&)            'F6 (n-D)

DECLARE FUNCTION F7(R(),Nd%,p%,j&)            'F7 (n-D)

DECLARE FUNCTION F8(R(),Nd%,p%,j&)            'F8 (n-D)

DECLARE FUNCTION F9(R(),Nd%,p%,j&)            'F9 (n-D)

DECLARE FUNCTION F10(R(),Nd%,p%,j&)           'F10 (n-D)

DECLARE FUNCTION F11(R(),Nd%,p%,j&)           'F11 (n-D)

DECLARE FUNCTION F12(R(),Nd%,p%,j&)           'F12 (n-D)

DECLARE FUNCTION u(Xi,a,k,m)                  'Auxiliary function for F12 & F13

DECLARE FUNCTION F13(R(),Nd%,p%,j&)           'F13 (n-D)

DECLARE FUNCTION F14(R(),Nd%,p%,j&)           'F14 (n-D)

DECLARE FUNCTION F15(R(),Nd%,p%,j&)           'F15 (n-D)

DECLARE FUNCTION F16(R(),Nd%,p%,j&)           'F16 (n-D)

DECLARE FUNCTION F17(R(),Nd%,p%,j&)           'F17 (n-D)

DECLARE FUNCTION F18(R(),Nd%,p%,j&)           'F18 (n-D)

DECLARE FUNCTION F19(R(),Nd%,p%,j&)           'F19 (n-D)

DECLARE FUNCTION F20(R(),Nd%,p%,j&)           'F20 (n-D)

DECLARE FUNCTION F21(R(),Nd%,p%,j&)           'F21 (n-D)

DECLARE FUNCTION F22(R(),Nd%,p%,j&)           'F22 (n-D)

DECLARE FUNCTION F23(R(),Nd%,p%,j&)           'F23 (n-D)

DECLARE FUNCTION F24(R(),Nd%,p%,j&)           'F24 (n-D)

DECLARE FUNCTION F25(R(),Nd%,p%,j&)           'F25 (n-D)

DECLARE FUNCTION F26(R(),Nd%,p%,j&)           'F26 (n-D)

DECLARE FUNCTION F27(R(),Nd%,p%,j&)           'F27 (n-D)

DECLARE FUNCTION ParrottP4(R(),Nd%,p%,j&)     'Parrott F4 (1-D)

DECLARE FUNCTION SGO(R(),Nd%,p%,j&)           'SGO Function (2-D)

DECLARE FUNCTION GoldsteinPrice(R(),Nd%,p%,j&) 'Goldstein-Price Function (2-D)

DECLARE FUNCTION StepFunction(R(),Nd%,p%,j&)  'Step Function (n-D)

DECLARE FUNCTION Schwefel226(R(),Nd%,p%,j&)   'Schwefel Prob. 2.26 (n-D)

DECLARE FUNCTION Colville(R(),Nd%,p%,j&)      'Colville Function (4-D)

DECLARE FUNCTION Griewank(R(),Nd%,p%,j&)      'Griewank (n-D)

DECLARE FUNCTION Himmelblau(R(),Nd%,p%,j&)    'Himmelblau (2-D)

DECLARE FUNCTION Rosenbrock(R(),Nd%,p%,j&)    'Rosenbrock (n-D)

DECLARE FUNCTION Sphere(R(),Nd%,p%,j&)        'Sphere n-D)

DECLARE FUNCTION HimmelblauNLO(R(),Nd%,p%,j&) 'Himmelblau NLO (5-D)

DECLARE FUNCTION PBM_1(R(),Nd%,p%,j&)         'PBM Benchmark #1

DECLARE FUNCTION PBM_2(R(),Nd%,p%,j&)         'PBM Benchmark #2

DECLARE FUNCTION PBM_3(R(),Nd%,p%,j&)         'PBM Benchmark #3

DECLARE FUNCTION PBM_4(R(),Nd%,p%,j&)         'PBM Benchmark #4
```

```
DECLARE FUNCTION PBM_5(R(),Nd%,p%,j&)          'PBM Benchmark #5

'------------------------------- SUB DECLARATIONS ------------------------------------

DECLARE SUB CopyBestMatrices(Np%,Nd%,Nt&,R(),M(),Rbest(),Mbest())

DECLARE SUB CheckNECFiles(NECfileError$)

DECLARE SUB GetTestFunctionNumber(FunctionName$)

DECLARE SUB FillArrayAij

DECLARE SUB Plot3DbestProbeTrajectories(NumTrajectories%,M(),R(),Np%,Nd%,LastStep&,FunctionName$)

DECLARE SUB Plot2DbestProbeTrajectories(NumTrajectories%,M(),R(),Np%,Nd%,LastStep&,FunctionName$)

DECLARE SUB Plot2DindividualProbeTrajectories(NumTrajectories%,M(),R(),Np%,Nd%,LastStep&,FunctionName$)

DECLARE SUB Show2Dprobes(R(),Np%,Nt&,j&,Frep,BestFitness,BestProbeNumber%,BestTimeStep&,FunctionName$,RepositionFactor$,Gamma)

DECLARE SUB Show3Dprobes(R(),Np%,Nd%,Nt&,j&,Frep,BestFitness,BestProbeNumber%,BestTimeStep&,FunctionName$,RepositionFactor$,Gamma)

DECLARE SUB StatusWindow(FunctionName$,StatusWindowHandle???)

DECLARE SUB PlotResults(FunctionName$,Nd%,Np%,BestFitnessOverall,BestNpNd%,BestGamma,Neval&&,Rbest(),Mbest(),BestProbeNumberOverall%,BestTimeStepOverall&,LastStepBestRun&)

DECLARE SUB
DisplayRunParameters(FunctionName$,Nd%,Np%,Nt&,G,DeltaT,Alpha,Beta,Frep,R(),A(),M(),PlaceInitialProbes$,InitialAcceleration$,RepositionFactor$,RunCFO$,ShrinkDS$,CheckForEarlyTermination$)

DECLARE SUB GetBestFitness(M(),Np%,StepNumber&,BestFitness,BestProbeNumber%,BestTimeStep&)

DECLARE SUB
Tabulate1DprobeCoordinates(Max1DprobesPlotted%,Nd%,Np%,LastStep&,G,DeltaT,Alpha,Beta,Frep,R(),M(),PlaceInitialProbes$,InitialAcceleration$,RepositionFactor$,FunctionName$,Gamma)

DECLARE SUB GetPlotAnnotation(PlotAnnotation$,Nd%,Np%,Nt&,G,DeltaT,Alpha,Beta,Frep,M(),PlaceInitialProbes$,InitialAcceleration$,RepositionFactor$,FunctionName$,Gamma)

DECLARE SUB ChangeRunParameters(NumProbesPerDimension%,Np%,Nd%,Nt&,G,Alpha,Beta,DeltaT,Frep,PlaceInitialProbes$,InitialAcceleration$,RepositionFactor$,FunctionName$)

DECLARE SUB CLEANUP

DECLARE SUB
Plot1DprobePositions(Max1DprobesPlotted%,Nd%,Np%,LastStep&,G,DeltaT,Alpha,Beta,Frep,R(),M(),PlaceInitialProbes$,InitialAcceleration$,RepositionFactor$,FunctionName$,Gamma)

DECLARE SUB DisplayMmatrix(Np%,Nt&,M())

DECLARE SUB DisplayMbestMatrix(Np%,Nt&,Mbest())

DECLARE SUB DisplayMmatrixThisTimeStep(Np%,j&,M())

DECLARE SUB DisplayAmatrix(Np%,Nd%,Nt&,A())

DECLARE SUB DisplayAmatrixThisTimeStep(Np%,Nd%,j&,A())

DECLARE SUB DisplayRmatrix(Np%,Nd%,Nt&,R())

DECLARE SUB DisplayRmatrixThisTimeStep(Np%,Nd%,j&,R(),Gamma)

DECLARE SUB DisplayXiMinMax(Nd%,XiMin(),XiMax())

DECLARE SUB DisplayRunParameters2(FunctionName$,Nd%,Np%,Nt&,G,DeltaT,Alpha,Beta,Frep,PlaceInitialProbes$,InitialAcceleration$,RepositionFactor$)

DECLARE SUB PlotBestProbeVsTimeStep(Nd%,Np%,LastStep&,G,DeltaT,Alpha,Beta,Frep,M(),PlaceInitialProbes$,InitialAcceleration$,RepositionFactor$,FunctionName$,Gamma)

DECLARE SUB PlotBestFitnessEvolution(Nd%,Np%,LastStep&,G,DeltaT,Alpha,Beta,Frep,M(),PlaceInitialProbes$,InitialAcceleration$,RepositionFactor$,FunctionName$,Gamma)

DECLARE SUB PlotAverageDistance(Nd%,Np%,LastStep&,G,DeltaT,Alpha,Beta,Frep,M(),PlaceInitialProbes$,InitialAcceleration$,RepositionFactor$,FunctionName$,R(),DiagLength,Gamma)

DECLARE SUB Plot2Dfunction(FunctionName$,R())

DECLARE SUB Plot1Dfunction(FunctionName$,R())

DECLARE SUB GetFunctionRunParameters(FunctionName$,Nd%,Np%,Nt&,G,DeltaT,Alpha,Beta,Frep,R(),A(),M(),DiagLength,PlaceInitialProbes$,InitialAcceleration$,RepositionFactor$)

DECLARE SUB InitialProbeDistribution(Np%,Nd%,Nt&,R(),PlaceInitialProbes$,Gamma)

DECLARE SUB RetrieveErrantProbes(Np%,Nd%,j&,R(),Frep)

DECLARE SUB
CFO(FunctionName$,Nd%,Nt&,R(),A(),M(),DiagLength,BestFitnessOverall,BestNpNd%,BestGamma,Neval&&,Rbest(),Mbest(),BestProbeNumberOverall%,BestTimeStepOverall&,LastStepBestRun&
) 'Self-contained CFO routine -> NO USER-SPECIFIED PARAMETERS

DECLARE SUB IFD(Np%,Nd%,Nt&,R(),Gamma)

DECLARE SUB ResetDecisionSpaceBoundaries(Nd%)

DECLARE SUB ThreeDplot(PlotFileName$,PlotTitle$,Annotation$,xCoord$,yCoord$,zCoord$, _
                       XaxisLabel$,YaxisLabel$,ZaxisLabel$,zMin$,zMax$,GnuPlotEXE$,A$)

DECLARE SUB ThreeDplot2(PlotFileName$,PlotTitle$,Annotation$,xCoord$,yCoord$,zCoord$,XaxisLabel$,_
                        YaxisLabel$,ZaxisLabel$,zMin$,zMax$,GnuPlotEXE$,A$,xStart$,xStop$,yStart$,yStop$)

DECLARE SUB ThreeDplot3(PlotFileName$,PlotTitle$,Annotation$,xCoord$,yCoord$,zCoord$, _
                        XaxisLabel$,YaxisLabel$,ZaxisLabel$,zMin$,zMax$,GnuPlotEXE$,xStart$,xStop$,yStart$,yStop$)

DECLARE SUB TwoDplot (PlotFileName$,PlotTitle$,xCoord$,yCoord$,XaxisLabel$,YaxisLabel$, _
                      LogXaxis$,LogYaxis$,xMin$,xMax$,yMin$,yMax$,xTics$,yTics$,GnuPlotEXE$,LineType$,Annotation$)

DECLARE SUB TwoDplot2Curves(PlotFileName1$,PlotFileName2$,PlotTitle$,Annotation$,xCoord$,yCoord$,XaxisLabel$,YaxisLabel$, _
                            LogXaxis$,LogYaxis$,xMin$,xMax$,yMin$,yMax$,xTics$,yTics$,GnuPlotEXE$,LineSize)

DECLARE SUB TwoDplot3curves(NumCurves%,PlotFileName1$,PlotFileName2$,PlotFileName3$,PlotTitle$,Annotation$,xCoord$,yCoord$,XaxisLabel$,YaxisLabel$, _
                            LogXaxis$,LogYaxis$,xMin$,xMax$,yMin$,yMax$,xTics$,yTics$,GnuPlotEXE$)

DECLARE SUB CreateGNUplotINIfile(PlotWindowULC_X%,PlotWindowULC_Y%,PlotWindowWidth%,PlotWindowHeight%)

DECLARE SUB Delay(NumSecs)

DECLARE SUB MathematicalConstants

DECLARE SUB AlphabetAndDigits

DECLARE SUB SpecialSymbols

DECLARE SUB EMconstants

DECLARE SUB ConversionFactors

DECLARE SUB ShowConstants

'------ FUNCTION DECLARATIONS -------

DECLARE FUNCTION SlopeRatio(M(),Np%,StepNumber&)

DECLARE CALLBACK FUNCTION DlgProc
```

```
DECLARE FUNCTION HasFITNESSsaturated$(Nsteps&,j&,Np%,Nd%,M(),R(),DiagLength)

DECLARE FUNCTION HasDAVGsaturated$(Nsteps&,j&,Np%,Nd%,M(),R(),DiagLength)

DECLARE FUNCTION OscillationInDavg$(j&,Np%,Nd%,M(),R(),DiagLength)

DECLARE FUNCTION DavgThisStep(j&,Np%,Nd%,M(),R(),DiagLength)

DECLARE FUNCTION NoSpaces$(X,NumDigits%)

DECLARE FUNCTION FormatFPS(X,Ndigits%)

DECLARE FUNCTION FormatInteger$(M%)

DECLARE FUNCTION TerminateNowForSaturation$(j&,Nd%,Np%,Nt&,G,DeltaT,Alpha,Beta,R(),A(),M())

DECLARE FUNCTION MagVector(V(),N&)

DECLARE FUNCTION UniformDeviate(u&&)

DECLARE FUNCTION RandomNum(a,b)

DECLARE FUNCTION GaussianDeviate(Mu,Sigma)

DECLARE FUNCTION UnitStep(X)

DECLARE FUNCTION Fibonacci&&(N%)

DECLARE FUNCTION ObjectiveFunction(R(),Nd%,p%,j&,FunctionName$)

DECLARE FUNCTION UnitStep(X)
'----------------------------------------------------------------------------------------------------
'----- MAIN PROGRAM ------
FUNCTION PBMAIN () AS LONG
'    ------ CFO Parameters -----
     LOCAL Nd&, Np%, Nt&

     LOCAL G, DeltaT, Alpha, Beta, Frep AS EXT

     LOCAL PlaceInitialProbes$, InitialAcceleration$, RepositionFactor$

     LOCAL R(), A(), M(), Rbest(), Mbest() AS EXT   'position, acceleration & fitness matrices

     LOCAL FunctionName$        'name of objective function
'    ------------------------- Miscellaneous Setup Parameters -----------------------
     LOCAL N%, i%, YN&, Neval&&, NevalTotal&, BestNpNd%, NumTrajectories%, Max1DprobesPlotted%, LastStepBestRun&, Pass%

     LOCAL A$, RunCFO$, CFOversion$, NECfileError$

     LOCAL BestGamma, BestFitnessThisRun, BestFitnessOverall, StartTime, StopTime AS EXT

     LOCAL BestProbeNum%, BestTimeStep&, BestProbeNumberOverall&, BestTimeStepOverall&, StatusWindowHandle???
'    -------------------- Global Constants --------------------
     REDIM Aij(1 TO 2, 1 TO 25) '(GLOBAL array for Shekel's Foxholes function)

     CALL FillArrayAij

     CALL MathematicalConstants 'NOTE: Calling order is important!!

     CALL AlphabetAndDigits

     CALL SpecialSymbols

     CALL EMconstants

     CALL ConversionFactors        ': CALL ShowConstants 'to verify constants have been set
'    -------------------------- General Setup --------------------------
     CFOversion$ = "CFO Ver. 02-27-2010"

     RANDOMIZE TIMER  'seed random number generator with program start time

     DESKTOP GET SIZE TO ScreenWidth&, ScreenHeight&  'get screen size (global variables)

     IF DIR$("wgnuplot.exe") = "" THEN

          MSGBOX("WARNING!  'wgnuplot.exe' not found.  Run terminated.") : EXIT FUNCTION

     END IF
'    ------------------------------------------------------------------ CFO RUN PARAMETERS ------------------------------------------------------------------
     CALL GetTestFunctionNumber(FunctionName$) : exit function 'DEBUG

     CALL GetFunctionRunParameters(FunctionName$,Nd%,Np%,Nt&,G,DeltaT,Alpha,Beta,Frep,R(),A(),M(),DiagLength,PlaceInitialProbes$,InitialAcceleration$,RepositionFactor$)
'NOTE: Parameters returned but not used in this version!!
     REDIM R(1 TO Np%, 1 TO Nd%, 0 TO Nt&), A(1 TO Np%, 1 TO Nd%, 0 TO Nt&), M(1 TO Np%, 0 TO Nt&) 'position, acceleration & fitness matrices

     REDIM Rbest(1 TO Np%, 1 TO Nd%, 0 TO Nt&), Mbest(1 TO Np%, 0 TO Nt&) 'overall best position & fitness matrices
'    -------- PLOT 1D AND 2D FUNCTIONS ON-SCREEN FOR VISUALIZATION --------
     IF Nd% = 2 AND INSTR(FunctionName$,"PBM_") > 0 THEN

          CALL CheckNECFiles(NECfileError$)

          IF NECfileError$ = "YES" THEN
               EXIT FUNCTION
          ELSE
               MSGBOX("Begin computing plot of function "+FunctionName$+"?  May take a while - be patient...")
          END IF

     END IF

     SELECT CASE Nd%
          CASE 1 : CALL Plot1Dfunction(FunctionName$,R()) : REDIM R(1 TO Np%, 1 TO Nd%, 0 TO Nt&) 'erases coordinate data in R()used to plot function
          CASE 2 : CALL Plot2Dfunction(FunctionName$,R()) : REDIM R(1 TO Np%, 1 TO Nd%, 0 TO Nt&) 'ditto
     END SELECT
'    ------------------------------------------------------------------ RUN CFO ------------------------------------------------------------------
--
     YN& = MSGBOX("RUN CFO ON FUNCTION " + FunctionName$ + "?"+CHR$(13)+CHR$(13)+"Get some coffee & sit back...",%MB_YESNO,"CONFIRM RUN") : IF YN& = %IDYES THEN RunCFO$ =
"YES"
```

```
    IF RunCFO$ = "YES" THEN

        StartTime = TIMER

        CALL
CFO(FunctionName$,Nd%,Nt&,R(),A(),M(),DiagLength,BestFitnessOverall,BestNpNd%,BestGamma,Neval&&,Rbest(),Mbest(),BestProbeNumberOverall%,BestTimeStepOverall&,LastStepBestRun&
)

        StopTime = TIMER

        CALL PlotResults(FunctionName$,Nd%,Np%,BestFitnessOverall,BestNpNd%,BestGamma,Neval&&,Rbest(),Mbest(),BestProbeNumberOverall%,BestTimeStepOverall&,LastStepBestRun&)

        MSGBOX(FunctionName$+CHR$(13)+"Total Function Evaluations = "+STR$(Neval&&)+CHR$(13)+"Runtime = "+STR$(ROUND((Stoptime-StartTime)/3600#,2))+" hrs")

    END IF

ExitPBMAIN:

END FUNCTION 'PBMAIN()

'=============================================================== CFO SUBROUTINE ================================================================
SUB
CFO(FunctionName$,Nd%,Nt&,R(),A(),M(),DiagLength,BestFitnessOverall,BestNpNd%,BestGamma,Neval&&,Rbest(),Mbest(),BestProbeNumberOverall%,BestTimeStepOverall&,LastStepBestRun&
)

LOCAL p%, i%, j& 'Standard Indices: Probe #, Coordinate #, Time Step #

LOCAL Np% 'Number of Probes

LOCAL MaxProbesPerDimension% 'Maximum # probes per dimension (depends on Nd%)

LOCAL k%, L% 'Dummy summation indices

LOCAL NumProbesPerDimension%, NumGammas% 'Probes/dimension on each probe line; # gamma points

LOCAL SumSQ, Denom, Numerator, Gamma, Frep, DeltaFrep AS EXT

LOCAL BestProbeNumber%, BestTimeStep&, LastStep&, BestProbeNumberThisRun%, BestTimeStepThisRun&

LOCAL BestFitness, BestFitnessThisRun AS EXT

LOCAL FitnessSaturation$

'--------------- Initia Parameter Values -----------------

Nt& = 1000 'set to a large value expecting early termination

IF FunctionName$ = "F7" THEN Nt& = 100 'to reduce runtime because this function contains random noise

LastStep& = Nt&

BestFitnessOverall = -1E4200 'very large negative number...

Neval&& = 0

DeltaFrep = 0.1##

SELECT CASE Nd% 'set Np%/Nd% based on Nd% to minimize run time

    CASE   1 TO  6 : MaxProbesPerDimension% = 14
    CASE   7 TO 10 : MaxProbesPerDimension% = 12
    CASE  11 TO 15 : MaxProbesPerDimension% = 10
    CASE  16 TO 20 : MaxProbesPerDimension% = 8
    CASE  21 TO 30 : MaxProbesPerDimension% = 6
    CASE ELSE      : MaxProbesPerDimension% = 4

END SELECT

'  ------------------------- Np/Nd LOOP --------------------

FOR NumProbesPerDimension% = 2 TO MaxProbesPerDimension% STEP 2

Np% = NumProbesPerDimension%*Nd%

'  ------------------------- GAMMA LOOP --------------------

FOR NumGammas% = 1 TO 11

Gamma = (NumGammas%-1)/10## 

REDIM R(1 TO Np%, 1 TO Nd%, 0 TO Nt&), A(1 TO Np%, 1 TO Nd%, 0 TO Nt&), M(1 TO Np%, 1 TO Nd%, 0 TO Nt&) 're-initializes Position Vector/Acceleration/Fitness matrices to zero

'STEP (A1) ------------ Compute Initial Probe Distribution (Step 0)----------------

    CALL IPD(Np%,Nd%,Nt&,R(),Gamma) 'Probe Line IPD intersecting on diagonal at a point determined by Gamma

'STEP (A2) ------------ Compute Initial Fitness Matrix (Step 0) ------------------------

    FOR p% = 1 TO Np% : M(p%,0) = ObjectiveFunction(R(),Nd%,p%,0,FunctionName$) : INCR Neval&& : NEXT p%

'STEP (A3) ------------ Set Initial Probe Accelerations to ZERO (Step 0)----------------

    FOR p% = 1 TO Np% : FOR i% = 1 TO Nd% : A(p%,i%,0) = 0## : NEXT i% : NEXT p%

'STEP (A4) ------------ Initialize Frep ----------------

    Frep = 0.5##

'  ------------------------------------------ LOOP ON TIME STEPS STARTING AT STEP #1 ==========================================

    BestFitnessThisRun = M(1,0)

    FOR j& = 1 TO Nt&

'STEP (B) ----------- Compute Probe Position Vectors for this Time Step --------

        FOR p% = 1 TO Np% : FOR i% = 1 TO Nd% : R(p%,i%,j&) = R(p%,i%,j&-1) + A(p%,i%,j&-1) : NEXT i% : NEXT p% 'note: factor of 1/2 combined with G=2 to produce unity coefficient

'STEP (C) ----------- Retrieve Errant Probes --------------

        CALL RetrieveErrantProbes(Np%,Nd%,j&,R(),Frep)

'STEP (D) ----------- Compute Fitness Matrix for Current Probe Distribution ---------

        FOR p% = 1 TO Np% : M(p%,j&) = ObjectiveFunction(R(),Nd%,p%,j&,FunctionName$) : INCR Neval&& : NEXT p%

'STEP (E) ----------- Compute Accelerations Based on Current Probe Distribution & Fitnesses ----------------

        FOR p% = 1 TO Np%

            FOR i% = 1 TO Nd%

                A(p%,i%,j&) = 0

                FOR k% = 1 TO Np%
```

```
                    IF k% <> p% THEN

                        SumSQ = 0##

                        FOR L% = 1 TO Nd%  : SumSQ = SumSQ + (R(k%,L%,j&)-R(p%,L%,j&))^2 : NEXT L% 'dummy index

                        IF SumSQ <> 0## THEN 'added to Ver. 2, 03-20-2010 [see discussion following eq.(1)]

                            Denom = SQR(SumSQ) : Numerator = UnitStep((M(k%,j&)-M(p%,j&)))*(M(k%,j&)-M(p%,j&))

                            A(p%,i%,j&) = A(p%,i%,j&) + (R(k%,i%,j&)-R(p%,i%,j&))*Numerator/Denom

                        END IF 'added to Ver. 2, 03-20-2010

                    END IF

                NEXT k% 'dummy index

            NEXT i% 'coord (dimension) #

        NEXT p% 'probe #

'  --------- Get Best Fitness & Corresponding Probe # and Time Step ---------

        CALL GetBestFitness(M(),Np%,j&,BestFitness,BestProbeNumber%,BestTimeStep&)

        IF BestFitness >= BestFitnessThisRun THEN

            BestFitnessThisRun = BestFitness : BestProbeNumberThisRun% = BestProbeNumber% : BestTimeStepThisRun& = BestTimeStep&

        END IF

'  ----- Increment Frep -----

        Frep = Frep + DeltaFrep

        IF Frep > 1## THEN Frep = 0.05## 'keep Frep in range [0.05,1]

'  --------- Starting at Step #20 Shrink Decision Space Around Best Probe Every 10th Step ----------

        IF j& MOD 10 = 0 AND j& >= 20 THEN

            FOR i% = 1 TO Nd%  : XiMin(i%) = XiMin(i%)+(R(BestProbeNumber%,i%,BestTimeStep&)-XiMin(i%))/2## : XiMax(i%) = XiMax(i%)-(XiMax(i%)-
R(BestProbeNumber%,i%,BestTimeStep&))/2## : NEXT i%

            CALL RetrieveErrantProbes(Np%,Nd%,j&,R(),Frep) 'TO RETRIEVE PROBES LYING OUTSIDE SHRUNKEN DS 'ADDED 02-07-2010

        END IF

'STEP (F) ----------- Check for Early Run Termination ---------

        IF HasFITNESSsaturated(25,j&,Np%,Nd%,M(),R(),DiagLength) = "YES" THEN

            LastStep& = j& : EXIT FOR

        END IF

    NEXT j& 'END TIME STEP LOOP

'---------------- Best Overall Fitness & Corresponding Parameters --------------------

    IF BestFitnessThisRun >= BestFitnessOverall THEN

        BestFitnessThisRun = BestFitnessThisRun : BestProbeNumberOverall% = BestProbeNumberThisRun% : BestTimeStepOverall& = BestTimeStepThisRun&

        BestNpNd% = NumProbesPerDimension% : BestGamma = Gamma : LastStepBestRun& = LastStep&

        CALL CopyBestMatrices(Np%,Nd%,Nt&,R(),M(),Rbest(),Mbest())

    END IF

'STEP (G) ----- Reset Decision Space Boundaries to Initial Values -----

    CALL ResetDecisionSpaceBoundaries(Nd%)

    NEXT NumGammas% 'END GAMMA LOOP

    NEXT NumProbesPerDimension% 'END Np/Nd LOOP

END SUB 'CFO()

'-------------

SUB PlotResults(FunctionName$,Nd%,Np%,BestFitnessOverall,BestNpNd%,BestGamma,Neval&,Rbest(),Mbest(),BestProbeNumberOverall%,BestTimeStepOverall&,LastStepBestRun&)

    LOCAL LastStep&, BestFitnessProbeNumber%, BestFitnessTimeStep&, NumTrajectories%, Max1DprobesPlotted%, i%

    LOCAL RepositionFactor$, PlaceInitialProbes$, InitialAcceleration$, A$, B$

    LOCAL G, DeltaT, Alpha, Beta, Frep AS EXT

        G = 2## : DeltaT = 1## : Alpha = 1## : Beta = 1## : Frep = 0.5## : RepositionFactor$ = "VARIABLE" : PlaceInitialProbes$ = "UNIFORM ON-AXIS " : InitialAcceleration$ =
"FIXED" 'THESE ARE NOW HARDWIRED IN THE CFO EQUATIONS

        B$ = "" : IF Nd% > 1 THEN B$ = "s"

        A$ = FunctionName$ + CHR$(13) +_
            "Best Fitness = " + REMOVE$(STR$(BestFitnessOverall),ANY" ")                + " returned by" + CHR$(13) +_
            "Probe # "        + REMOVE$(STR$(BestProbeNumberOverall%),ANY" ") +_
            " at Time Step " + REMOVE$(STR$(BestTimeStepOverall&),ANY" ")    + CHR$(13) + CHR$(13) + "P" + REMOVE$(STR$(BestProbeNumberOverall%),ANY" ") + " coordinate" + B$ +
":" + CHR$(13)

        FOR i% = 1 TO Nd% : A$ = A$ + STR$(i%)+"    "+REMOVE$(STR$(ROUND(Rbest(BestProbeNumberOverall%,i%,BestTimeStepOverall&),8)),ANY" ")+CHR$(13) : NEXT i%

        MSGBOX(A$)

'  ------------------------------------------------ PLOT EVOLUTION OF BEST FITNESS, AVG DISTANCE & BEST PROBE # ------------------------------------------------

    CALL
PlotBestFitnessEvolution(Nd%,Np%,LastStepBestRun&,G,DeltaT,Alpha,Beta,Frep,Mbest(),PlaceInitialProbes$,InitialAcceleration$,RepositionFactor$,FunctionName$,BestGamma)

    CALL
PlotAverageDistance(Nd%,Np%,LastStepBestRun&,G,DeltaT,Alpha,Beta,Frep,Mbest(),PlaceInitialProbes$,InitialAcceleration$,RepositionFactor$,FunctionName$,Rbest(),DiagLength,Bes
tGamma)

    CALL
PlotBestProbeVsTimeStep(Nd%,Np%,LastStepBestRun&,G,DeltaT,Alpha,Beta,Frep,Mbest(),PlaceInitialProbes$,InitialAcceleration$,RepositionFactor$,FunctionName$,BestGamma)

'  ------------------------------------------------ PLOT TRAJECTORIES OF BEST PROBES FOR 2/3-D FUNCTIONS ------------------------------------------------

    IF Nd% = 2 THEN

        NumTrajectories% = 10 : CALL Plot2DbestProbeTrajectories(NumTrajectories%,Mbest(),Rbest(),Np%,Nd%,LastStepBestRun&,FunctionName$)

        NumTrajectories% = 16 : CALL Plot2DindividualProbeTrajectories(NumTrajectories%,Mbest(),Rbest(),Np%,Nd%,LastStepBestRun&,FunctionName$)
```

```
        END IF

    IF Nd% = 3 THEN

        NumTrajectories% = 4 : CALL Plot3DbestProbeTrajectories(NumTrajectories%,Mbest(),Rbest(),Np%,Nd%,LastStepBestRun&,FunctionName$)

    END IF

'    ----------- For 1-D Objective Functions, Tabulate Probe Coordinates & if Np% =< Max1DprobesPlotted% Plot Evolution of Probe Positions -----------

    IF Nd% = 1 THEN

        Max1Dprobesplotted = 15

        CALL
Tabulate1DprobeCoordinates(Max1Dprobesplotted%,Nd%,Np%,LastStepBestRun&,G,DeltaT,Alpha,Beta,Frep,Rbest(),Mbest(),PlaceInitialProbes$,InitialAcceleration$,RepositionFactor$,FunctionName$,BestGamma)

        IF Np% =< Max1DprobesPlotted% THEN _
        CALL
Plot1DprobePositions(Max1Dprobesplotted%,Nd%,Np%,LastStepBestRun&,G,DeltaT,Alpha,Beta,Frep,Rbest(),Mbest(),PlaceInitialProbes$,InitialAcceleration$,RepositionFactor$,FunctionName$,BestGamma)

        CALL CLEANUP 'delete probe coordinate plot files, if any

    END IF

END SUB 'PlotResults()

'--------------------

SUB IPD(Np%,Nd%,Nt&,R(),Gamma)

LOCAL DeltaXi, DelX1, DelX2, Di AS EXT

LOCAL NumProbesPerDimension%, p%, i%, k%, NumX1points%, NumX2points%, x1pointNum%, x2pointNum%

            IF Nd% > 1 THEN

                NumProbesPerDimension% = Np%\Nd% 'even #

            ELSE

                NumProbesPerDimension% = Np%

            END IF

            FOR i% = 1 TO Nd%

                FOR p% = 1 TO Np%

                    R(p%,i%,0) = XiMin(i%) + Gamma*(XiMax(i%)-XiMin(i%))

                NEXT Np%

            NEXT i%

            FOR i% = 1 TO Nd% 'place probes probe line-by-probe line (i% is dimension number)

                DeltaXi = (XiMax(i%)-XiMin(i%))/(NumProbesPerDimension%-1)

                FOR k% = 1 TO NumProbesPerDimension%

                    p% = k% + NumProbesPerDimension%*(i%-1) 'probe #

                    R(p%,i%,0) = XiMin(i%) + (k%-1)*DeltaXi

                NEXT k%

            NEXT i%

END SUB 'IPD()

'----

FUNCTION HasFITNESSsaturated$(Nsteps&,j&,Np%,Nd%,M(),R(),DiagLength)

LOCAL A$

LOCAL k&, p%

LOCAL BestFitness, SumOfBestFitnesses, BestFitnessStepJ, FitnessSatTOL AS EXT

    A$ = "NO"

    FitnessSatTOL = 0.000001## 'tolerance for FITNESS saturation

    IF j& < Nsteps& + 10 THEN GOTO ExitHasFITNESSsaturated 'execute at least 10 steps after averaging interval before performing this check

    SumOfBestFitnesses = 0##

    FOR k& = j&-Nsteps&+1 TO j&

        BestFitness = M(k&,1)

        FOR p% = 1 TO Np%

            IF M(p%,k&) >= BestFitness THEN BestFitness = M(p%,k&)

        NEXT p%

        IF k& = j& THEN BestFitnessStepJ = BestFitness

        SumOfBestFitnesses = SumOfBestFitnesses + BestFitness

    NEXT k&

    IF ABS(SumOfBestFitnesses/Nsteps&-BestFitnessStepJ) =< FitnessSatTOL THEN A$ = "YES" 'saturation if (avg value - last value) are within TOL

ExitHasFITNESSsaturated:

    HasFITNESSsaturated$ = A$

END FUNCTION 'HasFITNESSsaturated$()

'------

SUB RetrieveErrantProbes(Np%,Nd%,j&,R(),Frep)

LOCAL p%, i%

    FOR p% = 1 TO Np%

        FOR i% = 1 TO Nd%

            IF R(p%,i%,j&) < XiMin(i%) THEN R(p%,i%,j&) = MAX(XiMin(i%) + Frep*(R(p%,i%,j&-1)-XiMin(i%)),XiMin(i%)) 'CHANGED 02-07-10
```

```
            IF R(p%,i%,j&) > XiMax(i%) THEN R(p%,i%,j&) = MIN(XiMax(i%) - Frep*(XiMax(i%)-R(p%,i%,j&-1)),XiMax(i%))

        NEXT i%

    NEXT p%

END SUB 'RetrieveErrantProbes()

'-----------------------------

SUB ResetDecisionSpaceBoundaries(Nd%)

    LOCAL i%

    FOR i% = 1 TO Nd% : XiMin(i%) = StartingXiMin(i%) : XiMax(i%) = StartingXiMax(i%) : NEXT i%

END SUB

'------

SUB CopyBestMatrices(Np%,Nd%,Nt&,R(),M(),Rbest(),Mbest())

LOCAL p%, i%, j&

REDIM Rbest(1 TO Np%, 1 TO Nd%, 0 TO Nt&), Mbest(1 TO Np%, 0 TO Nt&) 're-initializes Best Position Vetor/Fitness matrices to zero

    FOR p% = 1 TO Np%

        FOR i% = 1 TO Nd%

            FOR j& = 0 TO Nt&

                Rbest(p%,i%,j&) = R(p%,i%,j&) : Mbest(p%,j&) = M(p%,j&)

            NEXT j&

        NEXT i%

    NEXT p%

END SUB

'----------------------------------------------------------------------------------------------------------------------------------------

SUB CheckNECFiles(NECfileError$)

LOCAL N%

    NECfileError$ = "NO"

'    ----------------- NEC Files Required for PBM Antenna Benchmarks ------------------

    IF DIR$("n4l_2k1.exe") = "" THEN

        MSGBOX("WARNING!  'n4l_2k1.exe' not found.  Run terminated.") : NECfileError$ = "YES" : EXIT SUB

    END IF

    N% = FREEFILE : OPEN "ENDERR.DAT" FOR OUTPUT AS #N%  : PRINT #N%, "NO"      : CLOSE #N%

    N% = FREEFILE : OPEN "FILE_MSG.DAT" FOR OUTPUT AS #N% : PRINT #N%, "NO"     : CLOSE #N%

    N% = FREEFILE : OPEN "NHSCALE.DAT"  FOR OUTPUT AS #N% : PRINT #N%, "0.00001" : CLOSE #N%

END SUB

'------

FUNCTION SlopeRatio(M(),Np%,StepNumber&)

LOCAL p% 'probe #

LOCAL BestFitnessAtStepNumber, BestFitnessAtStepNumberMinus1, BestFitnessAtStepNumberMinus2, Z AS EXT

    IF StepNumber& < 5 THEN 'at least five steps to perform this test

        Z = 1## : GOTO ExitSlopeRatio 'assumes no slope change

    END IF

    BestFitnessAtStepNumber       = M(1,StepNumber&)   : FOR p% = 1 TO Np% : IF M(p%,StepNumber&)   >= BestFitnessAtStepNumber       THEN BestFitnessAtStepNumber       = M(p%,StepNumber&)   : NEXT p%

    BestFitnessAtStepNumberMinus1 = M(1,StepNumber&-1) : FOR p% = 1 TO Np% : IF M(p%,StepNumber&-1) >= BestFitnessAtStepNumberMinus1 THEN BestFitnessAtStepNumberMinus1 = M(p%,StepNumber&-1) : NEXT p%

    BestFitnessAtStepNumberMinus2 = M(1,StepNumber&-2) : FOR p% = 1 TO Np% : IF M(p%,StepNumber&-2) >= BestFitnessAtStepNumberMinus2 THEN BestFitnessAtStepNumberMinus2 = M(p%,StepNumber&-2) : NEXT p%

    Z = (BestFitnessAtStepNumber-BestFitnessAtStepNumberMinus1)/(BestFitnessAtStepNumberMinus1-BestFitnessAtStepNumberMinus2)

ExitSlopeRatio:

    SlopeRatio = Z

END FUNCTION

'------

SUB GetBestFitness(M(),Np%,StepNumber&,BestFitness,BestProbeNumber%,BestTimeStep&)

LOCAL p%, i&, A$

    BestFitness = M(1,0)

    FOR i& = 0 TO StepNumber&

        FOR p% = 1 TO Np%

            IF M(p%,i&) >= BestFitness THEN

                BestFitness = M(p%,i&) : BestProbeNumber% = p% : BestTimeStep& = i&

            END IF

        NEXT p%

    NEXT i&

END SUB

'======================================================== FUNCTION DEFINITIONS =========================================================

FUNCTION ObjectiveFunction(R(),Nd%,p%,j&,FunctionName$) 'Objective function to be MAXIMIZED is defined here

    SELECT CASE FunctionName$
```

```basic
        CASE "ParrottF4"    : ObjectiveFunction = ParrottF4(R(),Nd%,p%,j&)        'Parrott F4 (1-D)

        CASE "SGO"          : ObjectiveFunction = SGO(R(),Nd%,p%,j&)              'SGO Function (2-D)

        CASE "GP"           : ObjectiveFunction = GoldsteinPrice(R(),Nd%,p%,j&)   'Goldstein-Price Function (2-D)

        CASE "STEP"         : ObjectiveFunction = StepFunction(R(),Nd%,p%,j&)     'Step Function (n-D)

        CASE "SCHWEFEL_226"  : ObjectiveFunction = Schwefel226(R(),Nd%,p%,j&)     'Schwefel Prob. 2.26 (n-D)

        CASE "COLVILLE"     : ObjectiveFunction = Colville(R(),Nd%,p%,j&)         'Colville Function (4-D)

        CASE "GRIEWANK"     : ObjectiveFunction = Griewank(R(),Nd%,p%,j&)         'Griewank Function (n-D)

        CASE "HIMMELBLAU"   : ObjectiveFunction = Himmelblau(R(),Nd%,p%,j&)       'Himmelblau Function (2-D)

        CASE "ROSENBROCK"   : ObjectiveFunction = Rosenbrock(R(),Nd%,p%,j&)       'Rosenbrock Function (n-D)

        CASE "SPHERE"       : ObjectiveFunction = Sphere(R(),Nd%,j&)              'Sphere Function (n-D)

        CASE "HIMMELBLAUNLO" : ObjectiveFunction = HIMMELBLAUNLO(R(),Nd%,p%,j&)   'Himmelblau NLO (5-D)

'       ------------------------ GSO Paper Benchmark Functions -------------------------
        CASE "F1"           : ObjectiveFunction = F1(R(),Nd%,p%,j&)               'F1  (n-D)
        CASE "F2"           : ObjectiveFunction = F2(R(),Nd%,p%,j&)               'F2  (n-D)
        CASE "F3"           : ObjectiveFunction = F3(R(),Nd%,p%,j&)               'F3  (n-D)
        CASE "F4"           : ObjectiveFunction = F4(R(),Nd%,p%,j&)               'F4  (n-D)
        CASE "F5"           : ObjectiveFunction = F5(R(),Nd%,p%,j&)               'F5  (n-D)
        CASE "F6"           : ObjectiveFunction = F6(R(),Nd%,p%,j&)               'F6  (n-D)
        CASE "F7"           : ObjectiveFunction = F7(R(),Nd%,p%,j&)               'F7  (n-D)
        CASE "F8"           : ObjectiveFunction = F8(R(),Nd%,p%,j&)               'F8  (n-D)
        CASE "F9"           : ObjectiveFunction = F9(R(),Nd%,p%,j&)               'F9  (n-D)
        CASE "F10"          : ObjectiveFunction = F10(R(),Nd%,p%,j&)              'F10 (n-D)
        CASE "F11"          : ObjectiveFunction = F11(R(),Nd%,p%,j&)              'F11 (n-D)
        CASE "F12"          : ObjectiveFunction = F12(R(),Nd%,p%,j&)              'F12 (n-D)
        CASE "F13"          : ObjectiveFunction = F13(R(),Nd%,p%,j&)              'F13 (n-D)
        CASE "F14"          : ObjectiveFunction = F14(R(),Nd%,p%,j&)              'F14 (2-D)
        CASE "F15"          : ObjectiveFunction = F15(R(),Nd%,p%,j&)              'F15 (4-D)
        CASE "F16"          : ObjectiveFunction = F16(R(),Nd%,p%,j&)              'F16 (2-D)
        CASE "F17"          : ObjectiveFunction = F17(R(),Nd%,p%,j&)              'F17 (2-D)
        CASE "F18"          : ObjectiveFunction = F18(R(),Nd%,p%,j&)              'F18 (2-D)
        CASE "F19"          : ObjectiveFunction = F19(R(),Nd%,p%,j&)              'F19 (3-D)
        CASE "F20"          : ObjectiveFunction = F20(R(),Nd%,p%,j&)              'F20 (6-D)
        CASE "F21"          : ObjectiveFunction = F21(R(),Nd%,p%,j&)              'F21 (4-D)
        CASE "F22"          : ObjectiveFunction = F22(R(),Nd%,p%,j&)              'F22 (4-D)
        CASE "F23"          : ObjectiveFunction = F23(R(),Nd%,p%,j&)              'F23 (4-D)

'       ------------------------- PBM Antenna Benchmarks ------------------------------
        CASE "PBM_1"        : ObjectiveFunction = PBM_1(R(),Nd%,p%,j&)            'PBM_1 (2-D)
        CASE "PBM_2"        : ObjectiveFunction = PBM_2(R(),Nd%,p%,j&)            'PBM_2 (2-D)
        CASE "PBM_3"        : ObjectiveFunction = PBM_3(R(),Nd%,p%,j&)            'PBM_3 (2-D)
        CASE "PBM_4"        : ObjectiveFunction = PBM_4(R(),Nd%,p%,j&)            'PBM_4 (2-D)
        CASE "PBM_5"        : ObjectiveFunction = PBM_5(R(),Nd%,p%,j&)            'PBM_5 (2-D)

    END SELECT

END FUNCTION 'ObjectiveFunction()

'------

SUB GetFunctionRunParameters(FunctionName$,Nd%,Np%,Nt&,G,DeltaT,Alpha,Beta,Frep,R(),A(),M(),_
                             DiagLength,PlaceInitialProbes$,InitialAcceleration$,RepositionFactor$)

LOCAL i%, NumProbesPerDimension%, NN%, NumCollinearElements%

    SELECT CASE FunctionName$

        CASE "ParrottF4"

            Nd%                   = 1
            NumProbesPerDimension% = 3
            Np%                   = NumProbesPerDimension%*Nd%

            Nt&     = 500
            G       = 2##
            Alpha   = 2##
            Beta    = 2##
            DeltaT  = 1##
            Frep    = 0.9##

            PlaceInitialProbes$  = "UNIFORM ON-AXIS"
            InitialAcceleration$ = "ZERO"
            RepositionFactor$    = "FIXED"

            NumProbesPerDimension% = MAX(NumProbesPerDimension%,3) 'at least three for 1-D functions

            Np% = NumProbesPerDimension%*Nd%

            REDIM XiMin(1 TO Nd%), XiMax(1 TO Nd%) : XiMin(1) = 0## : XiMax(1) = 1##

            REDIM StartingXiMin(1 TO Nd%), StartingXiMax(1 TO Nd%) : FOR i% = 1 TO Nd% : StartingXiMin(i%) = XiMin(i%) : StartingXiMax(i%) = XiMax(i%) : NEXT i%

        CASE "SGO"

            Nd%                   = 2
            NumProbesPerDimension% = 4 '10 '4
            Np%                   = NumProbesPerDimension%*Nd%

            Nt&     = 500
            G       = 2##
            Alpha   = 2##
            Beta    = 2##
            DeltaT  = 1##
            Frep    = 0.5##

            PlaceInitialProbes$  = "UNIFORM ON-AXIS" '"2D GRID"
            InitialAcceleration$ = "ZERO"
            RepositionFactor$    = "VARIABLE"

            Np% = NumProbesPerDimension%*Nd%

            REDIM XiMin(1 TO Nd%), XiMax(1 TO Nd%) : FOR i% = 1 TO Nd% : XiMin(i%) = -50## : XiMax(i%) = 50## : NEXT i%

            REDIM StartingXiMin(1 TO Nd%), StartingXiMax(1 TO Nd%) : FOR i% = 1 TO Nd% : StartingXiMin(i%) = XiMin(i%) : StartingXiMax(i%) = XiMax(i%) : NEXT i%

            IF PlaceInitialProbes$ = "2D GRID" THEN

                Np% = NumProbesPerDimension%^2 : REDIM R(1 TO Np%, 1 TO Nd%, 0 TO Nt&) 'to create (Np/Nd) x (Np/Nd) grid

            END IF

        CASE "GP"

            Nd%                   = 2
            NumProbesPerDimension% = 4 '10
```

```
    Np%                     = NumProbesPerDimension%*Nd%

    Nt&      = 500
    G        = 2##
    Alpha    = 2## '0.2##
    Beta     = 2##
    DeltaT   = 1##
    Frep     = 0.5## '0.8## '0.9##

    PlaceInitialProbes$ = "UNIFORM ON-AXIS" '"2D GRID"
    InitialAcceleration$ = "ZERO"
    RepositionFactor$    = "VARIABLE"

    Np% = NumProbesPerDimension%*Nd%

    REDIM XiMin(1 TO Nd%), XiMax(1 TO Nd%) : FOR i% = 1 TO Nd% : XiMin(i%) = -100## : XiMax(i%) = 100## : NEXT i%

    REDIM StartingXiMin(1 TO Nd%), StartingXiMax(1 TO Nd%) : FOR i% = 1 TO Nd% : StartingXiMin(i%) = XiMin(i%) : StartingXiMax(i%) = XiMax(i%) : NEXT i%

    IF PlaceInitialProbes$ = "2D GRID" THEN

        Np% = NumProbesPerDimension%^2 : REDIM R(1 TO Np%, 1 TO Nd%, 0 TO Nt&) 'to create (Np/Nd) x (Np/Nd) grid

    END IF

CASE "STEP"

    Nd%                     = 2
    NumProbesPerDimension% = 4 '20
    Np%                     = NumProbesPerDimension%*Nd%

    Nt&      = 500
    G        = 2##
    Alpha    = 2##
    Beta     = 2##
    DeltaT   = 1##
    Frep     = 0.5##

    PlaceInitialProbes$ = "UNIFORM ON-AXIS"
    InitialAcceleration$ = "ZERO"
    RepositionFactor$    = "VARIABLE" '"FIXED"

'    CALL ChangeRunParameters(NumProbesPerDimension%,Np%,Nd%,Nt&,G,Alpha,Beta,DeltaT,Frep,PlaceInitialProbes$,InitialAcceleration$,RepositionFactor$,FunctionName$)

    Np% = NumProbesPerDimension%*Nd%

    REDIM XiMin(1 TO Nd%), XiMax(1 TO Nd%) : FOR i% = 1 TO Nd% : XiMin(i%) = -100## : XiMax(i%) = 100## : NEXT i%

'   REDIM XiMin(1 TO Nd%), XiMax(1 TO Nd%) : XiMin(1) = 72## : XiMax(1) = 78## : XiMin(2) = 27## : XiMax(2) = 33## 'use this to plot STEP detail

    REDIM StartingXiMin(1 TO Nd%), StartingXiMax(1 TO Nd%) : FOR i% = 1 TO Nd% : StartingXiMin(i%) = XiMin(i%) : StartingXiMax(i%) = XiMax(i%) : NEXT i%

    IF PlaceInitialProbes$ = "2D GRID" THEN

        Np% = NumProbesPerDimension%^2 : REDIM R(1 TO Np%, 1 TO Nd%, 0 TO Nt&) 'to create (Np/Nd) x (Np/Nd) grid

    END IF 'STEP

CASE "SCHWEFEL_226"

    Nd%                     = 30
    NumProbesPerDimension% = 4
    Np%                     = NumProbesPerDimension%*Nd%

    Nt&      = 500
    G        = 2##
    Alpha    = 2##
    Beta     = 2##
    DeltaT   = 1##
    Frep     = 0.5##

    PlaceInitialProbes$  = "UNIFORM ON-AXIS"
    InitialAcceleration$ = "ZERO"
    RepositionFactor$    = "VARIABLE"

    Np% = NumProbesPerDimension%*Nd%

    REDIM XiMin(1 TO Nd%), XiMax(1 TO Nd%) : FOR i% = 1 TO Nd% : XiMin(i%) = -500## : XiMax(i%) = 500## : NEXT i%

    REDIM StartingXiMin(1 TO Nd%), StartingXiMax(1 TO Nd%) : FOR i% = 1 TO Nd% : StartingXiMin(i%) = XiMin(i%) : StartingXiMax(i%) = XiMax(i%) : NEXT i%

    IF PlaceInitialProbes$ = "2D GRID" THEN

        Np% = NumProbesPerDimension%^2 : REDIM R(1 TO Np%, 1 TO Nd%, 0 TO Nt&) 'to create (Np/Nd) x (Np/Nd) grid

    END IF

CASE "COLVILLE"

    Nd%                     = 4
    NumProbesPerDimension% = 4 '14
    Np%                     = NumProbesPerDimension%*Nd%

    Nt&      = 500
    G        = 2##
    Alpha    = 2##
    Beta     = 2##
    DeltaT   = 1##
    Frep     = 0.5##

    PlaceInitialProbes$  = "UNIFORM ON-AXIS"
    InitialAcceleration$ = "ZERO"
    RepositionFactor$    = "VARIABLE"

    Nd% = 4 'cannot change dimensionality of Colville function!

    Np% = NumProbesPerDimension%*Nd%

    REDIM XiMin(1 TO Nd%), XiMax(1 TO Nd%) : FOR i% = 1 TO Nd% : XiMin(i%) = -10## : XiMax(i%) = 10## : NEXT i%

    REDIM StartingXiMin(1 TO Nd%), StartingXiMax(1 TO Nd%) : FOR i% = 1 TO Nd% : StartingXiMin(i%) = XiMin(i%) : StartingXiMax(i%) = XiMax(i%) : NEXT i%

    IF PlaceInitialProbes$ = "2D GRID" THEN

        Np% = NumProbesPerDimension%^2 : REDIM R(1 TO Np%, 1 TO Nd%, 0 TO Nt&) 'to create (Np/Nd) x (Np/Nd) grid

    END IF

CASE "GRIEWANK"

    Nd%                     = 2
    NumProbesPerDimension% = 4 '14
    Np%                     = NumProbesPerDimension%*Nd%

    Nt&      = 500
    G        = 2##
    Alpha    = 2##
```

```
        Beta    = 2##
        DeltaT  = 1##
        Frep    = 0.5##

        PlaceInitialProbes$  = "UNIFORM ON-AXIS"
        InitialAcceleration$ = "ZERO"
        RepositionFactor$    = "VARIABLE"

        Np% = NumProbesPerDimension%*Nd%

        REDIM XiMin(1 TO Nd%), XiMax(1 TO Nd%) : FOR i% = 1 TO Nd% : XiMin(i%) = -600## : XiMax(i%) = 600## : NEXT i%

        REDIM StartingXiMin(1 TO Nd%), StartingXiMax(1 TO Nd%) : FOR i% = 1 TO Nd% : StartingXiMin(i%) = XiMin(i%) : StartingXiMax(i%) = XiMax(i%) : NEXT i%

        IF PlaceInitialProbes$ = "2D GRID" THEN

            Np% = NumProbesPerDimension%^2 : REDIM R(1 TO Np%, 1 TO Nd%, 0 TO Nt&) 'to create (Np/Nd) x (Np/Nd) grid

        END IF

    CASE "HIMMELBLAU"

        Nd%                    = 2
        NumProbesPerDimension% = 4 '14
        Np%                    = NumProbesPerDimension%*Nd%

        Nt&     = 500
        G       = 2##
        Alpha   = 2##
        Beta    = 2##
        DeltaT  = 1##
        Frep    = 0.5##

        PlaceInitialProbes$  = "UNIFORM ON-AXIS"
        InitialAcceleration$ = "ZERO"
        RepositionFactor$    = "VARIABLE"

        Nd% = 2 'cannot change dimensionality of Himmelblau function!

        Np% = NumProbesPerDimension%*Nd%

        REDIM XiMin(1 TO Nd%), XiMax(1 TO Nd%) : FOR i% = 1 TO Nd% : XiMin(i%) = -6## : XiMax(i%) = 6## : NEXT i%

        REDIM StartingXiMin(1 TO Nd%), StartingXiMax(1 TO Nd%) : FOR i% = 1 TO Nd% : StartingXiMin(i%) = XiMin(i%) : StartingXiMax(i%) = XiMax(i%) : NEXT i%

        IF PlaceInitialProbes$ = "2D GRID" THEN

            Np% = NumProbesPerDimension%^2 : REDIM R(1 TO Np%, 1 TO Nd%, 0 TO Nt&) 'to create (Np/Nd) x (Np/Nd) grid

        END IF 'Himmelblau

    CASE "ROSENBROCK" '(n-D)

        Nd%                    = 2'30
        NumProbesPerDimension% = 4
        Np%                    = NumProbesPerDimension%*Nd%

        Nt&     = 250
        G       = 2##
        Alpha   = 2##
        Beta    = 2##
        DeltaT  = 1##
        Frep    = 0.5##

        PlaceInitialProbes$  = "UNIFORM ON-AXIS"
        InitialAcceleration$ = "ZERO"
        RepositionFactor$    = "VARIABLE"

        Np% = NumProbesPerDimension%*Nd%

        REDIM XiMin(1 TO Nd%), XiMax(1 TO Nd%) : FOR i% = 1 TO Nd% : XiMin(i%) = -2## : XiMax(i%) = 2## : NEXT i% :'XiMin(i%) = -6## : XiMax(i%) = 6## : NEXT i%

        REDIM StartingXiMin(1 TO Nd%), StartingXiMax(1 TO Nd%) : FOR i% = 1 TO Nd% : StartingXiMin(i%) = XiMin(i%) : StartingXiMax(i%) = XiMax(i%) : NEXT i%

        IF PlaceInitialProbes$ = "2D GRID" THEN

            Np% = NumProbesPerDimension%^2 : REDIM R(1 TO Np%, 1 TO Nd%, 0 TO Nt&) 'to create (Np/Nd) x (Np/Nd) grid

        END IF 'ROSENBROCK

    CASE "SPHERE" '(n-D)

        Nd%                    = 2'30
        NumProbesPerDimension% = 4
        Np%                    = NumProbesPerDimension%*Nd%

        Nt&     = 250
        G       = 2##
        Alpha   = 2##
        Beta    = 2##
        DeltaT  = 1##
        Frep    = 0.5##

        PlaceInitialProbes$  = "UNIFORM ON-AXIS"
        InitialAcceleration$ = "ZERO"
        RepositionFactor$    = "VARIABLE"

        Np% = NumProbesPerDimension%*Nd%

        REDIM XiMin(1 TO Nd%), XiMax(1 TO Nd%) : FOR i% = 1 TO Nd% : XiMin(i%) = -100## : XiMax(i%) = 100## : NEXT i%

        REDIM StartingXiMin(1 TO Nd%), StartingXiMax(1 TO Nd%) : FOR i% = 1 TO Nd% : StartingXiMin(i%) = XiMin(i%) : StartingXiMax(i%) = XiMax(i%) : NEXT i%

        IF PlaceInitialProbes$ = "2D GRID" THEN

            Np% = NumProbesPerDimension%^2 : REDIM R(1 TO Np%, 1 TO Nd%, 0 TO Nt&) 'to create (Np/Nd) x (Np/Nd) grid

        END IF 'SPHERE

    CASE "HIMMELBLAUNLO" '(5-D)

        Nd%                    = 5
        NumProbesPerDimension% = 4
        Np%                    = NumProbesPerDimension%*Nd%

        Nt&     = 250
        G       = 2##
        Alpha   = 2##
        Beta    = 2##
        DeltaT  = 1##
        Frep    = 0.5##

        PlaceInitialProbes$  = "UNIFORM ON-AXIS"
        InitialAcceleration$ = "ZERO"
        RepositionFactor$    = "VARIABLE"

        Nd% = 5 'cannot change dimensionality of HimmelblauNLO function!
```

```
            Np% = NumProbesPerDimension%*Nd%

            REDIM XiMin(1 TO Nd%), XiMax(1 TO Nd%)

            XiMin(1) = 78## : XiMax(1) = 102##
            XiMin(2) = 33## : XiMax(2) = 45##
            XiMin(3) = 27## : XiMax(3) = 45##
            XiMin(4) = 27## : XiMax(4) = 45##
            XiMin(5) = 27## : XiMax(5) = 45##

            REDIM StartingXiMin(1 TO Nd%), StartingXiMax(1 TO Nd%) : FOR i% = 1 TO Nd% : StartingXiMin(i%) = XiMin(i%) : StartingXiMax(i%) = XiMax(i%) : NEXT i%

            IF PlaceInitialProbes% = "2D GRID" THEN

                Np% = NumProbesPerDimension%^2 : REDIM R(1 TO Np%, 1 TO Nd%, 0 TO Nt&) 'to create (Np/Nd) x (Np/Nd) grid

            END IF 'HimmelblauNLO

  CASE "F1" '(n-D)

            Nd%                      = 30
            NumProbesPerDimension% = 2
            Np%                      = NumProbesPerDimension%*Nd%

            Nt&     = 1000
            G       = 2##
            Alpha   = 2##
            Beta    = 2##
            DeltaT  = 1##
            Frep    = 0.5##

            PlaceInitialProbes$  = "UNIFORM ON-AXIS"
            InitialAcceleration$ = "ZERO"
            RepositionFactor$    = "VARIABLE"

            Np% = NumProbesPerDimension%*Nd%

            REDIM XiMin(1 TO Nd%), XiMax(1 TO Nd%) : FOR i% = 1 TO Nd% : XiMin(i%) = -100## : XiMax(i%) = 100## : NEXT i%

            REDIM StartingXiMin(1 TO Nd%), StartingXiMax(1 TO Nd%) : FOR i% = 1 TO Nd% : StartingXiMin(i%) = XiMin(i%) : StartingXiMax(i%) = XiMax(i%) : NEXT i%

            IF PlaceInitialProbes$ = "2D GRID" THEN

                Np% = NumProbesPerDimension%^2 : REDIM R(1 TO Np%, 1 TO Nd%, 0 TO Nt&) 'to create (Np/Nd) x (Np/Nd) grid

            END IF 'F1

  CASE "F2" '(n-D)

            Nd%                      = 30
            NumProbesPerDimension% = 2
            Np%                      = NumProbesPerDimension%*Nd%

            Nt&     = 1000
            G       = 2##
            Alpha   = 2##
            Beta    = 2##
            DeltaT  = 1##
            Frep    = 0.5##

            PlaceInitialProbes$  = "UNIFORM ON-AXIS"
            InitialAcceleration$ = "ZERO"
            RepositionFactor$    = "VARIABLE"

            Np% = NumProbesPerDimension%*Nd%

            REDIM XiMin(1 TO Nd%), XiMax(1 TO Nd%) : FOR i% = 1 TO Nd% : XiMin(i%) = -10## : XiMax(i%) = 10## : NEXT i%

            REDIM StartingXiMin(1 TO Nd%), StartingXiMax(1 TO Nd%) : FOR i% = 1 TO Nd% : StartingXiMin(i%) = XiMin(i%) : StartingXiMax(i%) = XiMax(i%) : NEXT i%

            IF PlaceInitialProbes$ = "2D GRID" THEN

                Np% = NumProbesPerDimension%^2 : REDIM R(1 TO Np%, 1 TO Nd%, 0 TO Nt&) 'to create (Np/Nd) x (Np/Nd) grid

            END IF

  CASE "F3" '(n-D)

            Nd%                      = 30
            NumProbesPerDimension% = 2
            Np%                      = NumProbesPerDimension%*Nd%

            Nt&     = 1000
            G       = 2##
            Alpha   = 2##
            Beta    = 2##
            DeltaT  = 1##
            Frep    = 0.5##

            PlaceInitialProbes$  = "UNIFORM ON-AXIS"
            InitialAcceleration$ = "ZERO"
            RepositionFactor$    = "VARIABLE"

            Np% = NumProbesPerDimension%*Nd%

            REDIM XiMin(1 TO Nd%), XiMax(1 TO Nd%) : FOR i% = 1 TO Nd% : XiMin(i%) = -100## : XiMax(i%) = 100## : NEXT i%

            REDIM StartingXiMin(1 TO Nd%), StartingXiMax(1 TO Nd%) : FOR i% = 1 TO Nd% : StartingXiMin(i%) = XiMin(i%) : StartingXiMax(i%) = XiMax(i%) : NEXT i%

            IF PlaceInitialProbes$ = "2D GRID" THEN

                Np% = NumProbesPerDimension%^2 : REDIM R(1 TO Np%, 1 TO Nd%, 0 TO Nt&) 'to create (Np/Nd) x (Np/Nd) grid

            END IF

  CASE "F4" '(n-D)

            Nd%                      = 30
            NumProbesPerDimension% = 2
            Np%                      = NumProbesPerDimension%*Nd%

            Nt&     = 1000
            G       = 2##
            Alpha   = 2##
            Beta    = 2##
            DeltaT  = 1##
            Frep    = 0.5##

            PlaceInitialProbes$  = "UNIFORM ON-AXIS"
            InitialAcceleration$ = "ZERO"
            RepositionFactor$    = "VARIABLE"

            Np% = NumProbesPerDimension%*Nd%

            REDIM XiMin(1 TO Nd%), XiMax(1 TO Nd%) : FOR i% = 1 TO Nd% : XiMin(i%) = -100## : XiMax(i%) = 100## : NEXT i%
```

```
              REDIM StartingXiMin(1 TO Nd%), StartingXiMax(1 TO Nd%) : FOR i% = 1 TO Nd% : StartingXiMin(i%) = XiMin(i%) : StartingXiMax(i%) = XiMax(i%) : NEXT i%

              IF PlaceInitialProbes$ = "2D GRID" THEN

                  Np% = NumProbesPerDimension%^2 : REDIM R(1 TO Np%, 1 TO Nd%, 0 TO Nt&) 'to create (Np/Nd) x (Np/Nd) grid

              END IF

CASE "F5" '(n-D)

              Nd%                      = 30
              NumProbesPerDimension% = 2
              Np%                      = NumProbesPerDimension%*Nd%

              Nt&      = 1000
              G        = 2##
              Alpha    = 2##
              Beta     = 2##
              DeltaT   = 1##
              Frep     = 0.5##

              PlaceInitialProbes$  = "UNIFORM ON-AXIS"
              InitialAcceleration$ = "ZERO"
              RepositionFactor$    = "VARIABLE"

              Np% = NumProbesPerDimension%*Nd%

              REDIM XiMin(1 TO Nd%), XiMax(1 TO Nd%) : FOR i% = 1 TO Nd% : XiMin(i%) = -30## : XiMax(i%) = 30## : NEXT i%

              REDIM StartingXiMin(1 TO Nd%), StartingXiMax(1 TO Nd%) : FOR i% = 1 TO Nd% : StartingXiMin(i%) = XiMin(i%) : StartingXiMax(i%) = XiMax(i%) : NEXT i%

              IF PlaceInitialProbes$ = "2D GRID" THEN

                  Np% = NumProbesPerDimension%^2 : REDIM R(1 TO Np%, 1 TO Nd%, 0 TO Nt&) 'to create (Np/Nd) x (Np/Nd) grid

              END IF

CASE "F6" '(n-D) STEP

              Nd%                      = 30
              NumProbesPerDimension% = 2 '20
              Np%                      = NumProbesPerDimension%*Nd%

              Nt&      = 1000
              G        = 2##
              Alpha    = 2##
              Beta     = 2##
              DeltaT   = 1##
              Frep     = 0.5##

              PlaceInitialProbes$  = "UNIFORM ON-AXIS"
              InitialAcceleration$ = "ZERO"
              RepositionFactor$    = "VARIABLE" '"FIXED"

              Np% = NumProbesPerDimension%*Nd%

              REDIM XiMin(1 TO Nd%), XiMax(1 TO Nd%) : FOR i% = 1 TO Nd% : XiMin(i%) = -100## : XiMax(i%) = 100## : NEXT i%

              REDIM StartingXiMin(1 TO Nd%), StartingXiMax(1 TO Nd%) : FOR i% = 1 TO Nd% : StartingXiMin(i%) = XiMin(i%) : StartingXiMax(i%) = XiMax(i%) : NEXT i%

              IF PlaceInitialProbes$ = "2D GRID" THEN

                  Np% = NumProbesPerDimension%^2 : REDIM R(1 TO Np%, 1 TO Nd%, 0 TO Nt&) 'to create (Np/Nd) x (Np/Nd) grid

              END IF

CASE "F7" '(n-D)

              Nd%                      = 30
              NumProbesPerDimension% = 2 '20
              Np%                      = NumProbesPerDimension%*Nd%

              Nt&      = 100            'BECAUSE THIS FUNCTION HAS A RANDOM COMPONENT !!
              G        = 2##
              Alpha    = 2##
              Beta     = 2##
              DeltaT   = 1##
              Frep     = 0.5##

              PlaceInitialProbes$  = "UNIFORM ON-AXIS"
              InitialAcceleration$ = "ZERO"
              RepositionFactor$    = "VARIABLE" '"FIXED"

              Np% = NumProbesPerDimension%*Nd%

              REDIM XiMin(1 TO Nd%), XiMax(1 TO Nd%) : FOR i% = 1 TO Nd% : XiMin(i%) = -1.28## : XiMax(i%) = 1.28## : NEXT i%

              REDIM StartingXiMin(1 TO Nd%), StartingXiMax(1 TO Nd%) : FOR i% = 1 TO Nd% : StartingXiMin(i%) = XiMin(i%) : StartingXiMax(i%) = XiMax(i%) : NEXT i%

              IF PlaceInitialProbes$ = "2D GRID" THEN

                  Np% = NumProbesPerDimension%^2 : REDIM R(1 TO Np%, 1 TO Nd%, 0 TO Nt&) 'to create (Np/Nd) x (Np/Nd) grid

              END IF 'F7

CASE "F8" '(n-D)

              Nd%                      = 30
              NumProbesPerDimension% = 2 '4 '20
              Np%                      = NumProbesPerDimension%*Nd%

              Nt&      = 1000
              G        = 2##
              Alpha    = 2##
              Beta     = 2##
              DeltaT   = 1##
              Frep     = 0.5##

              PlaceInitialProbes$  = "UNIFORM ON-AXIS"
              InitialAcceleration$ = "ZERO"
              RepositionFactor$    = "VARIABLE" '"FIXED"

              Np% = NumProbesPerDimension%*Nd%

              REDIM XiMin(1 TO Nd%), XiMax(1 TO Nd%) : FOR i% = 1 TO Nd% : XiMin(i%) = -500## : XiMax(i%) = 500## : NEXT i%

              REDIM StartingXiMin(1 TO Nd%), StartingXiMax(1 TO Nd%) : FOR i% = 1 TO Nd% : StartingXiMin(i%) = XiMin(i%) : StartingXiMax(i%) = XiMax(i%) : NEXT i%

              IF PlaceInitialProbes$ = "2D GRID" THEN

                  Np% = NumProbesPerDimension%^2 : REDIM R(1 TO Np%, 1 TO Nd%, 0 TO Nt&) 'to create (Np/Nd) x (Np/Nd) grid

              END IF 'F8

CASE "F9" '(n-D)

              Nd%                      = 30
```

```
        NumProbesPerDimension% = 2 '4 '20
        Np%              = NumProbesPerDimension%*Nd%

        Nt&      = 1000
        G        = 2##
        Alpha    = 2##
        Beta     = 2##
        DeltaT   = 1##
        Frep     = 0.5##

        PlaceInitialProbes$ = "UNIFORM ON-AXIS"
        InitialAcceleration$ = "ZERO"
        RepositionFactor$    = "VARIABLE" '"FIXED"

        Np% = NumProbesPerDimension%*Nd%

        REDIM XiMin(1 TO Nd%), XiMax(1 TO Nd%) : FOR i% = 1 TO Nd% : XiMin(i%) = -5.12## : XiMax(i%) = 5.12## : NEXT i%

        REDIM StartingXiMin(1 TO Nd%), StartingXiMax(1 TO Nd%) : FOR i% = 1 TO Nd% : StartingXiMin(i%) = XiMin(i%) : StartingXiMax(i%) = XiMax(i%) : NEXT i%

        IF PlaceInitialProbes$ = "2D GRID" THEN

            Np% = NumProbesPerDimension%^2 : REDIM R(1 TO Np%, 1 TO Nd%, 0 TO Nt&) 'to create (Np/Nd) x (Np/Nd) grid

        END IF 'F9

    CASE "F10" '(n-D) Ackley's Function

        Nd%                  = 30
        NumProbesPerDimension% = 2 '4 '20
        Np%              = NumProbesPerDimension%*Nd%

        Nt&      = 1000
        G        = 2##
        Alpha    = 2##
        Beta     = 2##
        DeltaT   = 1##
        Frep     = 0.5##

        PlaceInitialProbes$ = "UNIFORM ON-AXIS"
        InitialAcceleration$ = "ZERO"
        RepositionFactor$    = "VARIABLE" '"FIXED"

        Np% = NumProbesPerDimension%*Nd%

        REDIM XiMin(1 TO Nd%), XiMax(1 TO Nd%) : FOR i% = 1 TO Nd% : XiMin(i%) = -32## : XiMax(i%) = 32## : NEXT i%

        REDIM StartingXiMin(1 TO Nd%), StartingXiMax(1 TO Nd%) : FOR i% = 1 TO Nd% : StartingXiMin(i%) = XiMin(i%) : StartingXiMax(i%) = XiMax(i%) : NEXT i%

        IF PlaceInitialProbes$ = "2D GRID" THEN

            Np% = NumProbesPerDimension%^2 : REDIM R(1 TO Np%, 1 TO Nd%, 0 TO Nt&) 'to create (Np/Nd) x (Np/Nd) grid

        END IF 'F10

    CASE "F11" '(n-D)

        Nd%                  = 30
        NumProbesPerDimension% = 2 '4 '20
        Np%              = NumProbesPerDimension%*Nd%

        Nt&      = 1000
        G        = 2##
        Alpha    = 2##
        Beta     = 2##
        DeltaT   = 1##
        Frep     = 0.5##

        PlaceInitialProbes$ = "UNIFORM ON-AXIS"
        InitialAcceleration$ = "ZERO"
        RepositionFactor$    = "VARIABLE" '"FIXED"

        Np% = NumProbesPerDimension%*Nd%

        REDIM XiMin(1 TO Nd%), XiMax(1 TO Nd%) : FOR i% = 1 TO Nd% : XiMin(i%) = -600## : XiMax(i%) = 600## : NEXT i%

        REDIM StartingXiMin(1 TO Nd%), StartingXiMax(1 TO Nd%) : FOR i% = 1 TO Nd% : StartingXiMin(i%) = XiMin(i%) : StartingXiMax(i%) = XiMax(i%) : NEXT i%

        IF PlaceInitialProbes$ = "2D GRID" THEN

            Np% = NumProbesPerDimension%^2 : REDIM R(1 TO Np%, 1 TO Nd%, 0 TO Nt&) 'to create (Np/Nd) x (Np/Nd) grid

        END IF 'F11

    CASE "F12" '(n-D) Penalized #1

        Nd%                  = 30
        NumProbesPerDimension% = 2 '4 '20
        Np%              = NumProbesPerDimension%*Nd%

        Nt&      = 1000
        G        = 2##
        Alpha    = 2##
        Beta     = 2##
        DeltaT   = 1##
        Frep     = 0.5##

        PlaceInitialProbes$ = "UNIFORM ON-AXIS"
        InitialAcceleration$ = "ZERO"
        RepositionFactor$    = "VARIABLE" '"FIXED"

        Np% = NumProbesPerDimension%*Nd%

        REDIM XiMin(1 TO Nd%), XiMax(1 TO Nd%) : FOR i% = 1 TO Nd% : XiMin(i%) = -50## : XiMax(i%) = 50## : NEXT i%
'       REDIM XiMin(1 TO Nd%), XiMax(1 TO Nd%) : FOR i% = 1 TO Nd% : XiMin(i%) = -5## : XiMax(i%) = 5## : NEXT i% 'use this interval for second run to improve
performance

        REDIM StartingXiMin(1 TO Nd%), StartingXiMax(1 TO Nd%) : FOR i% = 1 TO Nd% : StartingXiMin(i%) = XiMin(i%) : StartingXiMax(i%) = XiMax(i%) : NEXT i%

        IF PlaceInitialProbes$ = "2D GRID" THEN

            Np% = NumProbesPerDimension%^2 : REDIM R(1 TO Np%, 1 TO Nd%, 0 TO Nt&) 'to create (Np/Nd) x (Np/Nd) grid

        END IF 'F12

    CASE "F13" '(n-D) Penalized #2

        Nd%                  = 30
        NumProbesPerDimension% = 2 '4 '20
        Np%              = NumProbesPerDimension%*Nd%

        Nt&      = 1000
        G        = 2##
        Alpha    = 2##
        Beta     = 2##
        DeltaT   = 1##
```

```
        Frep    = 0.5##

        PlaceInitialProbes$ = "UNIFORM ON-AXIS"
        InitialAcceleration$ = "ZERO"
        RepositionFactor$    = "VARIABLE" '"FIXED"

        Np% = NumProbesPerDimension%*Nd%

        REDIM XiMin(1 TO Nd%), XiMax(1 TO Nd%) : FOR i% = 1 TO Nd% : XiMin(i%) = -50## : XiMax(i%) = 50## : NEXT i%

        REDIM StartingXiMin(1 TO Nd%), StartingXiMax(1 TO Nd%) : FOR i% = 1 TO Nd% : StartingXiMin(i%) = XiMin(i%) : StartingXiMax(i%) = XiMax(i%) : NEXT i%

        IF PlaceInitialProbes$ = "2D GRID" THEN

            Np% = NumProbesPerDimension%^2 : REDIM R(1 TO Np%, 1 TO Nd%, 0 TO Nt&) 'to create (Np/Nd) x (Np/Nd) grid

        END IF 'F13
CASE "F14" '(2-D) Shekel's Foxholes
        Nd%                  = 2
        NumProbesPerDimension% = 4 '20
        Np%                  = NumProbesPerDimension%*Nd%

        Nt&     = 1000
        G       = 2##
        Alpha   = 2##
        Beta    = 2##
        DeltaT  = 1##
        Frep    = 0.5##

        PlaceInitialProbes$ = "UNIFORM ON-AXIS"
        InitialAcceleration$ = "ZERO"
        RepositionFactor$    = "VARIABLE" '"FIXED"

        Np% = NumProbesPerDimension%*Nd%

        REDIM XiMin(1 TO Nd%), XiMax(1 TO Nd%) : FOR i% = 1 TO Nd% : XiMin(i%) = -65.536## : XiMax(i%) = 65.536## : NEXT i%

        REDIM StartingXiMin(1 TO Nd%), StartingXiMax(1 TO Nd%) : FOR i% = 1 TO Nd% : StartingXiMin(i%) = XiMin(i%) : StartingXiMax(i%) = XiMax(i%) : NEXT i%

        IF PlaceInitialProbes$ = "2D GRID" THEN

            Np% = NumProbesPerDimension%^2 : REDIM R(1 TO Np%, 1 TO Nd%, 0 TO Nt&) 'to create (Np/Nd) x (Np/Nd) grid

        END IF 'F14
CASE "F15" '(4-D) Kowalik's Function
        Nd%                  = 4
        NumProbesPerDimension% = 4 '20
        Np%                  = NumProbesPerDimension%*Nd%

        Nt&     = 1000
        G       = 2##
        Alpha   = 2##
        Beta    = 2##
        DeltaT  = 1##
        Frep    = 0.5##

        PlaceInitialProbes$ = "UNIFORM ON-AXIS"
        InitialAcceleration$ = "ZERO"
        RepositionFactor$    = "VARIABLE" '"FIXED"

        Np% = NumProbesPerDimension%*Nd%

        REDIM XiMin(1 TO Nd%), XiMax(1 TO Nd%) : FOR i% = 1 TO Nd% : XiMin(i%) = -5## : XiMax(i%) = 5## : NEXT i%

        REDIM StartingXiMin(1 TO Nd%), StartingXiMax(1 TO Nd%) : FOR i% = 1 TO Nd% : StartingXiMin(i%) = XiMin(i%) : StartingXiMax(i%) = XiMax(i%) : NEXT i%
CASE "F16" '(2-D) Camel Back
        Nd%                  = 2
        NumProbesPerDimension% = 4 '20
        Np%                  = NumProbesPerDimension%*Nd%

        Nt&     = 1000
        G       = 2##
        Alpha   = 2##
        Beta    = 2##
        DeltaT  = 1##
        Frep    = 0.5##

        PlaceInitialProbes$ = "UNIFORM ON-AXIS"
        InitialAcceleration$ = "ZERO"
        RepositionFactor$    = "VARIABLE" '"FIXED"

        Np% = NumProbesPerDimension%*Nd%

        REDIM XiMin(1 TO Nd%), XiMax(1 TO Nd%) : FOR i% = 1 TO Nd% : XiMin(i%) = -5## : XiMax(i%) = 5## : NEXT i%

        REDIM StartingXiMin(1 TO Nd%), StartingXiMax(1 TO Nd%) : FOR i% = 1 TO Nd% : StartingXiMin(i%) = XiMin(i%) : StartingXiMax(i%) = XiMax(i%) : NEXT i%

        IF PlaceInitialProbes$ = "2D GRID" THEN

            Np% = NumProbesPerDimension%^2 : REDIM R(1 TO Np%, 1 TO Nd%, 0 TO Nt&) 'to create (Np/Nd) x (Np/Nd) grid

        END IF 'F16
CASE "F17" '(2-D) Branin
        Nd%                  = 2
        NumProbesPerDimension% = 4 '20
        Np%                  = NumProbesPerDimension%*Nd%

        Nt&     = 1000
        G       = 2##
        Alpha   = 2##
        Beta    = 2##
        DeltaT  = 1##
        Frep    = 0.5##

        PlaceInitialProbes$ = "UNIFORM ON-AXIS"
        InitialAcceleration$ = "ZERO"
        RepositionFactor$    = "VARIABLE" '"FIXED"

        Np% = NumProbesPerDimension%*Nd%

        REDIM XiMin(1 TO Nd%), XiMax(1 TO Nd%) : XiMin(1) = -5## : XiMax(1) = 10## : XiMin(2) = 0## : XiMax(2) = 15##

        REDIM StartingXiMin(1 TO Nd%), StartingXiMax(1 TO Nd%) : FOR i% = 1 TO Nd% : StartingXiMin(i%) = XiMin(i%) : StartingXiMax(i%) = XiMax(i%) : NEXT i%

        IF PlaceInitialProbes$ = "2D GRID" THEN

            Np% = NumProbesPerDimension%^2 : REDIM R(1 TO Np%, 1 TO Nd%, 0 TO Nt&) 'to create (Np/Nd) x (Np/Nd) grid

        END IF 'F17
```

```
CASE "F18" '(2-D) Goldstein-Price

    Nd%                    = 2
    NumProbesPerDimension% = 4 '20
    Np%                    = NumProbesPerDimension%*Nd%

    Nt&      = 1000
    G        = 2##
    Alpha    = 2##
    Beta     = 2##
    DeltaT   = 1##
    Frep     = 0.5##

    PlaceInitialProbes$ = "UNIFORM ON-AXIS"
    InitialAcceleration$ = "ZERO"
    RepositionFactor$   = "VARIABLE" 'FIXED"

    Np% = NumProbesPerDimension%*Nd%

    REDIM XiMin(1 TO Nd%), XiMax(1 TO Nd%) : XiMin(1) = -2## : XiMax(1) = 2## : XiMin(2) = -2## : XiMax(2) = 2##

    REDIM StartingXiMin(1 TO Nd%), StartingXiMax(1 TO Nd%) : FOR i% = 1 TO Nd% : StartingXiMin(i%) = XiMin(i%) : StartingXiMax(i%) = XiMax(i%) : NEXT i%

    IF PlaceInitialProbes$ = "2D GRID" THEN

        Np% = NumProbesPerDimension%^2 : REDIM R(1 TO Np%, 1 TO Nd%, 0 TO Nt&) 'to create (Np/Nd) x (Np/Nd) grid

    END IF 'F18

CASE "F19" '(3-D) Hartman's Family #1

    Nd%                    = 3
    NumProbesPerDimension% = 4 '20
    Np%                    = NumProbesPerDimension%*Nd%

    Nt&      = 1000
    G        = 2##
    Alpha    = 2##
    Beta     = 2##
    DeltaT   = 1##
    Frep     = 0.5##

    PlaceInitialProbes$  = "UNIFORM ON-AXIS"
    InitialAcceleration$ = "ZERO"
    RepositionFactor$    = "VARIABLE" '"FIXED"

    Np% = NumProbesPerDimension%*Nd%

    REDIM XiMin(1 TO Nd%), XiMax(1 TO Nd%) : FOR i% = 1 TO Nd% : XiMin(i%) = 0## : XiMax(i%) = 1## : NEXT i%

    REDIM StartingXiMin(1 TO Nd%), StartingXiMax(1 TO Nd%) : FOR i% = 1 TO Nd% : StartingXiMin(i%) = XiMin(i%) : StartingXiMax(i%) = XiMax(i%) : NEXT i%

CASE "F20" '(6-D) Hartman's Family #2

    Nd%                    = 6
    NumProbesPerDimension% = 4 '20
    Np%                    = NumProbesPerDimension%*Nd%

    Nt&      = 1000
    G        = 2##
    Alpha    = 2##
    Beta     = 2##
    DeltaT   = 1##
    Frep     = 0.5##

    PlaceInitialProbes$  = "UNIFORM ON-AXIS"
    InitialAcceleration$ = "ZERO"
    RepositionFactor$    = "VARIABLE" '"FIXED"

    Np% = NumProbesPerDimension%*Nd%

    REDIM XiMin(1 TO Nd%), XiMax(1 TO Nd%) : FOR i% = 1 TO Nd% : XiMin(i%) = 0## : XiMax(i%) = 1## : NEXT i%

    REDIM StartingXiMin(1 TO Nd%), StartingXiMax(1 TO Nd%) : FOR i% = 1 TO Nd% : StartingXiMin(i%) = XiMin(i%) : StartingXiMax(i%) = XiMax(i%) : NEXT i%

CASE "F21" '(4-D) Shekel's Family m=5

    Nd%                    = 4
    NumProbesPerDimension% = 4 '20
    Np%                    = NumProbesPerDimension%*Nd%

    Nt&      = 1000
    G        = 2##
    Alpha    = 2##
    Beta     = 2##
    DeltaT   = 1##
    Frep     = 0.5##

    PlaceInitialProbes$  = "UNIFORM ON-AXIS"
    InitialAcceleration$ = "ZERO"
    RepositionFactor$    = "VARIABLE" '"FIXED"

    Np% = NumProbesPerDimension%*Nd%

    REDIM XiMin(1 TO Nd%), XiMax(1 TO Nd%) : FOR i% = 1 TO Nd% : XiMin(i%) = 0## : XiMax(i%) = 10## : NEXT i%

    REDIM StartingXiMin(1 TO Nd%), StartingXiMax(1 TO Nd%) : FOR i% = 1 TO Nd% : StartingXiMin(i%) = XiMin(i%) : StartingXiMax(i%) = XiMax(i%) : NEXT i%

CASE "F22" '(4-D) Shekel's Family m=7

    Nd%                    = 4
    NumProbesPerDimension% = 4 '20
    Np%                    = NumProbesPerDimension%*Nd%

    Nt&      = 1000
    G        = 2##
    Alpha    = 2##
    Beta     = 2##
    DeltaT   = 1##
    Frep     = 0.5##

    PlaceInitialProbes$  = "UNIFORM ON-AXIS"
    InitialAcceleration$ = "ZERO"
    RepositionFactor$    = "VARIABLE" '"FIXED"

    Np% = NumProbesPerDimension%*Nd%

    REDIM XiMin(1 TO Nd%), XiMax(1 TO Nd%) : FOR i% = 1 TO Nd% : XiMin(i%) = 0## : XiMax(i%) = 10## : NEXT i%

    REDIM StartingXiMin(1 TO Nd%), StartingXiMax(1 TO Nd%) : FOR i% = 1 TO Nd% : StartingXiMin(i%) = XiMin(i%) : StartingXiMax(i%) = XiMax(i%) : NEXT i%

CASE "F23" '(4-D) Shekel's Family m=10

    Nd%                    = 4
    NumProbesPerDimension% = 4 '20
    Np%                    = NumProbesPerDimension%*Nd%
```

```
        Nt&      = 1000
        G        = 2##
        Alpha    = 2##
        Beta     = 2##
        DeltaT   = 1##
        Frep     = 0.5##

        PlaceInitialProbes$  = "UNIFORM ON-AXIS"
        InitialAcceleration$ = "ZERO"
        RepositionFactor$    = "VARIABLE" '"FIXED"

        Np% = NumProbesPerDimension%*Nd%

        REDIM XiMin(1 TO Nd%), XiMax(1 TO Nd%) : FOR i% = 1 TO Nd% : XiMin(i%) = 0## : XiMax(i%) = 10## : NEXT i%

        REDIM StartingXiMin(1 TO Nd%), StartingXiMax(1 TO Nd%) : FOR i% = 1 TO Nd% : StartingXiMin(i%) = XiMin(i%) : StartingXiMax(i%) = XiMax(i%) : NEXT i%

  CASE "PBM_1" '2-D

        Nd%                     = 2
        NumProbesPerDimension%  = 2 '4 '20
        Np%                     = NumProbesPerDimension%*Nd%

        Nt&      = 100
        G        = 2##
        Alpha    = 2##
        Beta     = 2##
        DeltaT   = 1##
        Frep     = 0.5##

        PlaceInitialProbes$  = "UNIFORM ON-AXIS"
        InitialAcceleration$ = "ZERO"
        RepositionFactor$    = "VARIABLE" '"FIXED"

        Np% = NumProbesPerDimension%*Nd%

        REDIM XiMin(1 TO Nd%), XiMax(1 TO Nd%)

        XiMin(1) = 0.5## : XiMax(1) = 3## 'dipole length, L, in Wavelengths
        XiMin(2) = 0## :  XiMax(2) = Pi2 'polar angle, Theta, in Radians

        REDIM StartingXiMin(1 TO Nd%), StartingXiMax(1 TO Nd%) : FOR i% = 1 TO Nd% : StartingXiMin(i%) = XiMin(i%) : StartingXiMax(i%) = XiMax(i%) : NEXT i%

        NN% = FREEFILE : OPEN "INFILE.DAT" FOR OUTPUT AS #NN% : PRINT #NN%,"PBM1.NEC" : PRINT #NN%,"PBM1.OUT" : CLOSE #NN% 'NEC Input/Output Files

  CASE "PBM_2" '2-D

        AddNoiseToPBM2$ = "NO" '"YES" '"NO" '"YES"

        Nd%                     = 2
        NumProbesPerDimension%  = 4 '20
        Np%                     = NumProbesPerDimension%*Nd%

        Nt&      = 100
        G        = 2##
        Alpha    = 2##
        Beta     = 2##
        DeltaT   = 1##
        Frep     = 0.5##

        PlaceInitialProbes$  = "UNIFORM ON-AXIS"
        InitialAcceleration$ = "ZERO"
        RepositionFactor$    = "VARIABLE" '"FIXED"

        Np% = NumProbesPerDimension%*Nd%

        REDIM XiMin(1 TO Nd%), XiMax(1 TO Nd%)

        XiMin(1) = 5## : XiMax(1) = 15## 'dipole separation, D, in Wavelengths
        XiMin(2) = 0## : XiMax(2) = Pi   'polar angle, Theta, in Radians

        REDIM StartingXiMin(1 TO Nd%), StartingXiMax(1 TO Nd%) : FOR i% = 1 TO Nd% : StartingXiMin(i%) = XiMin(i%) : StartingXiMax(i%) = XiMax(i%) : NEXT i%

        NN% = FREEFILE : OPEN "INFILE.DAT" FOR OUTPUT AS #NN% : PRINT #NN%,"PBM2.NEC" : PRINT #NN%,"PBM2.OUT" : CLOSE #NN%

  CASE "PBM_3" '2-D

        Nd%                     = 2
        NumProbesPerDimension%  = 4 '20
        Np%                     = NumProbesPerDimension%*Nd%

        Nt&      = 100
        G        = 2##
        Alpha    = 2##
        Beta     = 2##
        DeltaT   = 1##
        Frep     = 0.5##

        PlaceInitialProbes$  = "UNIFORM ON-AXIS"
        InitialAcceleration$ = "ZERO"
        RepositionFactor$    = "VARIABLE" '"FIXED"

        Np% = NumProbesPerDimension%*Nd%

        REDIM XiMin(1 TO Nd%), XiMax(1 TO Nd%)

        XiMin(1) = 0## : XiMax(1) = 4## 'Phase Parameter, Beta (0-4)
        XiMin(2) = 0## : XiMax(2) = Pi  'polar angle, Theta, in Radians

        REDIM StartingXiMin(1 TO Nd%), StartingXiMax(1 TO Nd%) : FOR i% = 1 TO Nd% : StartingXiMin(i%) = XiMin(i%) : StartingXiMax(i%) = XiMax(i%) : NEXT i%

        NN% = FREEFILE : OPEN "INFILE.DAT" FOR OUTPUT AS #NN% : PRINT #NN%,"PBM3.NEC" : PRINT #NN%,"PBM3.OUT" : CLOSE #NN%

  CASE "PBM_4" '2-D

        Nd%                     = 2
        NumProbesPerDimension%  = 4 '6 '2 '4 '20
        Np%                     = NumProbesPerDimension%*Nd%

        Nt&      = 100
        G        = 2##
        Alpha    = 2##
        Beta     = 2##
        DeltaT   = 1##
        Frep     = 0.5##

        PlaceInitialProbes$  = "UNIFORM ON-AXIS"
        InitialAcceleration$ = "ZERO"
        RepositionFactor$    = "VARIABLE" '"FIXED"

        Np% = NumProbesPerDimension%*Nd%

        REDIM XiMin(1 TO Nd%), XiMax(1 TO Nd%)

        XiMin(1) = 0.5##   : XiMax(1) = 1.5## 'ARM LENGTH (NOT Total Length), wavelengths (0.5-1.5)
        XiMin(2) = Pi/18## : XiMax(2) = Pi/2## 'Inner angle, Alpha, in Radians (Pi/18-Pi/2)
```

```
              REDIM StartingXiMin(1 TO Nd%), StartingXiMax(1 TO Nd%) : FOR i% = 1 TO Nd% : StartingXiMin(i%) = XiMin(i%) : StartingXiMax(i%) = XiMax(i%) : NEXT i%

              NN% = FREEFILE : OPEN "INFILE.DAT" FOR OUTPUT AS #NN% : PRINT #NN%,"PBM4.NEC" : PRINT #NN%,"PBM4.OUT" : CLOSE #NN%

          CASE "PBM_5"

              NumCollinearElements%  = 6 '30 'EVEN or ODD: 6,7,10,13,16,24 used by PBM
              Nd%                    = NumCollinearElements% - 1
              NumProbesPerDimension% = 4 '20
              Np%                    = NumProbesPerDimension%*Nd%

              Nt&    = 100
              G      = 2##
              Alpha  = 2##
              Beta   = 2##
              DeltaT = 1##
              Frep   = 0.5##

              PlaceInitialProbes$  = "UNIFORM ON-AXIS"
              InitialAccelerations$ = "ZERO"
              RepositionFactor$    = "VARIABLE" '"FIXED"

              Nd% = NumCollinearElements% - 1

              Np% = NumProbesPerDimension%*Nd%

              REDIM XiMin(1 TO Nd%), XiMax(1 TO Nd%) : FOR i% = 1 TO Nd% : XiMin(i%) = 0.5## : XiMax(i%) = 1.5## : NEXT i%

              REDIM StartingXiMin(1 TO Nd%), StartingXiMax(1 TO Nd%) : FOR i% = 1 TO Nd% : StartingXiMin(i%) = XiMin(i%) : StartingXiMax(i%) = XiMax(i%) : NEXT i%

              NN% = FREEFILE : OPEN "INFILE.DAT" FOR OUTPUT AS #NN% : PRINT #NN%,"PBM5.NEC" : PRINT #NN%,"PBM5.OUT" : CLOSE #NN%

   ' ==========================================================================================
   ' NOTE - DON'T FORGET TO ADD NEW TEST FUNCTIONS TO FUNCTION ObjectiveFunction() ABOVE !!
   ' ==========================================================================================

   END SELECT

   IF Nd% > 100 THEN Nt& = MIN(Nt&,200) 'to avoid array dimensioning problems

   DiagLength = 0## : FOR i% = 1 TO Nd% : DiagLength = DiagLength + (XiMax(i%)-XiMin(i%))^2 : NEXT i% : DiagLength = SQR(DiagLength) 'compute length of decision space
   principal diagonal

END SUB 'GetFunctionRunParameters()

'-------------------------------

FUNCTION ParrottF4(R(),Nd%,p%,j&) 'Parrott F4 (1-D)

'MAXIMUM = 1 AT -0.0796875... WITH ZERO OFFSET (SEEMS TO WORK BEST WITH JUST 3 PROBES, BUT NOT ALLOWED IN THIS VERSION...)

'References:

'Beasley, D., D. R. Bull, and R. R. Martin, "A Sequential Niche Technique for Multimodal
'Function Optimization," Evol. Comp. (MIT Press), vol. 1, no. 2, 1993, pp. 101-125
'(online at http://citeseer.ist.psu.edu/beasley93sequential.html).

'Parrott, D., and X. Li, "Locating and Tracking Multiple Dynamic Optima by a Particle Swarm
'Model Using Speciation," IEEE Trans. Evol. Computation, vol. 10, no. 4, Aug. 2006, pp. 440-458.

LOCAL Z, x, offset AS EXT

    offset = 0##

    x = R(p%,1,j&)

    Z = EXP(-2##*LOG(2##)*((x-0.08##-offset)/0.854##)^2)*(SIN(5##*Pi*((x-offset)^0.75##-0.05##)))^6 'WARNING! This is a NATURAL LOG, NOT Log10!!!

    ParrottF4 = Z

END FUNCTION 'ParrottF4()

'----------------------------

FUNCTION SGO(R(),Nd%,p%,j&) 'SGO Function (2-D)

'MAXIMUM = -130.8323226... @ -(2.8362075...,-2.8362075...) WITH ZERO OFFSET.

'Reference:

'Hsiao, Y., Chuang, C., Jiang, J., and Chien, C., "A Novel Optimization Algorithm: Space
'Gravitational Optimization," Proc. of 2005 IEEE International Conference on Systems, Man,
'and Cybernetics, 3, 2323-2328. (2005)

    LOCAL x1, x2, Z, t1, t2, SGOx1offset, SGOx2offset AS EXT

    SGOx1offset = 0## : SGOx2offset = 0##

'   SGOx1offset = 40## : SGOx2offset = 10##

    x1 = R(p%,1,j&) - SGOx1offset : x2 = R(p%,2,j&) - SGOx2offset

    t1 = x1^4 - 16##*x1^2 + 0.5##*x1 : t2 = x2^4 - 16##*x2^2 + 0.5##*x2

    Z = t1 + t2

    SGO = -Z

END FUNCTION 'SGO()

'------------------

FUNCTION GoldsteinPrice(R(),Nd%,p%,j&) 'Goldstein-Price Function (2-D)

'MAXIMUM = -3 @ (0,-1) WITH ZERO OFFSET.

'Reference:

'Cui, Z., Zeng, J., and Sun, G. (2006) 'A Fast Particle Swarm Optimization,' Int'l. J.
'Innovative Computing, Information and Control, vol. 2, no. 6, December, pp. 1365-1380.

    LOCAL Z, x1, x2, offset1, offset2, t1, t2 AS EXT

    offset1 = 0## : offset2 = 0##

'   offset1 = 20## : offset2 = -10##

    x1 = R(p%,1,j&)-offset1 : x2 = R(p%,2,j&)-offset2

    t1 = 1##+(x1+x2+1##)^2*(19##-14##*x1+3##*x1^2-14##*x2+6##*x1*x2+3##*x2^2)

    t2 = 30##+(2##*x1-3##*x2)^2*(18##-32##*x1+12##*x1^2+48##*x2-36##*x1*x2+27##*x2^2)

    Z   = t1*t2
```

```
    GoldsteinPrice = -Z

END FUNCTION 'GoldsteinPrice()

'-----------

FUNCTION StepFunction(R(),Nd%,p%,j&) 'Step Function (n-D)

'MAXIMUM VALUE = 0 @ [Offset]^n.

'Reference:

'Yao, X., Liu, Y., and Lin, G., "Evolutionary Programming Made Faster,"
'IEEE Trans. Evolutionary Computation, Vol. 3, No. 2, 82-102, Jul. 1999.

    LOCAL Offset, Z AS EXT

    LOCAL i%

    Z = 0## : Offset = 0## '75.123## '0##

    FOR i% = 1 TO Nd%

        IF Nd% = 2 AND i% = 1 THEN Offset = 75 '75##

        IF Nd% = 2 AND i% = 2 THEN Offset = 35 '30 '35##

        Z = Z + INT((R(p%,i%,j&)-Offset) + 0.5##)^2

    NEXT i%

    StepFunction = -Z

END FUNCTION 'StepFunction()

'-----------

FUNCTION Schwefel226(R(),Nd%,p%,j&) 'Schwefel Problem 2.26 (n-D)

'MAXIMUM = 12,569.5 @ [420.8687]^30 (30-D CASE).

'Reference:

'Yao, X., Liu, Y., and Lin, G., "Evolutionary Programming Made Faster,"
'IEEE Trans. Evolutionary Computation, Vol. 3, No. 2, 82-102, Jul. 1999.

    LOCAL Z, Xi AS EXT

    LOCAL i%

    Z = 0##

    FOR i% = 1 TO Nd%

        Xi = R(p%,i%,j&)

        Z = Z + Xi*SIN(SQR(ABS(Xi)))

    NEXT i%

    Schwefel226 = Z

END FUNCTION 'SCHWEFEL226()

'-----------

FUNCTION Colville(R(),Nd%,p%,j&) 'Colville Function (4-D)

'MAXIMUM = 0 @ (1,1,1,1) WITH ZERO OFFSET.

'Reference:

'Doo-Hyun, and Se-Young, O., "A New Mutation Rule for Evolutionary Programming Motivated from
'Backpropagation Learning," IEEE Trans. Evolutionary Computation, Vol. 4, No. 2, pp. 188-190,
'July 2000.

    LOCAL Z, x1, x2, x3, x4, offset AS EXT

    offset = 0## '7.123##

    x1 = R(p%,1,j&)-offset : x2 = R(p%,2,j&)-offset : x3 = R(p%,3,j&)-offset : x4 = R(p%,4,j&)-offset

    Z = 100##*(x2-x1^2)^2 + (1##-x1)^2  + _
        90##*(x4-x3^2)^2 + (1##-x3)^2  + _
        10.1##*((x2-1##)^2 + (x4-1##)^2) + _
        19.8##*(x2-1##)*(x4-1##)

    Colville = -Z

END FUNCTION 'Colville()

'-----------

FUNCTION Griewank(R(),Nd%,p%,j&) 'Griewank (n-D)

'Max of zero at (0,...,0)

'Reference: Yao, X., Liu, Y., and Lin, G., "Evolutionary Programming Made Faster,"
'IEEE Trans. Evolutionary Computation, Vol. 3, No. 2, 82-102, Jul. 1999.

    LOCAL Offset, Sum, Prod, Z, Xi AS EXT

    LOCAL i%

    Sum = 0## : Prod = 1##

    Offset = 75.123##

    FOR i% = 1 TO Nd%

        Xi = R(p%,i%,j&) - Offset

        Sum = Sum + Xi^2

        Prod = Prod*COS(Xi/SQR(i%))

    NEXT i%

    Z = Sum/4000## - Prod + 1##

    Griewank = -Z

END FUNCTION 'Griewank()
```

```
'-----------
FUNCTION Himmelblau(R(),Nd%,p%,j&) 'Himmelblau (2-D)

    LOCAL Z, x1, x2, offset AS EXT

    offset = 0##

    x1 = R(p%,1,j&)-offset : x2 = R(p%,2,j&)-offset

    Z = 200## - (x1^2 + x2 -11##)^2 - (x1+x2^2-7##)^2

    Himmelblau = Z

END FUNCTION 'Himmelblau()
'-----------
FUNCTION Rosenbrock(R(),Nd%,p%,j&) 'Rosenbrock (n-D)

'MAXIMUM = 0 @ [1,...,1]^n (n-D CASE).

'Reference: Yao, X., Liu, Y., and Lin, G., "Evolutionary Programming Made Faster,"
'IEEE Trans. Evolutionary Computation, Vol. 3, No. 2, 82-102, Jul. 1999.

    LOCAL Z, Xi, Xi1 AS EXT

    LOCAL i%

    Z = 0##

    FOR i% = 1 TO Nd%-1

        Xi = R(p%,i%,j&) : Xi1 = R(p%,i%+1,j&)

        Z = Z + 100##*(Xi1-Xi^2)^2 + (Xi-1##)^2

    NEXT i%

    Rosenbrock = -Z

END FUNCTION 'ROSENBROCK()
'-----------
FUNCTION Sphere(R(),Nd%,p%,j&) 'Sphere (n-D)

'MAXIMUM = 0 @ [0,...,0]^n (n-D CASE).

'Reference: Yao, X., Liu, Y., and Lin, G., "Evolutionary Programming Made Faster,"
'IEEE Trans. Evolutionary Computation, Vol. 3, No. 2, 82-102, Jul. 1999.

    LOCAL Z, Xi, Xi1 AS EXT

    LOCAL i%

    Z = 0##

    FOR i% = 1 TO Nd%

        Xi  = R(p%,i%,j&)

        Z = Z + Xi^2

    NEXT i%

    Sphere = -Z

END FUNCTION 'SPHERE()
'-----------
FUNCTION HimmelblauNLO(R(),Nd%,p%,j&) 'Himmelblau non-linear optimization (5-D)

'MAXIMUM = 31025.5562644972 @ (78.0,33.0,27.0709971052,45.0,44.9692425501)

'Reference: "Constrained Optimization using CODEQ," Mahamed G.H. Omran & Ayed Salman,
'Chaos, Solitons and Tractals, 42(2009), 662-668

    LOCAL Z, x1, x2, x3, x4, x5, g1, g2, g3 AS EXT

    Z = 1E4200

    x1 = R(p%,1,j&) : x2 = R(p%,2,j&) : x3 = R(p%,3,j&) : x4 = R(p%,4,j&) : x5 = R(p%,5,j&)

    g1 = 85.334407## + 0.0056858##*x2*x5 + 0.00026##*x1*x4   - 0.0022053##*x3*x5

    g2 = 80.51249## + 0.0071317##*x2*x5 + 0.0029955##*x1*x2 + 0.0021813##*x3^2

    g3 = 9.300961## + 0.0047026##*x3*x5 + 0.0012547##*x1*x3 + 0.0019085##*x3*x4

    IF g1## < 0 OR g1 > 92## OR g2 < 90## OR g2 > 110## OR g3 < 20## OR g3 > 25## THEN GOTO ExitHimmelblauNLO

    Z = 5.3578547##*x3*x3 + 0.8356891##*x1*x5 + 37.29329##*x1 - 40792.141##

ExitHimmelblauNLO:

    HimmelblauNLO = -Z

END FUNCTION 'HimmelblauNLO()
'-----------
FUNCTION F1(R(),Nd%,p%,j&) 'F1 (n-D)

'MAXIMUM = ZERO (n-D CASE).

'Reference:

    LOCAL Z, Xi AS EXT

    LOCAL i%

    Z = 0##

    FOR i% = 1 TO Nd%

        Xi = R(p%,i%,j&)

        Z = Z + Xi^2

    NEXT i%

    F1 = -Z
```

```
END FUNCTION 'F1
'-----------

FUNCTION F2(R(),Nd%,p%,j&) 'F2 (n-D)
'MAXIMUM = ZERO (n-D CASE).
'Reference:
    LOCAL Sum, prod, Z, Xi AS EXT
    LOCAL i%
    Z = 0## : Sum = 0## : Prod = 1##
    FOR i% = 1 TO Nd%
        Xi = R(p%,i%,j&)
        Sum  = Sum+ ABS(Xi)
        Prod = Prod*ABS(Xi)
    NEXT i%
    Z = Sum + Prod
    F2 = -Z
END FUNCTION 'F2
'-----------

FUNCTION F3(R(),Nd%,p%,j&) 'F3 (n-D)
'MAXIMUM = ZERO (n-D CASE).
'Reference:
    LOCAL Z, Xk, Sum AS EXT
    LOCAL i%, k%
    Z = 0##
    FOR i% = 1 TO Nd%
        Sum = 0##
        FOR k% = 1 TO i%
            Xk = R(p%,k%,j&)
            Sum = Sum + Xk
        NEXT k%
        Z = Z + Sum^2
    NEXT i%
    F3 = -Z
END FUNCTION 'F3

'-----------

FUNCTION F4(R(),Nd%,p%,j&) 'F4 (n-D)
'MAXIMUM = ZERO (n-D CASE).
'Reference:
    LOCAL Z, Xi, MaxXi AS EXT
    LOCAL i%
    MaxXi = -1E4200
    FOR i% = 1 TO Nd%
        Xi = R(p%,i%,j&)
        IF ABS(Xi) >= MaxXi THEN MaxXi = ABS(Xi)
    NEXT i%
    F4 = -MaxXi
END FUNCTION 'F4
'-----------

FUNCTION F5(R(),Nd%,p%,j&) 'F5 (n-D)
'MAXIMUM = ZERO (n-D CASE).
'Reference:
    LOCAL Z, Xi, XiPlus1 AS EXT
    LOCAL i%
    Z = 0##
    FOR i% = 1 TO Nd%-1
        Xi      = R(p%,i%,j&)
        XiPlus1 = R(p%,i%+1,j&)
        Z = Z + (100##*(XiPlus1-Xi^2)^2+(Xi-1##))^2
    NEXT i%
    F5 = -Z
END FUNCTION 'F5
'-----------

FUNCTION F6(R(),Nd%,p%,j&) 'F6
'MAXIMUM VALUE = 0 @ [Offset]^n.
```

```
'Reference:

'Yao, X., Liu, Y., and Lin, G., "Evolutionary Programming Made Faster,"
'IEEE Trans. Evolutionary Computation, Vol. 3, No. 2, 82-102, Jul. 1999.

    LOCAL Z AS EXT

    LOCAL i%

    Z = 0##

    FOR i% = 1 TO Nd%

        Z = Z + INT(R(p%,i%,j&) + 0.5##)^2

    NEXT i%

    F6 = -Z

END FUNCTION 'F6

'-----------

FUNCTION F7(R(),Nd%,p%,j&) 'F7

'MAXIMUM VALUE = 0 @ [Offset]^n.

'Reference:

'Yao, X., Liu, Y., and Lin, G., "Evolutionary Programming Made Faster,"
'IEEE Trans. Evolutionary Computation, Vol. 3, No. 2, 82-102, Jul. 1999.

    LOCAL Z, Xi AS EXT

    LOCAL i%

    Z = 0##

    FOR i% = 1 TO Nd%

        Xi = R(p%,i%,j&)

        Z = Z + i%*Xi^4

    NEXT i%

    F7 = -Z - RandomNum(0##,1##)

END FUNCTION 'F7

'-----------

FUNCTION F8(R(),Nd%,p%,j&) '(n-D) F8 [Schwefel Problem 2.26]

'MAXIMUM = 12,569.5 @ [420.8687]^30 (30-D CASE).

'Reference:

'Yao, X., Liu, Y., and Lin, G., "Evolutionary Programming Made Faster,"
'IEEE Trans. Evolutionary Computation, Vol. 3, No. 2, 82-102, Jul. 1999.

    LOCAL Z, Xi AS EXT

    LOCAL i%

    Z = 0##

    FOR i% = 1 TO Nd%

        Xi = R(p%,i%,j&)

        Z = Z - Xi*SIN(SQR(ABS(Xi)))

    NEXT i%

    F8 = -Z

END FUNCTION 'F8

'-----------

FUNCTION F9(R(),Nd%,p%,j&) '(n-D) F9 [Rastrigin]

'MAXIMUM = ZERO (n-D CASE).

'Reference:

'Yao, X., Liu, Y., and Lin, G., "Evolutionary Programming Made Faster,"
'IEEE Trans. Evolutionary Computation, Vol. 3, No. 2, 82-102, Jul. 1999.

    LOCAL Z, Xi AS EXT

    LOCAL i%

    Z = 0##

    FOR i% = 1 TO Nd%

        Xi = R(p%,i%,j&)

        Z = Z + (Xi^2 - 10##*COS(TwoPi*Xi) + 10##)^2

    NEXT i%

    F9 = -Z

END FUNCTION 'F9

'-----------

FUNCTION F10(R(),Nd%,p%,j&) '(n-D) F10 [Ackley's Function]

'MAXIMUM = ZERO (n-D CASE).

'Reference:

'Yao, X., Liu, Y., and Lin, G., "Evolutionary Programming Made Faster,"
'IEEE Trans. Evolutionary Computation, Vol. 3, No. 2, 82-102, Jul. 1999.

    LOCAL Z, Xi, Sum1, Sum2 AS EXT

    LOCAL i%

    Z = 0## : Sum1 = 0## : Sum2 = 0##
```

```
    FOR i% = 1 TO Nd%

        Xi = R(p%,i%,j&)

        Sum1 = Sum1 + Xi^2

        Sum2 = Sum2 + COS(TwoPi*Xi)

    NEXT i%

    Z = -20##*EXP(-0.2##*SQR(Sum1/Nd%)) - EXP(Sum2/Nd%) + 20## + e

    F10 = -Z

END FUNCTION 'F10

'-----------

FUNCTION F11(R(),Nd%,p%,j&) '(n-D) F11

'MAXIMUM = ZERO (n-D CASE).

'Reference:

'Yao, X., Liu, Y., and Lin, G., "Evolutionary Programming Made Faster,"
'IEEE Trans. Evolutionary Computation, Vol. 3, No. 2, 82-102, Jul. 1999.

    LOCAL Z, Xi, Sum, Prod AS EXT

    LOCAL i%

    Z = 0## : Sum = 0## : Prod = 1##

    FOR i% = 1 TO Nd%

        Xi   = R(p%,i%,j&)

        Sum  = Sum + (Xi-100##)^2

        Prod = Prod*COS((Xi-100##)/SQR(i%))

    NEXT i%

    Z = Sum/4000## - Prod + 1##

    F11 = -Z

END FUNCTION 'F11

'-----

FUNCTION u(Xi,a,k,m)

LOCAL Z AS EXT

    Z = 0##

    SELECT CASE Xi

        CASE > a  : Z = k*(Xi-a)^m

        CASE < -a : Z = k*(-Xi-a)^m

    END SELECT

    u = Z

END FUNCTION

'-----------

FUNCTION F12(R(),Nd%,p%,j&) '(n-D) F12, Penalized #1

'Ref: Yao(1999).  Max=0 @ (-1,-1,...,-1), -50=<Xi=<50.

    LOCAL Offset, Sum1, Sum2, Z, X1, Y1, Xn, Yn, Xi, Yi, XiPlus1, YiPlus1 AS EXT

    LOCAL i%, m%, A0

    X1 = R(p%,1,j&)   : Y1 = 1## + (X1+1##)/4##

    Xn = R(p%,Nd%,j&) : Yn = 1## + (Xn+1##)/4##

    Sum1 = 0##

    FOR i% = 1 TO Nd%-1

        Xi      = R(p%,i%,j&)  : Yi      = 1## + (Xi+1##)/4##

        XiPlus1 = R(p%,i%+1,j&): YiPlus1 = 1## + (XiPlus1+1##)/4##

        Sum1 = Sum1 + (Yi-1##)^2*(1##+10##*(SIN(Pi*YiPlus1))^2)

    NEXT i%

    Sum1 = Sum1 + 10##*(SIN(Pi*Y1))^2 + (Yn-1##)^2

    Sum1 = Pi*Sum1/Nd%

    Sum2 = 0##

    FOR i% = 1 TO Nd%

        Xi = R(p%,i%,j&)

        Sum2 = Sum2 + u(Xi,10##,100##,4##)

    NEXT i%

    Z = Sum1 + Sum2

    F12 = -Z

END FUNCTION 'F12()

'------------------

FUNCTION F13(R(),Nd%,p%,j&) '(n-D) F13, Penalized #2

'Ref: Yao(1999).  Max=0 @ (1,1,...,1), -50=<Xi=<50.

    LOCAL Offset, Sum1, Sum2, Z, Xi, Xn, XiPlus1, X1 AS EXT

    LOCAL i%, m%, A0

    X1 = R(p%,1,j&) : Xn = R(p%,Nd%,j&)
```

```
    Sum1 = 0##

    FOR i% = 1 TO Nd%-1

        Xi  = R(p%,i%,j&) : XiPlus1 = R(p%,i%+1,j&)

        Sum1 = Sum1 + (Xi-1##)^2*(1##+(SIN(3##*Pi*XiPlus1))^2)

    NEXT i%

    Sum1 = Sum1 + (SIN(Pi*3##*X1))^2 +(Xn-1##)^2*(1##+(SIN(TwoPi*Xn))^2)

    Sum2 = 0##

    FOR i% = 1 TO Nd%

        Xi = R(p%,i%,j&)

        Sum2 = Sum2 + u(Xi,5##,100##,4##)

    NEXT i%

    Z = Sum1/10## + Sum2

    F13 = -Z

END FUNCTION 'F13()

'-----------------

SUB FillArrayAij  'needed for function F14, Shekel's Foxholes

    Aij(1,1)=-32## : Aij(1,2)=-16## : Aij(1,3)=0## : Aij(1,4)=16## : Aij(1,5)=32##
    Aij(1,6)=-32## : Aij(1,7)=-16## : Aij(1,8)=0## : Aij(1,9)=16## : Aij(1,10)=32##
    Aij(1,11)=-32## : Aij(1,12)=-16## : Aij(1,13)=0## : Aij(1,14)=16## : Aij(1,15)=32##
    Aij(1,16)=-32## : Aij(1,17)=-16## : Aij(1,18)=0## : Aij(1,19)=16## : Aij(1,20)=32##
    Aij(1,21)=-32## : Aij(1,22)=-16## : Aij(1,23)=0## : Aij(1,24)=16## : Aij(1,25)=32##

    Aij(2,1)=-32## : Aij(2,2)=-32## : Aij(2,3)=-32## : Aij(2,4)=-32## : Aij(2,5)=-32##
    Aij(2,6)=-16## : Aij(2,7)=-16## : Aij(2,8)=-16## : Aij(2,9)=-16## : Aij(2,10)=-16##
    Aij(2,11)=0## : Aij(2,12)=0## : Aij(2,13)=0## : Aij(2,14)=0## : Aij(2,15)=0##
    Aij(2,16)=16## : Aij(2,17)=16## : Aij(2,18)=16## : Aij(2,19)=16## : Aij(2,20)=16##
    Aij(2,21)=32## : Aij(2,22)=32## : Aij(2,23)=32## : Aij(2,24)=32## : Aij(2,25)=32##

END SUB

'-----

FUNCTION F14(R(),Nd%,p%,j&) 'F14 (2-D) Shekel's Foxholes (INVERTED...)

    LOCAL Sum1, Sum2, Z, Xi AS EXT

    LOCAL i%, jj%

    Sum1 = 0##

    FOR jj% = 1 TO 25

        Sum2 = 0##

        FOR i% = 1 TO 2

            Xi = R(p%,i%,j&)

            Sum2 = Sum2 + (Xi-Aij(i%,jj%))^6

        NEXT i%

        Sum1 = Sum1 + 1##/(jj%+Sum2)

    NEXT j%

    Z = 1##/(0.002##+Sum1)

    F14 = -Z

END FUNCTION 'F14

'-----------

FUNCTION F16(R(),Nd%,p%,j&) 'F16 (2-D) 6-Hump Camel-Back

    LOCAL x1, x2, Z AS EXT

    x1 = R(p%,1,j&) : x2 = R(p%,2,j&)

    Z = 4##*x1^2 - 2.1##*x1^4 + x1^6/3## + x1*x2 - 4*x2^2 + 4*x2^4

    F16 = -Z

END FUNCTION 'F16

'-----------

FUNCTION F15(R(),Nd%,p%,j&) 'F15 (4-D) Kowalik's Function

'Global maximum = -0.0003075 @ (0.1928,0.1908,0.1231,0.1358)

    LOCAL x1, x2, x3, x4, Num, Denom, Z, Aj(), Bj() AS EXT

    LOCAL jj%

    REDIM Aj(1 TO 11), Bj(1 TO 11)

    Aj(1)  = 0.1957## : Bj(1)  = 1##/0.25##
    Aj(2)  = 0.1947## : Bj(2)  = 1##/0.50##
    Aj(3)  = 0.1735## : Bj(3)  = 1##/1.00##
    Aj(4)  = 0.1600## : Bj(4)  = 1##/2.00##
    Aj(5)  = 0.0844## : Bj(5)  = 1##/4.00##
    Aj(6)  = 0.0627## : Bj(6)  = 1##/6.00##
    Aj(7)  = 0.0456## : Bj(7)  = 1##/8.00##
    Aj(8)  = 0.0342## : Bj(8)  = 1##/10.0##
    Aj(9)  = 0.0323## : Bj(9)  = 1##/12.0##
    Aj(10) = 0.0235## : Bj(10) = 1##/14.0##
    Aj(11) = 0.0246## : Bj(11) = 1##/16.0##

    Z = 0##

    x1 = R(p%,1,j&) : x2 = R(p%,2,j&) : x3 = R(p%,3,j&) : x4 = R(p%,4,j&)

    FOR jj% = 1 TO 11

        Num  = x1*(Bj(jj%)^2+Bj(jj%)*x2)

        Denom = Bj(jj%)^2+Bj(jj%)*x3+x4

        Z = Z + (Aj(jj%)-Num/Denom)^2
```

```
    NEXT jj%

    F15 = -Z

END FUNCTION 'F15

'-----------

FUNCTION F17(R(),Nd%,p%,j&) 'F17, (2-D) Branin

'Global maximum = -0.398 @ (-3.142,12.275), (3.142,2.275), (9.425,2.425)

    LOCAL x1, x2, Z AS EXT

    x1 = R(p%,1,j&) : x2 = R(p%,2,j&)

    Z = (x2-5.1##*x1^2/(4##*Pi^2)+5##*x1/Pi-6##)^2 + 10##*(1##-1##/(8##*Pi))*COS(x1) + 10##

    F17 = -Z

END FUNCTION 'F17

'-----------

FUNCTION F18(R(),Nd%,p%,j&) 'Goldstein-Price 2-D Test Function

'Global maximum = -3 @ (0,-1)

    LOCAL Z, x1, x2, t1, t2 AS EXT

    x1 = R(p%,1,j&) : x2 = R(p%,2,j&)

    t1 = 1##+(x1+x2+1##)^2*(19##-14##*x1+3##*x1^2-14##*x2+6##*x1*x2+3##*x2^2)

    t2 = 30##+(2##*x1-3##*x2)^2*(18##-32##*x1+12##*x1^2+48##*x2-36##*x1*x2+27##*x2^2)

    Z = t1*t2

    F18 = -Z

END FUNCTION 'F18()

'-----------

FUNCTION F19(R(),Nd%,p%,j&) 'F19 (3-D) Hartman's Family #1

'Global maximum = 3.86 @ (0.114,0.556,0.852)

    LOCAL Xi, Z, Sum, Aji(), Cj(), Pji() AS EXT

    LOCAL i%, jj%, m%

    REDIM Aji(1 TO 4, 1 TO 3), Cj(1 TO 4), Pji(1 TO 4, 1 TO 3)

    Aji(1,1) = 3.0## : Aji(1,2) = 10## : Aji(1,3) = 30## : Cj(1) = 1.0##
    Aji(2,1) = 0.1## : Aji(2,2) = 10## : Aji(2,3) = 35## : Cj(2) = 1.2##
    Aji(3,1) = 3.0## : Aji(3,2) = 10## : Aji(3,3) = 30## : Cj(3) = 3.0##
    Aji(4,1) = 0.1## : Aji(4,2) = 10## : Aji(4,3) = 35## : Cj(4) = 3.2##

    Pji(1,1) = 0.36890## : Pji(1,2) = 0.1170## : Pji(1,3) = 0.2673##
    Pji(2,1) = 0.46990## : Pji(2,2) = 0.4387## : Pji(2,3) = 0.7470##
    Pji(3,1) = 0.10910## : Pji(3,2) = 0.8732## : Pji(3,3) = 0.5547##
    Pji(4,1) = 0.03815## : Pji(4,2) = 0.5743## : Pji(4,3) = 0.8828##

    Z = 0##

    FOR jj% = 1 TO 4

        Sum = 0##

        FOR i% = 1 TO 3

            Xi = R(p%,i%,j&)

            Sum = Sum + Aji(jj%,i%)*(Xi-Pji(jj%,i%))^2

        NEXT i%

        Z = Z + Cj(jj%)*EXP(-Sum)

    NEXT jj%

    F19 = Z

END FUNCTION 'F19

'-----------

FUNCTION F20(R(),Nd%,p%,j&) 'F20 (6-D) Hartman's Family #2

'Global maximum = 3.32 @ (0.201,0.150,0.477,0.275,0.311,0.657)

    LOCAL Xi, Z, Sum, Aji(), Cj(), Pji() AS EXT

    LOCAL i%, jj%, m%

    REDIM Aji(1 TO 4, 1 TO 6), Cj(1 TO 4), Pji(1 TO 4, 1 TO 6)

    Aji(1,1) = 10.0## : Aji(1,2) = 3.00## : Aji(1,3) = 17.0## : Cj(1) = 1.0##
    Aji(2,1) = 0.05## : Aji(2,2) = 10.0## : Aji(2,3) = 17.0## : Cj(2) = 1.2##
    Aji(3,1) = 3.00## : Aji(3,2) = 3.50## : Aji(3,3) = 1.70## : Cj(3) = 3.0##
    Aji(4,1) = 17.0## : Aji(4,2) = 8.00## : Aji(4,3) = 0.05## : Cj(4) = 3.2##

    Aji(1,4) = 3.5## : Aji(1,5) = 1.7## : Aji(1,6) = 8##
    Aji(2,4) = 0.1## : Aji(2,5) = 8## : Aji(2,6) = 14##
    Aji(3,4) = 10## : Aji(3,5) = 17## : Aji(3,6) = 8##
    Aji(4,4) = 10## : Aji(4,5) = 0.1## : Aji(4,6) = 14##

    Pji(1,1) = 0.1312## : Pji(1,2) = 0.1696## : Pji(1,3) = 0.5569##
    Pji(2,1) = 0.23290## : Pji(2,2) = 0.4135## : Pji(2,3) = 0.8307##
    Pji(3,1) = 0.23480## : Pji(3,2) = 0.1415## : Pji(3,3) = 0.3522##
    Pji(4,1) = 0.40470## : Pji(4,2) = 0.8828## : Pji(4,3) = 0.8732##

    Pji(1,4) = 0.01240## : Pji(1,5) = 0.8283## : Pji(1,6) = 0.5886##
    Pji(2,4) = 0.37360## : Pji(2,5) = 0.1004## : Pji(2,6) = 0.9991##
    Pji(3,4) = 0.28830## : Pji(3,5) = 0.3047## : Pji(3,6) = 0.6650##
    Pji(4,4) = 0.57430## : Pji(4,5) = 0.1091## : Pji(4,6) = 0.0381##

    Z = 0##

    FOR jj% = 1 TO 4

        Sum = 0##

        FOR i% = 1 TO 6
```

```
        Xi = R(p%,i%,j&)

        Sum = Sum + Aji(jj%,i%)*(Xi-Pji(jj%,i%))^2

      NEXT i%

    Z = Z + Cj(jj%)*EXP(-Sum)

  NEXT jj%

  F20 = Z

END FUNCTION 'F20

'----------

FUNCTION F21(R(),Nd%,p%,j&) 'F21 (4-D) Shekel's Family m=5

'Global maximum = 10

    LOCAL Xi, Z, Sum, Aji(), Cj() AS EXT

    LOCAL i%, jj%, m%

    m% = 5 : REDIM Aji(1 TO m%, 1 TO 4), Cj(1 TO m%)

    Aji(1,1) =  4## : Aji(1,2)  =   4## : Aji(1,3) =  4## : Aji(1,4)  =   4## : Cj(1)  = 0.1##
    Aji(2,1) =  1## : Aji(2,2)  =   1## : Aji(2,3) =  1## : Aji(2,4)  =   1## : Cj(2)  = 0.2##
    Aji(3,1) =  8## : Aji(3,2)  =   8## : Aji(3,3) =  8## : Aji(3,4)  =   8## : Cj(3)  = 0.2##
    Aji(4,1) =  6## : Aji(4,2)  =   6## : Aji(4,3) =  6## : Aji(4,4)  =   6## : Cj(4)  = 0.4##
    Aji(5,1) =  3## : Aji(5,2)  =   7## : Aji(5,3) =  3## : Aji(5,4)  =   7## : Cj(5)  = 0.4##

    Z = 0##

    FOR jj% = 1 TO m%  'NOTE:  Index jj% is used to avoid same variable name as j&

        Sum = 0##

        FOR i% = 1 TO 4 'Shekel's family is 4-D only

            Xi = R(p%,i%,j&)

            Sum = Sum + (Xi-Aji(jj%,i%))^2

        NEXT i%

      Z = Z + 1##/(Sum + Cj(jj%))

    NEXT jj%

    F21 = Z

END FUNCTION 'F21

'----------

FUNCTION F22(R(),Nd%,p%,j&) 'F22 (4-D) Shekel's Family m=7

'Global maximum = 10

    LOCAL Xi, Z, Sum, Aji(), Cj() AS EXT

    LOCAL i%, jj%, m%

    m% = 7 : REDIM Aji(1 TO m%, 1 TO 4), Cj(1 TO m%)

    Aji(1,1) =  4## : Aji(1,2)  =   4## : Aji(1,3) =  4## : Aji(1,4)  =   4## : Cj(1)  = 0.1##
    Aji(2,1) =  1## : Aji(2,2)  =   1## : Aji(2,3) =  1## : Aji(2,4)  =   1## : Cj(2)  = 0.2##
    Aji(3,1) =  8## : Aji(3,2)  =   8## : Aji(3,3) =  8## : Aji(3,4)  =   8## : Cj(3)  = 0.2##
    Aji(4,1) =  6## : Aji(4,2)  =   6## : Aji(4,3) =  6## : Aji(4,4)  =   6## : Cj(4)  = 0.4##
    Aji(5,1) =  3## : Aji(5,2)  =   7## : Aji(5,3) =  3## : Aji(5,4)  =   7## : Cj(5)  = 0.4##
    Aji(6,1) =  2## : Aji(6,2)  =   9## : Aji(6,3) =  2## : Aji(6,4)  =   9## : Cj(6)  = 0.6##
    Aji(7,1) =  5## : Aji(7,2)  =   5## : Aji(7,3) =  3## : Aji(7,4)  =   3## : Cj(7)  = 0.3##

    Z = 0##

    FOR jj% = 1 TO m%  'NOTE:  Index jj% is used to avoid same variable name as j&

        Sum = 0##

        FOR i% = 1 TO 4 'Shekel's family is 4-D only

            Xi = R(p%,i%,j&)

            Sum = Sum + (Xi-Aji(jj%,i%))^2

        NEXT i%

      Z = Z + 1##/(Sum + Cj(jj%))

    NEXT jj%

    F22 = Z

END FUNCTION 'F22

'----------

FUNCTION F23(R(),Nd%,p%,j&) 'F23 (4-D) Shekel's Family m=10

'Global maximum = 10

    LOCAL Xi, Z, Sum, Aji(), Cj() AS EXT

    LOCAL i%, jj%, m%

    m% = 10 : REDIM Aji(1 TO m%, 1 TO 4), Cj(1 TO m%)

    Aji(1,1)  =  4## : Aji(1,2)   =    4## : Aji(1,3)  =  4## : Aji(1,4)   =    4## : Cj(1)  = 0.1##
    Aji(2,1)  =  1## : Aji(2,2)   =    1## : Aji(2,3)  =  1## : Aji(2,4)   =    1## : Cj(2)  = 0.2##
    Aji(3,1)  =  8## : Aji(3,2)   =    8## : Aji(3,3)  =  8## : Aji(3,4)   =    8## : Cj(3)  = 0.2##
    Aji(4,1)  =  6## : Aji(4,2)   =    6## : Aji(4,3)  =  6## : Aji(4,4)   =    6## : Cj(4)  = 0.4##
    Aji(5,1)  =  3## : Aji(5,2)   =    7## : Aji(5,3)  =  3## : Aji(5,4)   =    7## : Cj(5)  = 0.4##
    Aji(6,1)  =  2## : Aji(6,2)   =    9## : Aji(6,3)  =  2## : Aji(6,4)   =    9## : Cj(6)  = 0.6##
    Aji(7,1)  =  5## : Aji(7,2)   =    5## : Aji(7,3)  =  3## : Aji(7,4)   =    3## : Cj(7)  = 0.3##
    Aji(8,1)  =  8## : Aji(8,2)   =    1## : Aji(8,3)  =  8## : Aji(8,4)   =    1## : Cj(8)  = 0.7##
    Aji(9,1)  =  6## : Aji(9,2)   =    2## : Aji(9,3)  =  6## : Aji(9,4)   =    2## : Cj(9)  = 0.5##
    Aji(10,1) =  7## : Aji(10,2)  = 3.6## : Aji(10,3) =  7## : Aji(10,4)  = 3.6## : Cj(10) = 0.5##

    Z = 0##

    FOR jj% = 1 TO m%  'NOTE:  Index jj% is used to avoid same variable name as j&

        Sum = 0##

        FOR i% = 1 TO 4 'Shekel's family is 4-D only
```

```
        Xi = R(p%,i%,j&)

        Sum = Sum + (Xi-Aji(jj%,i%))^2

    NEXT i%

    Z = Z + I##/(Sum + Cj(jj%))

  NEXT jj%

  F23 = Z

END FUNCTION 'F23
'======================================================= END FUNCTION DEFINITIONS =======================================================

SUB Plot2UbestProbeTrajectories(NumTrajectories%,M(),R(),Np%,Nd%,LastStep&,FunctionName$)

LOCAL TrajectoryNumber%, ProbeNumber%, StepNumber&, N%, M%, ProcID???

LOCAL MaximumFitness, MinimumFitness AS EXT

LOCAL BestProbeThisStep%()

LOCAL BestFitnessThisStep(), TempFitness() AS EXT

LOCAL Annotation$, xCoord$, yCoord$, GnuPlotEXE$, PlotWithLines$

    Annotation$   = ""

    PlotWithLines$ = "YES" '"NO"

    NumTrajectories% = MIN(Np%,NumTrajectories%)

    GnuPlotEXE$ = "wgnuplot.exe"
'    --------------- Get Min/Max Fitnesses -----------------

    MaximumFitness = M(1,0) : MinimumFitness = M(1,0)  'Note:  M(p%,j&)

    FOR StepNumber% = 0 TO LastStep&

        FOR ProbeNumber% = 1 TO Np%

            IF M(ProbeNumber%,StepNumber%) >= MaximumFitness THEN MaximumFitness = M(ProbeNumber%,StepNumber&)

            IF M(ProbeNumber%,StepNumber%) =< MinimumFitness THEN MinimumFitness = M(ProbeNumber%,StepNumber&)

        NEXT ProbeNumber%

    NEXT StepNumber%
'    ------------ Copy Fitness Array M() into TempFitness to Preserve M() ----------------

    REDIM TempFitness(1 TO Np%, 0 TO LastStep&)

    FOR StepNumber& = 0 TO LastStep&

        FOR ProbeNumber% = 1 TO Np%

            TempFitness(ProbeNumber%,StepNumber&) = M(ProbeNumber%,StepNumber&)

        NEXT ProbeNumber%

    NEXT StepNumber%
'    ------------ LOOP ON TRAJECTORIES -----------

    FOR TrajectoryNumber% = 1 TO NumTrajectories%
'        --------------- Get Trajectory Coordinate Data ----------------

        REDIM BestFitnessThisStep(0 TO LastStep&), BestProbeThisStep(0 TO LastStep&)

        FOR StepNumber& = 0 TO LastStep&

            BestFitnessThisStep(StepNumber&) = TempFitness(1,StepNumber%)

            FOR ProbeNumber% = 1 TO Np%

                IF TempFitness(ProbeNumber%,StepNumber&) >= BestFitnessThisStep(StepNumber&) THEN

                    BestFitnessThisStep(StepNumber&) = TempFitness(ProbeNumber%,StepNumber&)

                    BestProbeThisStep%(StepNumber&)  = ProbeNumber%

                END IF

            NEXT ProbeNumber%

        NEXT StepNumber&
'    ----- Create Plot Data File -----

    N% = FREEFILE

    SELECT CASE TrajectoryNumber%

        CASE 1  : OPEN "t1"  FOR OUTPUT AS #N%
        CASE 2  : OPEN "t2"  FOR OUTPUT AS #N%
        CASE 3  : OPEN "t3"  FOR OUTPUT AS #N%
        CASE 4  : OPEN "t4"  FOR OUTPUT AS #N%
        CASE 5  : OPEN "t5"  FOR OUTPUT AS #N%
        CASE 6  : OPEN "t6"  FOR OUTPUT AS #N%
        CASE 7  : OPEN "t7"  FOR OUTPUT AS #N%
        CASE 8  : OPEN "t8"  FOR OUTPUT AS #N%
        CASE 9  : OPEN "t9"  FOR OUTPUT AS #N%
        CASE 10 : OPEN "t10" FOR OUTPUT AS #N%

    END SELECT
'    ------------ Write Plot File Data ------------

    FOR StepNumber& = 0 TO LastStep&

        PRINT #N%, USING$("######.######## ######.########",R(BestProbeThisStep%(StepNumber&),1,StepNumber%),R(BestProbeThisStep%(StepNumber&),2,StepNumber%))

        TempFitness(BestProbeThisStep%(StepNumber&),StepNumber%) = MinimumFitness 'so that same max will not be found for next trajectory

    NEXT StepNumber%

    CLOSE #N%

    NEXT TrajectoryNumber%
'    ------------------------- Plot Trajectories -------------------------
```

```
CALL CreateGNUplotINIfile(0.13##*ScreenWidth&,0.18##*ScreenHeight&,0.7##*ScreenHeight&,0.7##*ScreenHeight&)

Annotation$ = ""

N% = FREEFILE

OPEN "cmd2d.gp" FOR OUTPUT AS #N%

    PRINT #N%, "set xrange ["+REMOVE$(STR$(XiMin(1)),ANY" ")+":"+REMOVE$(STR$(XiMax(1)),ANY" ")+"]"
    PRINT #N%, "set yrange ["+REMOVE$(STR$(XiMin(2)),ANY" ")+":"+REMOVE$(STR$(XiMax(2)),ANY" ")+"]"

    'PRINT #N%, "set label "      + Quote$ + Annotation$ + Quote$ + " at graph " + xCoord$ + "," + yCoord$
    PRINT #N%, "set grid xtics " + "10"
    PRINT #N%, "set grid ytics " + "10"
    PRINT #N%, "set grid mxtics"
    PRINT #N%, "set grid mytics"
    PRINT #N%, "show grid"
    PRINT #N%, "set title " + Quote$ + "2D "+ FunctionName$+" TRAJECTORIES OF PROBES WITH BEST\nFITNESSES (ORDERED BY FITNESS)" + "\n" + RunID$ + Quote$
    PRINT #N%, "set xlabel " + Quote$ + "x1\n\n"                        + Quote$
    PRINT #N%, "set ylabel " + Quote$ + "\nx2"                          + Quote$

    IF PlotWithLines$ = "YES" THEN

        SELECT CASE NumTrajectories%

            CASE 1 : PRINT #N%, "plot "+Quote$+"t1"+Quote$+" w 1 lw 3"
            CASE 2 : PRINT #N%, "plot "+Quote$+"t1"+Quote$+" w 1 lw 3,"+Quote$+"t2"+Quote$+" w 1"
            CASE 3 : PRINT #N%, "plot "+Quote$+"t1"+Quote$+" w 1 lw 3,"+Quote$+"t2"+Quote$+" w 1,"+Quote$+"t3"+Quote$+" w 1"
            CASE 4 : PRINT #N%, "plot "+Quote$+"t1"+Quote$+" w 1 lw 3,"+Quote$+"t2"+Quote$+" w 1,"+Quote$+"t3"+Quote$+" w 1,"+Quote$+"t4"+Quote$+" w 1"
            CASE 5 : PRINT #N%, "plot "+Quote$+"t1"+Quote$+" w 1 lw 3,"+Quote$+"t2"+Quote$+" w 1,"+Quote$+"t3"+Quote$+" w 1,"+Quote$+"t4"+Quote$+" w 1,"+Quote$+"t4"+Quote$+" w 1"+
l,"+Quote$+"t5"+Quote$+" w 1"
            CASE 6 : PRINT #N%, "plot "+Quote$+"t1"+Quote$+" w 1 lw 3,"+Quote$+"t2"+Quote$+" w 1,"+Quote$+"t3"+Quote$+" w 1,"+Quote$+"t4"+Quote$+" w
l,"+Quote$+"t5"+Quote$+" w 1,"+Quote$+"t6"+Quote$+" w 1"
            CASE 7 : PRINT #N%, "plot "+Quote$+"t1"+Quote$+" w 1 lw 3,"+Quote$+"t2"+Quote$+" w 1,"+Quote$+"t3"+Quote$+" w 1,"+Quote$+"t4"+Quote$+" w
l,"+Quote$+"t5"+Quote$+" w 1,"+Quote$+"t6"+Quote$+" w 1,"+_
                                                        Quote$+"t7"+Quote$+" w 1"
            CASE 8 : PRINT #N%, "plot "+Quote$+"t1"+Quote$+" w 1 lw 3,"+Quote$+"t2"+Quote$+" w 1,"+Quote$+"t3"+Quote$+" w 1,"+Quote$+"t4"+Quote$+" w
l,"+Quote$+"t5"+Quote$+" w 1,"+Quote$+"t6"+Quote$+" w 1,"+_
                                                        Quote$+"t7"+Quote$+" w 1,"    +Quote$+"t8"+Quote$+" w 1"
            CASE 9 : PRINT #N%, "plot "+Quote$+"t1"+Quote$+" w 1 lw 3,"+Quote$+"t2"+Quote$+" w 1,"+Quote$+"t3"+Quote$+" w 1,"+Quote$+"t4"+Quote$+" w
l,"+Quote$+"t5"+Quote$+" w 1,"+Quote$+"t6"+Quote$+" w 1,"+_
                                                        Quote$+"t7"+Quote$+" w 1,"    +Quote$+"t8"+Quote$+" w 1,"+Quote$+"t9"+Quote$+" w 1"
            CASE 10 : PRINT #N%, "plot "+Quote$+"t1"+Quote$+" w 1 lw 3,"+Quote$+"t2"+Quote$+" w 1,"+Quote$+"t3"+Quote$+" w 1,"+Quote$+"t4"+Quote$+" w
l,"+Quote$+"t5"+Quote$+" w 1,"+Quote$+"t6"+Quote$+" w 1,"+_
                                                        Quote$+"t7"+Quote$+" w 1,"    +Quote$+"t8"+Quote$+" w 1,"+Quote$+"t9"+Quote$+" w 1,"+Quote$+"t10"+Quote$+" w 1"

        END SELECT

    ELSE

        SELECT CASE NumTrajectories%

            CASE 1 : PRINT #N%, "plot "+Quote$+"t1"+Quote$+" lw 2"
            CASE 2 : PRINT #N%, "plot "+Quote$+"t1"+Quote$+" lw 2,"+Quote$+"t2"+Quote$
            CASE 3 : PRINT #N%, "plot "+Quote$+"t1"+Quote$+" lw 2,"+Quote$+"t2"+Quote$+" ,"+Quote$+"t3"+Quote$
            CASE 4 : PRINT #N%, "plot "+Quote$+"t1"+Quote$+" lw 2,"+Quote$+"t2"+Quote$+" ,"+Quote$+"t3"+Quote$+" ,"+Quote$+"t4"+Quote$
            CASE 5 : PRINT #N%, "plot "+Quote$+"t1"+Quote$+" lw 2,"+Quote$+"t2"+Quote$+" ,"+Quote$+"t3"+Quote$+" ,"+Quote$+"t4"+Quote$+" ,"+Quote$+"t5"+Quote$
            CASE 6 : PRINT #N%, "plot "+Quote$+"t1"+Quote$+" lw 2,"+Quote$+"t2"+Quote$+" ,"+Quote$+"t3"+Quote$+" ,"+Quote$+"t4"+Quote$+" ,"+Quote$+"t5"+Quote$+"
,"+Quote$+"t6"+Quote$
            CASE 7 : PRINT #N%, "plot "+Quote$+"t1"+Quote$+" lw 2,"+Quote$+"t2"+Quote$+" ,"+Quote$+"t3"+Quote$+" ,"+Quote$+"t4"+Quote$+" ,"+Quote$+"t5"+Quote$+"
,"+Quote$+"t6"+Quote$+" ,"+_
                                                        Quote$+"t7"+Quote$
            CASE 8 : PRINT #N%, "plot "+Quote$+"t1"+Quote$+" lw 2,"+Quote$+"t2"+Quote$+" ,"+Quote$+"t3"+Quote$+" ,"+Quote$+"t4"+Quote$+" ,"+Quote$+"t5"+Quote$+"
,"+Quote$+"t6"+Quote$+" ,"+_
                                                        Quote$+"t7"+Quote$+" ,"    +Quote$+"t8"+Quote$
            CASE 9 : PRINT #N%, "plot "+Quote$+"t1"+Quote$+" lw 2,"+Quote$+"t2"+Quote$+" ,"+Quote$+"t3"+Quote$+" ,"+Quote$+"t4"+Quote$+" ,"+Quote$+"t5"+Quote$+"
,"+Quote$+"t6"+Quote$+" ,"+_
                                                        Quote$+"t7"+Quote$+" ,"    +Quote$+"t8"+Quote$+" ,"+Quote$+"t9"+Quote$
            CASE 10 : PRINT #N%, "plot "+Quote$+"t1"+Quote$+" lw 2,"+Quote$+"t2"+Quote$+" ,"+Quote$+"t3"+Quote$+" ,"+Quote$+"t4"+Quote$+" ,"+Quote$+"t5"+Quote$+"
,"+Quote$+"t6"+Quote$+" ,"+_
                                                        Quote$+"t7"+Quote$+" ,"    +Quote$+"t8"+Quote$+" ,"+Quote$+"t9"+Quote$+" ,"+Quote$+"t10"+Quote$

        END SELECT

    END IF

    CLOSE #N%

    ProcID??? = SHELL(GnuPlotEXE$+" cmd2d.gp -") : CALL Delay(0.5##)

END SUB 'Plot2DbestProbeTrajectories()

'-----

SUB Plot2DindividualProbeTrajectories(NumTrajectories%,M(),R(),Np%,Nd%,LastStep&,FunctionName$)

LOCAL ProbeNumber%, StepNumber&, N%, ProcID???

LOCAL Annotation$, xCoord$, yCoord$, GnuPlotEXE$, PlotWithLines$

    NumTrajectories% = MIN(Np%,NumTrajectories%)

    Annotation$    = ""

    PlotWithLines$ = "YES" '"NO"

    GnuPlotEXE$ = "wgnuplot.exe"

'   ------------ LOOP ON PROBES ----------------

    FOR ProbeNumber% = 1 TO MIN(NumTrajectories%,Np%)

'   ----- Create Plot Data File -----

        N% = FREEFILE

        SELECT CASE ProbeNumber%
            CASE 1 : OPEN "p1"  FOR OUTPUT AS #N%
            CASE 2 : OPEN "p2"  FOR OUTPUT AS #N%
            CASE 3 : OPEN "p3"  FOR OUTPUT AS #N%
            CASE 4 : OPEN "p4"  FOR OUTPUT AS #N%
            CASE 5 : OPEN "p5"  FOR OUTPUT AS #N%
            CASE 6 : OPEN "p6"  FOR OUTPUT AS #N%
            CASE 7 : OPEN "p7"  FOR OUTPUT AS #N%
            CASE 8 : OPEN "p8"  FOR OUTPUT AS #N%
            CASE 9 : OPEN "p9"  FOR OUTPUT AS #N%
            CASE 10 : OPEN "p10" FOR OUTPUT AS #N%
            CASE 11 : OPEN "p11" FOR OUTPUT AS #N%
            CASE 12 : OPEN "p12" FOR OUTPUT AS #N%
            CASE 13 : OPEN "p13" FOR OUTPUT AS #N%
            CASE 14 : OPEN "p14" FOR OUTPUT AS #N%
            CASE 15 : OPEN "p15" FOR OUTPUT AS #N%
            CASE 16 : OPEN "p16" FOR OUTPUT AS #N%

        END SELECT

'   ------------ Write Plot File Data ------------
```

```
    FOR StepNumber% = 0 TO LastSteps%

        PRINT #N%, USINGS("######.######## ######.########",R(ProbeNumber%,1,StepNumber&),R(ProbeNumber%,2,StepNumber&))

    NEXT StepNumber%

    CLOSE #N%

NEXT ProbeNumber%
'   -------------------------------------- Plot Trajectories --------------------------------------
'usage: CALL CreateGNUplotINIfile(PlotWindowULC_X%,PlotWindowULC_Y%,PlotWindowWidth%,PlotWindowHeight%)

    CALL CreateGNUplotINIfile(0.17##*ScreenWidth&,0.22##*ScreenHeight&,0.7##*ScreenHeight&,0.7##*ScreenHeight&)

    Annotation$ = ""

    N% = FREEFILE

    OPEN "cmd2d.gp" FOR OUTPUT AS #N%

        PRINT #N%, "set xrange ["+REMOVE$(STR$(XiMin(1)),ANY"" )+":"+REMOVE$(STR$(XiMax(1)),ANY" ")+"]"
        PRINT #N%, "set yrange ["+REMOVE$(STR$(XiMin(2)),ANY" ")+":"+REMOVE$(STR$(XiMax(2)),ANY" ")+"]"

        PRINT #N%, "set grid xtics " + "10"
        PRINT #N%, "set grid ytics " + "10"
        PRINT #N%, "set grid mxtics"
        PRINT #N%, "set grid mytics"
        PRINT #N%, "show grid"
        PRINT #N%, "set title " + Quote$ + "2D "+ FunctionName$+" INDIVIDUAL PROBE TRAJECTORIES\n(ORDERED BY PROBE #)" + "\n" + RunID$ + Quote$
        PRINT #N%, "set xlabel " + Quote$ + "x1\n\n"                          + Quote$
        PRINT #N%, "set ylabel " + Quote$ + "\nx2"                            + Quote$

    IF PlotWithLines$ = "YES" THEN

        SELECT CASE NumTrajectories%

            CASE 1 : PRINT #N%, "plot "+Quote$+"p1"  +Quote$+" w l lw 1"
            CASE 2 : PRINT #N%, "plot "+Quote$+"p1"  +Quote$+" w l lw 1,"+Quote$+"p2"+Quote$+" w l"
            CASE 3 : PRINT #N%, "plot "+Quote$+"p1"  +Quote$+" w l lw 1,"+Quote$+"p2"+Quote$+" w l,"+Quote$+"p3"+Quote$+" w l"
            CASE 4 : PRINT #N%, "plot "+Quote$+"p1"  +Quote$+" w l lw 1,"+quote$+"p2"+Quote$+" w l,"+Quote$+"p3"+Quote$+" w l,"+Quote$+"p4"+Quote$+" w l"
            CASE 5 : PRINT #N%, "plot "+Quote$+"p1"  +Quote$+" w l lw 1,"+Quote$+"p2"+Quote$+" w l,"+Quote$+"p3"+Quote$+" w l,"+Quote$+"p4"+Quote$+" w
l,"+Quote$+"p5"+Quote$+" w l"
            CASE 6 : PRINT #N%, "plot "+Quote$+"p1"  +Quote$+" w l lw 1,"+Quote$+"p2"+Quote$+" w l,"+Quote$+"p3"+Quote$+" w l,"+Quote$+"p4"+Quote$+" w
l,"+Quote$+"p5"+Quote$+" w l,"+Quote$+"p6"+Quote$+" w l"
            CASE 7 : PRINT #N%, "plot "+Quote$+"p1"  +Quote$+" w l lw 1,"+Quote$+"p2"+Quote$+" w l,"+Quote$+"p3"+Quote$+" w l,"+Quote$+"p4"+Quote$+" w
l,"+Quote$+"p5"+Quote$+" w l,"+Quote$+"p6"+Quote$+" w l,"+_
                                               Quote$+"p7"  +Quote$+" w l"
            CASE 8 : PRINT #N%, "plot "+Quote$+"p1"  +Quote$+" w l lw 1,"+Quote$+"p2"+Quote$+" w l,"+Quote$+"p3"+Quote$+" w l,"+Quote$+"p4"+Quote$+" w
l,"+Quote$+"p5"+Quote$+" w l,"+Quote$+"p6"+Quote$+" w l,"+_
                                               Quote$+"p7"  +Quote$+" w l,"        +Quote$+"p8"+Quote$+" w l"
            CASE 9 : PRINT #N%, "plot "+Quote$+"p1"  +Quote$+" w l lw 1,"+Quote$+"p2"+Quote$+" w l,"+Quote$+"p3"+Quote$+" w l,"+Quote$+"p4"+Quote$+" w
l,"+Quote$+"p5"+Quote$+" w l,"+Quote$+"p6"+Quote$+" w l,"+_
                                               Quote$+"p7"  +Quote$+" w l,"        +Quote$+"p8"+Quote$+" w l,"+Quote$+"p9"+Quote$+" w l"
            CASE 10 : PRINT #N%, "plot "+Quote$+"p1"  +Quote$+" w l lw 1,"+Quote$+"p2"  +Quote$+" w l,"+Quote$+"p3"  +Quote$+" w l,"+Quote$+"p4"  +Quote$+" w
l,"+Quote$+"p5"+Quote$+" w l,"+Quote$+"p6"+Quote$+" w l,"+_
                                                Quote$+"p7"  +Quote$+" w l,"        +Quote$+"p8"+Quote$+" w l,"+Quote$+"p9"+Quote$+" w l,"+Quote$+"p10"+Quote$+" w l"
            CASE 11 : PRINT #N%, "plot "+Quote$+"p1"  +Quote$+" w l lw 1,"+Quote$+"p2"  +Quote$+" w l,"+Quote$+"p3"  +Quote$+" w l,"+Quote$+"p4"  +Quote$+" w
l,"+Quote$+"p5"+Quote$+" w l,"+Quote$+"p6"+Quote$+" w l,"+_
                                                Quote$+"p7"  +Quote$+" w l,"        +Quote$+"p8"+Quote$+" w l,"+Quote$+"p9"+Quote$+" w l,"+Quote$+"p10"+Quote$+" w
l,"+Quote$+"p11"+Quote$+" w l"
            CASE 12 : PRINT #N%, "plot "+Quote$+"p1"  +Quote$+" w l lw 1,"+Quote$+"p2"  +Quote$+" w l,"+Quote$+"p3"  +Quote$+" w l,"+Quote$+"p4"  +Quote$+" w
l,"+Quote$+"p5"+Quote$+" w l,"+Quote$+"p6"+Quote$+" w l,"+_
                                                Quote$+"p7"  +Quote$+" w l,"        +Quote$+"p8"+Quote$+" w l,"+Quote$+"p9"+Quote$+" w l,"+Quote$+"p10"+Quote$+" w
l,"+Quote$+"p11"+Quote$+" w l,"+Quote$+"p12"+Quote$+" w l"
            CASE 13 : PRINT #N%, "plot "+Quote$+"p1"  +Quote$+" w l lw 1,"+Quote$+"p2"  +Quote$+" w l,"+Quote$+"p3"  +Quote$+" w l,"+Quote$+"p4"  +Quote$+" w
l,"+Quote$+"p5"+Quote$+" w l,"+Quote$+"p6"+Quote$+" w l,"+_
                                                Quote$+"p7"  +Quote$+" w l,"        +Quote$+"p8"+Quote$+" w l,"+Quote$+"p9"+Quote$+" w l,"+Quote$+"p10"+Quote$+" w
l,"+Quote$+"p11"+Quote$+" w l,"+Quote$+"p12"+Quote$+" w l,"+_
                                                Quote$+"p13"+Quote$+" w l"
            CASE 14 : PRINT #N%, "plot "+Quote$+"p1"  +Quote$+" w l lw 1,"+Quote$+"p2"  +Quote$+" w l,"+Quote$+"p3"  +Quote$+" w l,"+Quote$+"p4"  +Quote$+" w
l,"+Quote$+"p5"+Quote$+" w l,"+Quote$+"p6"+Quote$+" w l,"+_
                                                Quote$+"p7"  +Quote$+" w l,"        +Quote$+"p8"+Quote$+" w l,"+Quote$+"p9"+Quote$+" w l,"+Quote$+"p10"+Quote$+" w
l,"+Quote$+"p11"+Quote$+" w l,"+Quote$+"p12"+Quote$+" w l,"+_
                                                Quote$+"p13"  +Quote$+" w l,"       +Quote$+"p14"+Quote$+" w l"
            CASE 15 : PRINT #N%, "plot "+Quote$+"p1"  +Quote$+" w l lw 1,"+Quote$+"p2"  +Quote$+" w l,"+Quote$+"p3"  +Quote$+" w l,"+Quote$+"p4"  +Quote$+" w
l,"+Quote$+"p5"+Quote$+" w l,"+Quote$+"p6"+Quote$+" w l,"+_
                                                Quote$+"p7"  +Quote$+" w l,"        +Quote$+"p8"+Quote$+" w l,"+Quote$+"p9"+Quote$+" w l,"+Quote$+"p10"+Quote$+" w
l,"+Quote$+"p11"+Quote$+" w l,"+Quote$+"p12"+Quote$+" w l,"+_
                                                Quote$+"p13"  +Quote$+" w l,"       +Quote$+"p14"+Quote$+" w l,"+Quote$+"p15"+Quote$+" w l"
            CASE 16 : PRINT #N%, "plot "+Quote$+"p1"  +Quote$+" w l lw 1,"+Quote$+"p2"  +Quote$+" w l,"+Quote$+"p3"  +Quote$+" w l,"+Quote$+"p4"  +Quote$+" w
l,"+Quote$+"p5"+Quote$+" w l,"+Quote$+"p6"+Quote$+" w l,"+_
                                                Quote$+"p7"  +Quote$+" w l,"        +Quote$+"p8"+Quote$+" w l,"+Quote$+"p9"+Quote$+" w l,"+Quote$+"p10"+Quote$+" w
l,"+Quote$+"p11"+Quote$+" w l,"+Quote$+"p12"+Quote$+" w l,"+_
                                                Quote$+"p13"  +Quote$+" w l,"       +Quote$+"p14"+Quote$+" w l,"+Quote$+"p15"+Quote$+" w l,"+Quote$+"p16"+Quote$+" w l"
        END SELECT

    ELSE

        SELECT CASE NumTrajectories%

            CASE 1 : PRINT #N%, "plot "+Quote$+"p1"+Quote$+" lw 1"
            CASE 2 : PRINT #N%, "plot "+Quote$+"p1"+Quote$+" lw 1,"+Quote$+"p2"+Quote$
            CASE 3 : PRINT #N%, "plot "+Quote$+"p1"+Quote$+" lw 1,"+Quote$+"p2"+Quote$+" ,"+Quote$+"p3"+Quote$
            CASE 4 : PRINT #N%, "plot "+Quote$+"p1"+Quote$+" lw 1,"+Quote$+"p2"+Quote$+" ,"+Quote$+"p3"+Quote$+" ,"+Quote$+"p4"+Quote$
            CASE 5 : PRINT #N%, "plot "+Quote$+"p1"+Quote$+" lw 1,"+Quote$+"p2"+Quote$+" ,"+Quote$+"p3"+Quote$+" ,"+Quote$+"p4"+Quote$+" ,"+Quote$+"p5"+Quote$
            CASE 6 : PRINT #N%, "plot "+Quote$+"p1"+Quote$+" lw 1,"+Quote$+"p2"+Quote$+" ,"+Quote$+"p3"+Quote$+" ,"+Quote$+"p4"+Quote$+" ,"+Quote$+"p5"+Quote$+"
,"+Quote$+"p6"+Quote$
            CASE 7 : PRINT #N%, "plot "+Quote$+"p1"+Quote$+" lw 1,"+Quote$+"p2"+Quote$+" ,"+Quote$+"p3"+Quote$+" ,"+Quote$+"p4"+Quote$+" ,"+Quote$+"p5"+Quote$+"
,"+Quote$+"p6"+Quote$+" ,"+_
                                               Quote$+"p7"+Quote$
            CASE 8 : PRINT #N%, "plot "+Quote$+"p1"+Quote$+" lw 1,"+Quote$+"p2"+Quote$+" ,"+Quote$+"p3"+Quote$+" ,"+Quote$+"p4"+Quote$+" ,"+Quote$+"p5"+Quote$+"
,"+Quote$+"p6"+Quote$+" ,"+_
                                               Quote$+"p7"+Quote$+" ,"        +Quote$+"p8"+Quote$
            CASE 9 : PRINT #N%, "plot "+Quote$+"p1"+Quote$+" lw 1,"+Quote$+"p2"+Quote$+" ,"+Quote$+"p3"+Quote$+" ,"+Quote$+"p4"+Quote$+" ,"+Quote$+"p5"+Quote$+"
,"+Quote$+"p6"+Quote$+" ,"+_
                                               Quote$+"p7"+Quote$+" ,"        +Quote$+"p8"+Quote$+" ,"+Quote$+"p9"+Quote$
            CASE 10 : PRINT #N%, "plot "+Quote$+"p1"+Quote$+" lw 1,"  +Quote$+"p2" +Quote$+" ,"+Quote$+"p3"+Quote$+" ,"+Quote$+"p4"  +Quote$+" ,"+Quote$+"p5"+Quote$+"
,"+Quote$+"p6"+Quote$+" ,"+_
                                                Quote$+"p7"+Quote$+" ,"        +Quote$+"p8"+Quote$+" ,"+Quote$+"p9"+Quote$+" ,"+Quote$+"p10"  +Quote$
            CASE 11 : PRINT #N%, "plot "+Quote$+"p1"+Quote$+" lw 1,"  +Quote$+"p2" +Quote$+" ,"+Quote$+"p3"+Quote$+" ,"+Quote$+"p4"  +Quote$+" ,"+Quote$+"p5" +Quote$+"
,"+Quote$+"p6"+Quote$+" ,"+_
                                                Quote$+"p7"+Quote$+" ,"        +Quote$+"p8"+Quote$+" ,"+Quote$+"p9"+Quote$+" ,"+Quote$+"p10"+Quote$+" ,"+Quote$+"p11"+Quote$
            CASE 12 : PRINT #N%, "plot "+Quote$+"p1"+Quote$+" lw 1,"  +Quote$+"p2" +Quote$+" ,"+Quote$+"p3"+Quote$+" ,"+Quote$+"p4"  +Quote$+" ,"+Quote$+"p5"  +Quote$+"
,"+Quote$+"p6" +Quote$+" ,"+_
```

```
                                                Quote$+"p7"+Quote$+" ,"    +Quote$+"p8" +Quote$+" ,"+Quote$+"p9"+Quote$+" ,"+Quote$+"p10" +Quote$+" ,"+Quote$+"p11"+Quote$+"
,"+Quote$+"p12"+Quote$

            CASE 13 : PRINT #N%, "plot "+Quote$+"p1" +Quote$+" lw 1,"+Quote$+"p2" +Quote$+" ,"+Quote$+"p3"+Quote$+" ,"+Quote$+"p4" +Quote$+" ,"+Quote$+"p5" +Quote$+" ,"
+Quote$+"p6" +Quote$+" ,"+_
+Quote$+"p12"+Quote$+" ,"+_                      Quote$+"p7" +Quote$+" ,"    +Quote$+"p8" +Quote$+" ,"+Quote$+"p9"+Quote$+" ,"+Quote$+"p10" +Quote$+" ,"+Quote$+"p11"+Quote$+" ,"
                                                Quote$+"p13"+Quote$

            CASE 14 : PRINT #N%, "plot "+Quote$+"p1" +Quote$+" lw 1,"+Quote$+"p2" +Quote$+" ,"+Quote$+"p3"+Quote$+" ,"+Quote$+"p4"  +Quote$+" ,"+Quote$+"p5" +Quote$+" ,"
+Quote$+"p6" +Quote$+" ,"+_
+Quote$+"p12"+Quote$+" ,"+_                      Quote$+"p7" +Quote$+" ,"    +Quote$+"p8" +Quote$+" ,"+Quote$+"p9"+Quote$+" ,"+Quote$+"p10" +Quote$+" ,"+Quote$+"p11" +Quote$+" ,"
                                                Quote$+"p13"+Quote$+" ,"    +Quote$+"p14"+Quote$

            CASE 15 : PRINT #N%, "plot "+Quote$+"p1" +Quote$+" lw 1,"+Quote$+"p2" +Quote$+" ,"+Quote$+"p3" +Quote$+" ,"+Quote$+"p4" +Quote$+" ,"+Quote$+"p5" +Quote$+" ,"
+Quote$+"p6" +Quote$+" ,"+_
+Quote$+"p12"+Quote$+" ,"+_                      Quote$+"p7" +Quote$+" ,"    +Quote$+"p8" +Quote$+" ,"+Quote$+"p9" +Quote$+" ,"+Quote$+"p10"+Quote$+" ,"+Quote$+"p11" +Quote$+" ,"
                                                Quote$+"p13"+Quote$+" ,"    +Quote$+"p14"+Quote$+" ,"+Quote$+"p15"+Quote$

            CASE 16 : PRINT #N%, "plot "+Quote$+"p1" +Quote$+" lw 1,"+Quote$+"p2" +Quote$+" ,"+Quote$+"p3" +Quote$+" ,"+Quote$+"p4" +Quote$+" ,"+Quote$+"p5" +Quote$+" ,"
+Quote$+"p6" +Quote$+" ,"+_
+Quote$+"p12"+Quote$+" ,"+_                      Quote$+"p7" +Quote$+" ,"    +Quote$+"p8" +Quote$+" ,"+Quote$+"p9" +Quote$+" ,"+Quote$+"p10"+Quote$+" ,"+Quote$+"p11" +Quote$+" ,"
                                                Quote$+"p13"+Quote$+" ,"    +Quote$+"p14"+Quote$+" ,"+Quote$+"p15"+Quote$+" ,"+Quote$+"p16"+Quote$
        END SELECT

    END IF

  CLOSE #N%

  ProcID??? = SHELL(GnuPlotEXE$+" cmd2d.gp -") : CALL Delay(0.5##)

END SUB 'Plot2DindividualProbeTrajectories()

'----

SUB Plot3DbestProbeTrajectories(NumTrajectories%,M(),R(),Np%,Nd%,LastStep&,FunctionName$) 'XYZZY

LOCAL TrajectoryNumber%, ProbeNumber%, StepNumber&, N%, M%, ProcID???

LOCAL MaximumFitness, MinimumFitness AS EXT

LOCAL BestProbeThisStep%()

LOCAL BestFitnessThisStep(), TempFitness() AS EXT

LOCAL Annotation$, xCoord$, yCoord$, zCoord$, GnuPlotEXE$, PlotWithLines$

  Annotation$      = ""
  PlotWithLines$   = "NO" '"YES" '"NO"
  NumTrajectories% = MIN(Np%,NumTrajectories%)
  GnuPlotEXE$ = "wgnuplot.exe"

'  --------------- Get Min/Max Fitnesses ----------------
  MaximumFitness = M(1,0) : MinimumFitness = M(1,0)  'Note:  M(p%,j&)
  FOR StepNumber% = 0 TO LastStep&

    FOR ProbeNumber% = 1 TO Np%

      IF M(ProbeNumber%,StepNumber&) >= MaximumFitness THEN MaximumFitness = M(ProbeNumber%,StepNumber&)
      IF M(ProbeNumber%,StepNumber&) =< MinimumFitness THEN MinimumFitness = M(ProbeNumber%,StepNumber&)

    NEXT ProbeNumber%
  NEXT StepNumber%
'  ------------- Copy Fitness Array M() into TempFitness to Preserve M() ---------------
  REDIM TempFitness(1 TO Np%, 0 TO LastStep&)
  FOR StepNumber& = 0 TO LastStep&

    FOR ProbeNumber% = 1 TO Np%

      TempFitness(ProbeNumber%,StepNumber%) = M(ProbeNumber%,StepNumber%)
    NEXT ProbeNumber%
  NEXT StepNumber%
'  ------------- LOOP ON TRAJECTORIES ----------
  FOR TrajectoryNumber% = 1 TO NumTrajectories%
'      --------------- Get Trajectory Coordinate Data ----------------
      REDIM BestFitnessThisStep(0 TO LastStep&), BestProbeThisStep%(0 TO LastStep&)
      FOR StepNumber& = 0 TO LastStep&

        BestFitnessThisStep(StepNumber&) = TempFitness(1,StepNumber&)

        FOR ProbeNumber% = 1 TO Np%

          IF TempFitness(ProbeNumber%,StepNumber&) >= BestFitnessThisStep(StepNumber&) THEN

            BestFitnessThisStep(StepNumber&) = TempFitness(ProbeNumber%,StepNumber&)

            BestProbeThisStep%(StepNumber&) = ProbeNumber%

          END IF
        NEXT ProbeNumber%
      NEXT StepNumber&
'  ----- Create Plot Data File -----
    N% = FREEFILE

    SELECT CASE TrajectoryNumber%
        CASE 1 : OPEN "t1" FOR OUTPUT AS #N%
        CASE 2 : OPEN "t2" FOR OUTPUT AS #N%
        CASE 3 : OPEN "t3" FOR OUTPUT AS #N%
        CASE 4 : OPEN "t4" FOR OUTPUT AS #N%
        CASE 5 : OPEN "t5" FOR OUTPUT AS #N%
        CASE 6 : OPEN "t6" FOR OUTPUT AS #N%
        CASE 7 : OPEN "t7" FOR OUTPUT AS #N%
```

```
            CASE 8  : OPEN "t8"  FOR OUTPUT AS #N%
            CASE 9  : OPEN "t9"  FOR OUTPUT AS #N%
            CASE 10 : OPEN "t10" FOR OUTPUT AS #N%

    END SELECT

'   ----------- Write Plot File Data ------------

    FOR StepNumber& = 0 TO LastStep&

        PRINT #N%, USING$("######.######## ######.########
######.########",R(BestProbeThisStep%(StepNumber&),1,StepNumber&),R(BestProbeThisStep%(StepNumber&),2,StepNumber&),R(BestProbeThisStep%(StepNumber&),3,StepNumber&))+CHR$(13)

        TempFitness(BestProbeThisStep%(StepNumber&),StepNumber&) = MinimumFitness 'so that same max will not be found for next trajectory

    NEXT StepNumber%

    CLOSE #N%

    NEXT TrajectoryNumber%

'   ----------------------- Plot Trajectories --------------------------

    'CALL CreateGNUplotINIfile(0.1##*ScreenWidth&,0.25##*ScreenHeight&,0.6##*ScreenHeight&,0.6##*ScreenHeight&)

    Annotation$ = ""

    N% = FREEFILE

    OPEN "cmd3d.gp" FOR OUTPUT AS #N%

    PRINT #N%, "set pm3d"
    PRINT #N%, "show pm3d"
    PRINT #N%, "set hidden3d"
    PRINT #N%, "set view 45, 45, 1, 1"

    PRINT #N%, "unset colorbox"

    PRINT #N%, "set xrange [" + REMOVE$(STR$(XiMin(1)),ANY"" ) + ":" + REMOVE$(STR$(XiMax(1)),ANY"" ) + "]"
    PRINT #N%, "set yrange [" + REMOVE$(STR$(XiMin(2)),ANY"" ) + ":" + REMOVE$(STR$(XiMax(2)),ANY"" ) + "]"
    PRINT #N%, "set zrange [" + REMOVE$(STR$(XiMin(3)),ANY"" ) + ":" + REMOVE$(STR$(XiMax(3)),ANY"" ) + "]"

    PRINT #N%, "set grid xtics ytics ztics"
    PRINT #N%, "show grid"
    PRINT #N%, "set title " + Quote$ + "3D " + FunctionName$ + " PROBE TRAJECTORIES" + "\n" + RunID$ + Quote$
    PRINT #N%, "set xlabel " + Quote$ + "x1"                        + Quote$
    PRINT #N%, "set ylabel " + Quote$ + "x2"                        + Quote$
    PRINT #N%, "set zlabel " + Quote$ + "x3"                        + Quote$

    IF PlotWithLines$ = "YES" THEN

        SELECT CASE NumTrajectories%

            CASE 1  : PRINT #N%, "splot "+Quote$+"t1"+Quote$+" w l lw 1"
            CASE 2  : PRINT #N%, "splot "+Quote$+"t1"+Quote$+" w l lw 3,"+Quote$+"t2"+Quote$+" w l"
            CASE 3  : PRINT #N%, "splot "+Quote$+"t1"+Quote$+" w l lw 3,"+Quote$+"t2"+Quote$+" w l,"+Quote$+"t3"+Quote$+" w l"
            CASE 4  : PRINT #N%, "splot "+Quote$+"t1"+Quote$+" w l lw 3,"+Quote$+"t2"+Quote$+" w l,"+Quote$+"t3"+Quote$+" w l,"+Quote$+"t4"+Quote$+" w l"
            CASE 5  : PRINT #N%, "splot "+Quote$+"t1"+Quote$+" w l lw 3,"+Quote$+"t2"+Quote$+" w l,"+Quote$+"t3"+Quote$+" w l,"+Quote$+"t4"+Quote$+" w
l,"+Quote$+"t5"+Quote$+" w l"
            CASE 6  : PRINT #N%, "splot "+Quote$+"t1"+Quote$+" w l lw 3,"+Quote$+"t2"+Quote$+" w l,"+Quote$+"t3"+Quote$+" w l,"+Quote$+"t4"+Quote$+" w
l,"+Quote$+"t5"+Quote$+" w l,"+Quote$+"t6"+Quote$+" w l"
            CASE 7  : PRINT #N%, "splot "+Quote$+"t1"+Quote$+" w l lw 3,"+Quote$+"t2"+Quote$+" w l,"+Quote$+"t3"+Quote$+" w l,"+Quote$+"t4"+Quote$+" w
l,"+Quote$+"t5"+Quote$+" w l,"+Quote$+"t6"+Quote$+" w l,"+_
                                 Quote$+"t7"+Quote$+" w l"
            CASE 8  : PRINT #N%, "splot "+Quote$+"t1"+Quote$+" w l lw 3,"+Quote$+"t2"+Quote$+" w l,"+Quote$+"t3"+Quote$+" w l,"+Quote$+"t4"+Quote$+" w
l,"+Quote$+"t5"+Quote$+" w l,"+Quote$+"t6"+Quote$+" w l,"+_
                                 Quote$+"t7"+Quote$+" w l,"      +Quote$+"t8"+Quote$+" w l"
            CASE 9  : PRINT #N%, "splot "+Quote$+"t1"+Quote$+" w l lw 3,"+Quote$+"t2"+Quote$+" w l,"+Quote$+"t3"+Quote$+" w l,"+Quote$+"t4"+Quote$+" w
l,"+Quote$+"t5"+Quote$+" w l,"+Quote$+"t6"+Quote$+" w l,"+_
                                 Quote$+"t7"+Quote$+" w l,"      +Quote$+"t8"+Quote$+" w l,"+Quote$+"t9"+Quote$+" w l"
            CASE 10 : PRINT #N%, "splot "+Quote$+"t1"+Quote$+" w l lw 3,"+Quote$+"t2"+Quote$+" w l,"+Quote$+"t3"+Quote$+" w l,"+Quote$+"t4"+Quote$+" w
l,"+Quote$+"t5"+Quote$+" w l,"+Quote$+"t6"+Quote$+" w l,"+_
                                 Quote$+"t7"+Quote$+" w l,"      +Quote$+"t8"+Quote$+" w l,"+Quote$+"t9"+Quote$+" w l,"+Quote$+"t10"+Quote$+" w l"
        END SELECT

    ELSE

        SELECT CASE NumTrajectories%

            CASE 1  : PRINT #N%, "splot "+Quote$+"t1"+Quote$+" lw 2"
            CASE 2  : PRINT #N%, "splot "+Quote$+"t1"+Quote$+" lw 2,"+Quote$+"t2"+Quote$
            CASE 3  : PRINT #N%, "splot "+Quote$+"t1"+Quote$+" lw 2,"+Quote$+"t2"+Quote$+" ,"+Quote$+"t3"+Quote$
            CASE 4  : PRINT #N%, "splot "+Quote$+"t1"+Quote$+" lw 2,"+Quote$+"t2"+Quote$+" ,"+Quote$+"t3"+Quote$+" ,"+Quote$+"t4"+Quote$
            CASE 5  : PRINT #N%, "splot "+Quote$+"t1"+Quote$+" lw 2,"+Quote$+"t2"+Quote$+" ,"+Quote$+"t3"+Quote$+" ,"+Quote$+"t4"+Quote$+" ,"+Quote$+"t5"+Quote$
            CASE 6  : PRINT #N%, "splot "+Quote$+"t1"+Quote$+" lw 2,"+Quote$+"t2"+Quote$+" ,"+Quote$+"t3"+Quote$+" ,"+Quote$+"t4"+Quote$+" ,"+Quote$+"t5"+Quote$+"
,"+Quote$+"t6"+Quote$
            CASE 7  : PRINT #N%, "splot "+Quote$+"t1"+Quote$+" lw 2,"+Quote$+"t2"+Quote$+" ,"+Quote$+"t3"+Quote$+" ,"+Quote$+"t4"+Quote$+" ,"+Quote$+"t5"+Quote$+"
,"+Quote$+"t6"+Quote$+" ,"+_
                                 Quote$+"t7"+Quote$+" ,"   +Quote$+"t8"+Quote$
            CASE 8  : PRINT #N%, "splot "+Quote$+"t1"+Quote$+" lw 2,"+Quote$+"t2"+Quote$+" ,"+Quote$+"t3"+Quote$+" ,"+Quote$+"t4"+Quote$+" ,"+Quote$+"t5"+Quote$+"
,"+Quote$+"t6"+Quote$+" ,"+_
                                 Quote$+"t7"+Quote$+" ,"   +Quote$+"t8"+Quote$
            CASE 9  : PRINT #N%, "splot "+Quote$+"t1"+Quote$+" lw 2,"+Quote$+"t2"+Quote$+" ,"+Quote$+"t3"+Quote$+" ,"+Quote$+"t4"+Quote$+" ,"+Quote$+"t5"+Quote$+"
,"+Quote$+"t6"+Quote$+" ,"+_
                                 Quote$+"t7"+Quote$+" ,"   +Quote$+"t8"+Quote$+" ,"+Quote$+"t9"+Quote$
            CASE 10 : PRINT #N%, "splot "+Quote$+"t1"+Quote$+" lw 2,"+Quote$+"t2"+Quote$+" ,"+Quote$+"t3"+Quote$+" ,"+Quote$+"t4"+Quote$+" ,"+Quote$+"t5"+Quote$+"
,"+Quote$+"t6"+Quote$+" ,"+_
                                 Quote$+"t7"+Quote$+" ,"   +Quote$+"t8"+Quote$+" ,"+Quote$+"t9"+Quote$+" ,"+Quote$+"t10"+Quote$
        END SELECT

    END IF

    CLOSE #N%

    ProcID??? = SHELL(GnuPlotEXE$+" cmd3d.gp -") : CALL Delay(0.5##)

END SUB 'Plot3DbestProbeTrajectories()

'-----------

FUNCTION HasDAVGsaturated$(Nsteps&,j&,Np%,Nd%,M(),R(),DiagLength)

LOCAL A$

LOCAL k&

LOCAL SumOfDavg, DavgStepJ AS EXT

LOCAL DavgSatTOL AS EXT

    A$ = "NO"

    DavgSatTOL = 0.0005## 'tolerance for DAVG saturation

    IF j& < Nsteps& + 10 THEN GOTO ExitHasDAVGsaturated 'execute at least 10 steps after averaging interval before performing this check

    DavgStepJ = DavgThisStep(j&,Np%,Nd%,M(),R(),DiagLength)
```

```
        SumOfDavg = 0##

        FOR k& = j&-Nsteps&+1 TO j& 'check this step and previous (Nsteps&-1) steps

                SumOfDavg = SumOfDavg + DavgThisStep(k&,Np%,Nd%,M(),R(),DiagLength)

        NEXT k&

        IF ABS(SumOfDavg/Nsteps&-DavgStepJ) =< DavgSatTOL THEN A$ = "YES" 'saturation if (avg value - last value) are within TOL

ExitHasDAVGsaturated:

        HasDAVGsaturated$ = A$

END FUNCTION 'HasDAVGsaturated$()

'-----------

FUNCTION OscillationInDavg$(j&,Np%,Nd%,M(),R(),DiagLength)

LOCAL A$

LOCAL k&, NumSlopeChanges%

        A$ = "NO"

        NumSlopeChanges% = 0

        IF j& < 15 THEN GOTO ExitDavgOscillation 'wait at least 15 steps

        FOR k& = j&-10 TO j&-1 'check previous ten steps

                IF (DavgThisStep(k&,Np%,Nd%,M(),R(),DiagLength)-DavgThisStep(k&-1,Np%,Nd%,M(),R(),DiagLength))* _
                (DavgThisStep(k&+1,Np%,Nd%,M(),R(),DiagLength)-DavgThisStep(k&,Np%,Nd%,M(),R(),DiagLength)) < 0## THEN INCR NumSlopeChanges%

        NEXT j&

        IF NumSlopeChanges% >= 3 THEN A$ = "YES"

ExitDavgOscillation:

        OscillationInDavg$ = A$

END FUNCTION 'OscillationInDavg()

'------

FUNCTION DavgThisStep(j&,Np%,Nd%,M(),R(),DiagLength)

LOCAL BestFitness, TotalDistanceAllProbes, SumSQ AS EXT

LOCAL p%, k&, N%, i%, BestProbeNumber%, BestTimeStep&

'   ----------- Best Probe #, etc. -----------

        FOR k& = 0 TO j&

                BestFitness = M(1,k&)

                FOR p% = 1 TO Np%

                        IF M(p%,k&) >= BestFitness THEN

                                BestFitness = M(p%,k&) : BestProbeNumber% = p% : BestTimeStep& = k&

                        END IF

                NEXT p% 'probe #

        NEXT k& 'time step

'   --------- Average Distance to Best Probe -----------

        TotalDistanceAllProbes = 0##

        FOR p% = 1 TO Np%

                SumSQ = 0##

                FOR i% = 1 TO Nd%

                        SumSQ = SumSQ + (R(BestProbeNumber%,i%,BestTimeStep&)-R(p%,i%,j&))^2 'do not exclude p%=BestProbeNumber%(j&) from sum because it adds zero

                NEXT i%

                TotalDistanceAllProbes = TotalDistanceAllProbes + SQR(SumSQ)

        NEXT p%

        DavgThisStep = TotalDistanceAllProbes/(DiagLength*(Np%-1)) 'but exclude best prove from average

END FUNCTION 'DavgThisStep()

'-----------

SUB PlotBestFitnessEvolution(Nd%,Np%,LastStep&,G,DeltaT,Alpha,Beta,Frep,M(),PlaceInitialProbes$,InitialAcceleration$,RepositionFactor$,FunctionName$,Gamma)

LOCAL BestFitness(), GlobalBestFitness AS EXT

LOCAL PlotAnnotation$, PlotTitle$

LOCAL p%, j&, N%

        REDIM BestFitness(0 TO LastStep&)

        CALL GetPlotAnnotation(PlotAnnotation$,Nd%,Np%,LastStep&,G,DeltaT,Alpha,Beta,Frep,M(),PlaceInitialProbes$,InitialAcceleration$,RepositionFactor$,FunctionName$,Gamma)

        GlobalBestFitness = M(1,0)

        FOR j& = 0 TO LastStep&

                BestFitness(j&) = M(1,j&)

                FOR p% = 1 TO Np%

                        IF M(p%,j&) >= BestFitness(j&)   THEN BestFitness(j&)   = M(p%,j&)

                        IF M(p%,j&) >= GlobalBestFitness THEN GlobalBestFitness = M(p%,j&)

                NEXT p% 'probe #

        NEXT j& 'time step

        N% = FREEFILE

        OPEN "Fitness" FOR OUTPUT AS #N%
```

```
            FOR j& = 0 TO LastStep&

                PRINT #N%, USING$("###### #######.######",j&,BestFitness(j&))

            NEXT j&

    CLOSE #N%

    PlotAnnotation$ = PlotAnnotation$ + "Best Fitness = " + REMOVE$(STR$(ROUND(GlobalBestFitness,8)),ANY" ")

    PlotTitle$ = "Best Fitness vs Time Step\n" + "[" + REMOVE$(STR$(Np%),ANY" ") + " probes, "+REMOVE$(STR$(LastStep&),ANY" ")+" time steps]"

    CALL CreateGNUplotINIfile(0.1##*ScreenWidth&,0.1##*ScreenHeight&,0.6##*ScreenWidth&,0.6##*ScreenHeight&)

    CALL TwoDplot("Fitness","Best Fitness","0.7","0.7","Time Step\n\n.",".\n\nBest Fitness(X)", _
                  "","","","","","","","wgnuplot.exe"," with lines linewidth 2",PlotAnnotation$)

END SUB 'PlotBestFitnessEvolution()

'------

SUB PlotAverageDistance(Nd%,Np%,LastStep&,G,DeltaT,Alpha,Beta,Frep,M(),PlaceInitialProbes$,InitialAcceleration$,RepositionFactor$,FunctionName$,R(),DiagLength,Gamma)

LOCAL Davg(), BestFitness(), TotalDistanceAllProbes, SumSQ AS EXT

LOCAL PlotAnnotation$, PlotTitle$

LOCAL p%, j&, N%, i%, BestProbeNumber%(), BestTimeStep&()

    REDIM Davg(0 TO LastStep&), BestFitness(0 TO LastStep&), BestProbeNumber%(0 TO LastStep&), BestTimeStep&(0 TO LastStep&)

    CALL GetPlotAnnotation(PlotAnnotation$,Nd%,Np%,LastStep&,G,DeltaT,Alpha,Beta,Frep,M(),PlaceInitialProbes$,InitialAcceleration$,RepositionFactor$,FunctionName$,Gamma)

'  ----------- Best Probe #, etc. -----------

    FOR j& = 0 TO LastStep&

        BestFitness(j&) = M(1,j&)

        FOR p% = 1 TO Np%

            IF M(p%,j&) >= BestFitness(j&) THEN

                BestFitness(j&) = M(p%,j&) : BestProbeNumber%(j&) = p% : BestTimeStep&(j&) = j& 'only probe number is used at this time, but other data are computed for
possible future use.

            END IF

        NEXT p% 'probe #

    NEXT j& 'time step

    N% = FREEFILE

'  --------- Average Distance to Best Probe -----------

    FOR j& = 0 TO LastStep&

        TotalDistanceAllProbes = 0##

        FOR p% = 1 TO Np%

            SumSQ = 0##

            FOR i% = 1 TO Nd%

                SumSQ = SumSQ + (R(BestProbeNumber%(j&),i%,j&)-R(p%,i%,j&))^2 'do not exclude p%=BestProbeNumber%(j&) from sum because it adds zero

            NEXT i%

            TotalDistanceAllProbes = TotalDistanceAllProbes + SQR(SumSQ)

        NEXT p%

        Davg(j&) = TotalDistanceAllProbes/(DiagLength*(Np%-1)) 'but exclude best prove from average

    NEXT j&

'  ----------- Create Plot Data File -----------

    OPEN "Davg" FOR OUTPUT AS #N%

        FOR j& = 0 TO LastStep&

            PRINT #N%, USING$("###### #######.######",j&,Davg(j&))

        NEXT j&

    CLOSE #N%

    PlotTitle$ = "Average Distance of " + REMOVE$(STR$(Np%-1),ANY" ") + " Probes to Best Probe\nNormalized to Size of Decision Space\n" + _
                 "[" + REMOVE$(STR$(Np%),ANY" ") + " probes, " + REMOVE$(STR$(LastStep&),ANY" ") + " time steps]"

    CALL CreateGNUplotINIfile(0.2##*ScreenWidth&,0.2##*ScreenHeight&,0.6##*ScreenWidth&,0.6##*ScreenHeight&)

    CALL TwoDplot("Davg",PlotTitle$,"0.7","0.9","Time Step\n\n.",".\n\n<D>/Ldiag", _
                  "","","","","","","","wgnuplot.exe"," with lines linewidth 2",PlotAnnotation$)

END SUB 'PlotAverageDistance()

'------

SUB GetPlotAnnotation(PlotAnnotation$,Nd%,Np%,LastStep&,G,DeltaT,Alpha,Beta,Frep,M(),PlaceInitialProbes$,InitialAcceleration$,RepositionFactor$,FunctionName$,Gamma)

LOCAL A$

    A$ = "" : IF PlaceInitialProbes$ = "UNIFORM ON-AXIS" AND Nd% > 1 THEN A$ = "("+REMOVE$(STR$(Np%/Nd%),ANY" ") + "/axis)"

    PlotAnnotation$ = RunID$ + "\n" + _
                      FunctionName$ + " Function" + " ("+ FormatInteger$(Nd%) + "-D)  \n"    +_
                      FormatInteger$(Np%) + " probes"        + A$ + "\n" +_
                      "G = " + FormatFP$(G,2)               + "\n" +_
                      "Alpha = "     + FormatFP$(Alpha,1)   + "\n" +_
                      "Beta = "      + FormatFP$(Beta,1)    + "\n" +_
                      "DelT = "      + FormatFP$(DeltaT,1)  + "\n" +_
                      "Gamma = "     + FormatFP$(Gamma,3)   + "\n" +_
                      "Init Probes " + PlaceInitialProbes$  + "\n" +_
                      "Init Accel " + InitialAcceleration$ + "\n" +_
                      "Frep "        + RepositionFactor$ + "\n"

END SUB

'------

SUB PlotBestProbeVsTimeStep(Nd%,Np%,LastStep&,G,DeltaT,Alpha,Beta,Frep,M(),PlaceInitialProbes$,InitialAcceleration$,RepositionFactor$,FunctionName$,Gamma)
```

```
LOCAL BestFitness AS EXT

LOCAL PlotAnnotation$, PlotTitle$

LOCAL p%, j&, N%, BestProbeNumber%()

    REDIM BestProbeNumber%(0 TO LastStep&)

    CALL GetPlotAnnotation(PlotAnnotation$,Nd%,Np%,LastStep&,G,DeltaT,Alpha,Beta,Frep,M())PlaceInitialProbes$,InitialAcceleration$,RepositionFactor$,FunctionName$,Gamma)

    FOR j& = 0 TO LastStep&

        Bestfitness = M(1,j&)

        FOR p% = 1 TO Np%

            IF M(p%,j&) >= BestFitness THEN

                BestFitness = M(p%,j&) : BestProbeNumber%(j&) = p%

            END IF

        NEXT p% 'probe #

    NEXT j& 'time step

    N% = FREEFILE

    OPEN "Best Probe" FOR OUTPUT AS #N%

        FOR j& = 0 TO LastStep&

            PRINT #N%, USING$("###### ####",j&,BestProbeNumber%(j&)

        NEXT j&

    CLOSE #N%

    PlotTitle$ = "Best Probe Number vs Time Step\n" + "[" +REMOVE$(STR$(Np%),ANY" ") + " probes, " + REMOVE$(STR$(LastStep&),ANY" ") + " time steps]"

    CALL CreateGNUplotINIfile(0.15#*ScreenWidth&,0.15#*ScreenHeight&,0.6##*ScreenWidth&,0.6##*ScreenHeight&)

'USAGE: CALL
TwoDplot(PlotFileName$,PlotTitle$,xCoord$,yCoord$,XaxisLabel$,YaxisLabel$,LogXaxis$,LogYaxis$,xMin$,xMax$,yMin$,yMax$,xTics$,yTics$,GnuPlotEXE$,LineType$,Annotation$)

    CALL TwoDplot("Best Probe",PlotTitle$,"0.7","0.7","Time Step\n\n",".\n\nBest Probe #","","","","0",NoSpaces$(Np%+1,0),"","","wgnuplot.exe"," pt 8 ps .5 lw
1",PlotAnnotation$) 'pt, pointtype; ps, pointsize; lw, linewidth

END SUB 'PlotBestProbeVsTimeStep()

'------

FUNCTION FormatInteger$(M%) : FormatInteger$ = REMOVE$(STR$(M%),ANY" ") : END FUNCTION

'------

FUNCTION FormatFP$(X,Ndigits%)

LOCAL A$

    IF X = 0## THEN

        A$ = "0." : GOTO ExitFormatFP

    END IF

    A$ = REMOVE$(STR$(ROUND(ABS(X),Ndigits%)),ANY" ")

    IF ABS(X) < 1## THEN

        IF X > 0## THEN

            A$ = "0" + A$

        ELSE

            A$ = "-0" + A$

        END IF

    ELSE

        IF X < 0## THEN A$ = "-" + A$

    END IF

ExitFormatFP:

    FormatFP$ = A$

END FUNCTION

'-----------

SUB InitialProbeDistribution(Np%,Nd%,Nt&,R(),PlaceInitialProbes$,Gamma)

LOCAL DeltaXi, DelX1, DelX2, Di AS EXT

LOCAL NumProbesPerDimension%, p%, i%, k%, NumX1points%, NumX2points%, x1pointNum%, x2pointNum%, A$

    SELECT CASE PlaceInitialProbes$

        CASE "UNIFORM ON-AXIS"

            IF Nd% > 1 THEN

                NumProbesPerDimension% = Np%\Nd% 'even #

            ELSE

                NumProbesPerDimension% = Np%

            END IF

            FOR i% = 1 TO Nd%

                FOR p% = 1 TO Np%

                    R(p%,i%,0) = XiMin(i%) + Gamma*(XiMax(i%)-XiMin(i%))

                NEXT Np%

            NEXT i%

            FOR i% = 1 TO Nd% 'place probes probe line-by-probe line (i% is dimension number)
```

```
                DeltaXi = (XiMax(i%)-XiMin(i%))/(NumProbesPerDimension%-1)

                FOR k% = 1 TO NumProbesPerDimension%

                    p% = k% + NumProbesPerDimension%*(i%-1) 'probe #

                    R(p%,i%,0) = XiMin(i%) + (k%-1)*DeltaXi

                NEXT k%

            NEXT i%

        CASE "UNIFORM ON-DIAGONAL"

            FOR p% = 1 TO Np%

                FOR i% = 1 TO Nd%

                    DeltaXi = (XiMax(i%)-XiMin(i%))/(Np%-1)

                    R(p%,i%,0) = XiMin(i%) + (p%-1)*DeltaXi

                NEXT i%

            NEXT p%

        CASE "2D GRID"

            NumProbesPerDimension% = SQR(Np%) : NumX1points% = NumProbesPerDimension% : NumX2points% = NumX1points% 'broken down for possible future use

            DelX1 = (XiMax(1)-XiMin(1))/(NumX1points%-1)

            DelX2 = (XiMax(2)-XiMin(2))/(NumX2points%-1)

            FOR x1pointNum% = 1 TO NumX1points%

                FOR x2pointNum% = 1 TO NumX2points%

                    p% = NumX1points%*(x1pointNum%-1)+x2pointNum% 'probe #

                    R(p%,1,0) = XiMin(1) + DelX1*(x1pointNum%-1) 'x1 coord
                    R(p%,2,0) = XiMin(2) + DelX2*(x2pointNum%-1) 'x2 coord

                NEXT x2pointNum%

            NEXT x1pointNum%

        CASE "RANDOM"

            FOR p% = 1 TO Np%

                FOR i% = 1 TO Nd%

                    R(p%,i%,0) = XiMin(i%) + RandomNum(0##,1##)*(XiMax(i%)-XiMin(i%))

                NEXT i%

            NEXT p%

    END SELECT

END SUB 'InitialProbeDistribution()

'------

SUB ChangeRunParameters(NumProbesPerDimension%,Np%,Nd%,Nt&,G,Alpha,Beta,DeltaT,Frep,PlaceInitialProbes$,InitialAcceleration$,RepositionFactor$,FunctionName$) 'THIS PROCEDURE
NOT USED

LOCAL A$, DefaultValue$

    A$ = INPUTBOX$("# dimensions?","Change # Dimensions ("+FunctionName$+")",NoSpaces$(Nd%+0,0)) : Nd%    = VAL(A$) : IF Nd% < 1 OR Nd% > 500 THEN Nd% = 2

    IF Nd% > 1 THEN NumProbesPerDimension% = 2*((NumProbesPerDimension%+1)\2) 'require an even # probes on each probe line to avoid overlapping at origin (in symmetrical
spaces at least...)

    IF Nd% = 1 THEN NumProbesPerDimension% = MAX(NumProbesPerDimension%,3)    'at least 3 probes on x-axis for 1-D functions

    Np% = NumProbesPerDimension%*Nd%

    A$ = INPUTBOX$("# time steps?","Change # Steps ("+FunctionName$+")",NoSpaces$(Nt&+0,0)) : Nt&    = VAL(A$) : IF Nt& < 3                             THEN Nt& = 50

    A$ = INPUTBOX$("Grav Const G?","Change G ("+FunctionName$+")",NoSpaces$(G,2))            : G     = VAL(A$) : IF G < -100## OR G > 100## THEN G  = 2##

    A$ = INPUTBOX$("Alpha?","Change Alpha ("+FunctionName$+")",NoSpaces$(Alpha,2))           : Alpha  = VAL(A$) : IF Alpha < -50## OR Alpha > 50## THEN Alpha = 2##'

    A$ = INPUTBOX$("Beta?","Change Beta ("+FunctionName$+")",NoSpaces$(Beta,2))              : Beta   = VAL(A$) : IF Beta  < -50## OR Beta  > 50## THEN Beta  = 2##'

    A$ = INPUTBOX$("Delta T","Change Delta-T ("+FunctionName$+")",NoSpaces$(DeltaT,2))       : DeltaT = VAL(A$) : IF DeltaT =< 0##                 THEN DeltaT = 1##

    A$ = INPUTBOX$("Frep [0-1]?","Change Frep ("+FunctionName$+")",NoSpaces$(Frep,3))        : Frep   = VAL(A$) : IF Frep < 0##   OR Frep > 1##    THEN Frep = 0.5##

'   ------------ Initial Probe Distribution ------------

    SELECT CASE PlaceInitialProbes$
        CASE "UNIFORM ON-AXIS"     : DefaultValue$ = "1"
        CASE "UNIFORM ON-DIAGONAL" : DefaultValue$ = "2"
        CASE "2D GRID"             : DefaultValue$ = "3"
        CASE "RANDOM"              : DefaultValue$ = "4"
    END SELECT

    A$ = INPUTBOX$("Initial Probes?"+CHR$(13)+"1 - UNIFORM ON-AXIS"+CHR$(13)+"2 - UNIFORM ON-DIAGONAL"+CHR$(13)+"3 - 2D GRID"+CHR$(13)+"4 - RANDOM","Initial Probe
Distribution ("+FunctionName$+")",DefaultValue$)

    IF VAL(A$) < 1 OR VAL(A$)  > 4 THEN A$ = "1"

    SELECT CASE VAL(A$)
        CASE 1 : PlaceInitialProbes$ = "UNIFORM ON-AXIS"
        CASE 2 : PlaceInitialProbes$ = "UNIFORM ON-DIAGONAL"
        CASE 3 : PlaceInitialProbes$ = "2D GRID"
        CASE 4 : PlaceInitialProbes$ = "RANDOM"
    END SELECT

    IF Nd% = 1  AND PlaceInitialProbes$ = "UNIFORM ON-DIAGONAL" THEN PlaceInitialProbes$ = "UNIFORM ON-AXIS" 'cannot do diagonal in 1-D space

    IF Nd% <> 2 AND PlaceInitialProbes$ = "2D GRID" THEN PlaceInitialProbes$ = "UNIFORM ON-AXIS" '2D grid is available only in 2 dimensions!

'   ------------- Initial Acceleration ----------------

    SELECT CASE InitialAcceleration$
        CASE "ZERO"   : DefaultValue$ = "1"
        CASE "FIXED"  : DefaultValue$ = "2"
        CASE "RANDOM" : DefaultValue$ = "3"
    END SELECT

    A$ = INPUTBOX$("Initial Acceleration?"+CHR$(13)+"1 - ZERO"+CHR$(13)+"2 - FIXED"+CHR$(13)+"3 - RANDOM","Initial Acceleration ("+FunctionName$+")",DefaultValue$)
```

```
    IF VAL(A$) < 1 OR VAL(A$) > 3 THEN A$ = "1"

    SELECT CASE VAL(A$)
        CASE 1 : InitialAcceleration$ = "ZERO"
        CASE 2 : InitialAcceleration$ = "FIXED"
        CASE 3 : InitialAcceleration$ = "RANDOM"
    END SELECT

'   ----------- Reposition Factor ---------------

    SELECT CASE RepositionFactor$
        CASE "FIXED"    : DefaultValue$ = "1"
        CASE "VARIABLE" : DefaultValue$ = "2"
        CASE "RANDOM"   : DefaultValue$ = "3"
    END SELECT

    A$ = INPUTBOX$("Reposition Factor?"+CHR$(13)+"1 - FIXED"+CHR$(13)+"2 - VARIABLE"+CHR$(13)+"3 - RANDOM","Retrieve Probes ("+FunctionName$+")",DefaultValue$)

    IF VAL(A$) < 1 OR VAL(A$) > 3 THEN A$ = "1"

    SELECT CASE VAL(A$)
        CASE 1 : RepositionFactor$ = "FIXED"
        CASE 2 : RepositionFactor$ = "VARIABLE"
        CASE 3 : RepositionFactor$ = "RANDOM"
    END SELECT

END SUB 'ChangeRunParameters()

'------

FUNCTION NoSpaces$(X,NumDigits%) :  NoSpaces$ = REMOVE$(STR$(X,NumDigits%),ANY" ") : END FUNCTION

'-----------

FUNCTION TerminateNowForSaturation$(j&,Nd&,Np%,Nt&,G,DeltaT,Alpha,Beta,R(),A(),M())

LOCAL A$, i&, p%, NumStepsForAveraging&

LOCAL BestFitness, AvgFitness, FitnessTOL AS EXT 'terminate if avg fitness does not change over NumStepsForAveraging& time steps

    FitnessTOL = 0.00001## : NumStepsForAveraging& = 10

    A$ = "NO"

    IF j& >= NumStepsForAveraging&+10 THEN 'wait until step 10 to start checking for fitness saturation

        AvgFitness = 0##

        FOR i& = j&-NumStepsForAveraging&+1 TO j& 'avg fitness over current step & previous NumStepsForAveraging&-1 steps

            BestFitness = M(1,i&)

            FOR p% = 1 TO Np%

                IF M(p%,i&) >= BestFitness THEN BestFitness = M(p%,i&)

            NEXT p%

            AvgFitness = AvgFitness + BestFitness

        NEXT i&

        AvgFitness = AvgFitness/NumStepsForAveraging&

        IF ABS(AvgFitness-BestFitness) < FitnessTOL THEN A$ = "YES" 'compare avg fitness to best fitness at this step

    END IF

    TerminateNowForSaturation$ = A$

END FUNCTION 'TerminateNowForSaturation$()

'-----------

FUNCTION MagVector(V(),N%) 'returns magnitude of Nx1 column vector V

LOCAL SumSQ AS EXT

LOCAL i%

    SumSQ = 0## : FOR i% = 1 TO N% : SumSQ = SumSQ + V(i%)^2 : NEXT i% : MagVector = SQR(SumSQ)

END FUNCTION 'MagVector()

'---

FUNCTION UnitStep(X)

LOCAL Z AS EXT

    IF X < 0## THEN

        Z = 0##

    ELSE

        Z = 1##

    END IF

    UnitStep = Z

END FUNCTION 'UnitStep()

'---

SUB Plot1Dfunction(FunctionName$,R()) 'plots 1D function on-screen

LOCAL NumPoints%, i%, N%

LOCAL DeltaX, X AS EXT

    NumPoints% = 32001

    DeltaX = (XiMax(1)-XiMin(1))/(NumPoints%-1)

    N% = FREEFILE

    SELECT CASE FunctionName$

        CASE "ParrottF4" 'PARROTT F4 FUNCTION

            OPEN "ParrottF4" FOR OUTPUT AS #N%

                FOR i% = 1 TO NumPoints%

                    R(1,1,0) = XiMin(1) + (i%-1)*DeltaX
```

```
                    PRINT #N%, USINGS$("#.#####  #.#####",R(1,1,0),ParrottF4(R(),1,1,0))

                NEXT i%

            CLOSE #N%

            CALL CreateGNUplotINIfile(0.2##*ScreenWidth&,0.2##*ScreenHeight&,0.6##*ScreenWidth&,0.6##*ScreenHeight&)

            CALL TwoDplot("ParrottF4","Parrott F4 Function","0.7","0.7","X\n\n.",".\n\nParrott F4(X)","","","0","1","0","1","","","wgnuplot.exe," with lines linewidth
2","")

        END SELECT

END SUB

'------

SUB CLEANUP 'probe coordinate plot files

    IF DIR$("P1")  <> "" THEN KILL "P1"
    IF DIR$("P2")  <> "" THEN KILL "P2"
    IF DIR$("P3")  <> "" THEN KILL "P3"
    IF DIR$("P4")  <> "" THEN KILL "P4"
    IF DIR$("P5")  <> "" THEN KILL "P5"
    IF DIR$("P6")  <> "" THEN KILL "P6"
    IF DIR$("P7")  <> "" THEN KILL "P7"
    IF DIR$("P8")  <> "" THEN KILL "P8"
    IF DIR$("P9")  <> "" THEN KILL "P9"
    IF DIR$("P10") <> "" THEN KILL "P10"
    IF DIR$("P11") <> "" THEN KILL "P11"
    IF DIR$("P12") <> "" THEN KILL "P12"
    IF DIR$("P13") <> "" THEN KILL "P13"
    IF DIR$("P14") <> "" THEN KILL "P14"
    IF DIR$("P15") <> "" THEN KILL "P15"

END SUB

'------

SUB Plot2Dfunction(FunctionName$,R())

LOCAL AS

LOCAL NumPoints%, i%, k%, N%

LOCAL DelX1, DelX2, Z AS EXT

    SELECT CASE FunctionName$

        CASE "PBM_1","PBM_2","PBM_3","PBM_4","PBM_5" : NumPoints = 25

        CASE ELSE : NumPoints% = 100

    END SELECT

    N% = FREEFILE : OPEN "TwoDplot.DAT" FOR OUTPUT AS #N%

    DelX1 = (XiMax(1)-XiMin(1))/(NumPoints%-1) : DelX2 = (XiMax(2)-XiMin(2))/(NumPoints%-1)

    FOR i% = 1 TO NumPoints%

        R(1,1,0) = XiMin(1) + (i%-1)*DelX1 'x1 value

        FOR k% = 1 TO NumPoints%

            R(1,2,0) = XiMin(2) + (k%-1)*DelX2 'x2 value

            Z = ObjectiveFunction(R(),2,1,0,FunctionName$)

            PRINT #N%, USINGS$("######.###### ######.###### ######.######^^^",R(1,1,0),R(1,2,0),Z)

        NEXT k%

        PRINT #N%, ""

    NEXT i%

    CLOSE #N%

    CALL CreateGNUplotINIfile(0.1##*ScreenWidth&,0.1##*ScreenHeight&,0.6##*ScreenWidth&,0.6##*ScreenHeight&)

    AS = "" : IF INSTR(FunctionName$,"PBM_") > 0 THEN AS = "Coarse "

    CALL ThreeDplot2("TwoDplot.DAT",A$+"Plot of "+FunctionName$+" Function","","0.6","0.6","1.2", _
                    "x1","x2","z=F(x1,x2)","","","wgnuplot.exe","","","","")

END SUB

'------

    SUB TwoDplot3curves(NumCurves%,PlotFileName1$,PlotFileName2$,PlotFileName3$,PlotTitle$,Annotation$,xCoord$,yCoord$,XaxisLabel$,YaxisLabel$, _
                        LogXaxis$,LogYaxis$,xMin$,xMax$,yMin$,yMax$,xTics$,yTics$,GnuPlotEXE$)

        LOCAL N%

        LOCAL LineSize$

        LineSize$ = "2"

        N% = FREEFILE

        OPEN "cmd2d.gp" FOR OUTPUT AS #N%

            IF LogXaxis$ = "YES" AND LogYaxis$ = "NO"  THEN PRINT #N%, "set logscale x"
            IF LogXaxis$ = "NO" AND LogYaxis$ = "YES" THEN PRINT #N%, "set logscale y"
            IF LogXaxis$ = "YES" AND LogYaxis$ = "YES" THEN PRINT #N%, "set logscale xy"

            IF xMin$ <> "" AND xMax$ <> "" THEN  PRINT #N%, "set xrange ["+xMin$+":"+xMax$+"]"

            IF yMin$ <> "" AND yMax$ <> "" THEN  PRINT #N%, "set yrange ["+yMin$+":"+yMax$+"]"

            PRINT #N%, "set label "+Quote$+Annotation$+Quote$+" at graph "+xCoord$+","+yCoord$
            PRINT #N%, "set grid xtics"
            PRINT #N%, "set grid ytics"
            PRINT #N%, "set xtics "+xTics$
            PRINT #N%, "set ytics "+yTics$
            PRINT #N%, "set grid mxtics"
            PRINT #N%, "set grid mytics"
            PRINT #N%, "set title " +Quote$+PlotTitle$+Quote$
            PRINT #N%, "set xlabel "+Quote$+XaxisLabel$+Quote$
            PRINT #N%, "set ylabel "+Quote$+YaxisLabel$+Quote$

            SELECT CASE NumCurves%

            CASE 1
            PRINT #N%, "plot " + Quote$ + PlotFileName1$ + Quote$ + " with lines linewidth " + LineSize$
```

```
                CASE 2
                    PRINT #N%, "plot " + Quote$ + PlotFileName1$ + Quote$ + " with lines linewidth " + LineSize$+", " + _
                                        Quote$ + PlotFileName2$ + Quote$ + " with lines linewidth " + LineSize$
                CASE 3
                    PRINT #N%, "plot " + Quote$ + PlotFileName1$ + Quote$ + " with lines linewidth " + LineSize$+", " + _
                                        Quote$ + PlotFileName2$ + Quote$ + " with lines linewidth " + LineSize$+", " + _
                                        Quote$ + PlotFileName3$ + Quote$ + " with lines linewidth " + LineSize$
                END SELECT

        CLOSE #N%

        SHELL(GnuPlotEXE$+" cmd2d.gp -")

        CALL Delay(0.3)

    END SUB 'TwoDplot3Curves()

'---

FUNCTION Fibonacci&&(N%) 'RETURNS Nth FIBONACCI NUMBER

LOCAL i%, Fn&&, Fn1&&, Fn2&&

LOCAL A$

    IF N% > 91 OR N% < 0 THEN

        MSGBOX("ERROR!  Fibonacci argument"+STR$(N%)+" > 91.  Out of range or < 0...") : EXIT FUNCTION

    END IF

    SELECT CASE N%

        CASE 0: Fn&& = 1

        CASE ELSE

            Fn&& = 0 : Fn2&& = 1 : i% = 0

            FOR i% = 1 TO N%

                Fn&& = Fn1&& + Fn2&&

                Fn1&& = Fn2&&

                Fn2&& = Fn&&

            NEXT i% 'LOOP

    END SELECT

    Fibonacci&& = Fn&&

END FUNCTION 'Fibonacci&&()

'-----------

FUNCTION RandomNum(a,b) 'Returns random number X, a=< X < b.

    RandomNum = a + (b-a)*RND

END FUNCTION 'RandomNum()

'-----------

FUNCTION GaussianDeviate(Mu,Sigma) 'returns NORMAL (Gaussian) random deviate with mean Mu and standard deviation Sigma (variance = Sigma^2)

'Refs: (1) Press, W.H., Flannery, B.P., Teukolsky, S.A., and Vetterling, W.T., "Numerical Recipes: The Art of Scientific Computing,"
'           §7.2, Cambridge University Press, Cambridge, UK, 1986.
'      (2) Shinzato, T., "Box Muller Method," 2007, http://www.sp.dis.titech.ac.jp/~shinzato/boxmuller.pdf

LOCAL s, t, Z AS EXT

    s = RND : t = RND

    Z = Mu + Sigma*SQR(-2#*LOG(s))*COS(TwoPi*t)

    GaussianDeviate = Z

END FUNCTION 'GaussianDeviate()

'-----------

    SUB ContourPlot(PlotFileName$,PlotTitle$,Annotation$,xCoord$,yCoord$,zCoord$, _
                    XaxisLabel$,YaxisLabel$,ZaxisLabel$,zMin$,zMax$,GnuPlotEXE$,A$)

        LOCAL N%

        N% = FREEFILE

        OPEN "cmd3d.gp" FOR OUTPUT AS #N%

            PRINT #N%, "show surface"
            PRINT #N%, "set hidden3d"
            IF zMin$ <> "" AND zMax$ <> "" THEN  PRINT #N%, "set zrange ["+zMin$+":"+zMax$+"]"
            PRINT #N%, "set label "+Quote$+AnnoTation$+Quote$+" at graph "+xCoord$+","+yCoord$+","+zCoord$
            PRINT #N%, "show label"
            PRINT #N%, "set grid xtics ytics ztics"
            PRINT #N%, "show grid"
            PRINT #N%, "set title "+Quote$+PlotTitle$+Quote$
            PRINT #N%, "set xlabel "+Quote$+XaxisLabel$+Quote$
            PRINT #N%, "set ylabel "+Quote$+YaxisLabel$+Quote$
            PRINT #N%, "set zlabel "+Quote$+ZaxisLabel$+Quote$
            PRINT #N%, "splot "+Quote$+PlotFileName$+Quote$+A$  '" notitle with linespoints" 'A$'" notitle with lines"
        CLOSE #N%

        SHELL(GnuPlotEXE$+" cmd3d.gp -")

    END SUB 'ContourPlot()

'---

    SUB ThreeDplot(PlotFileName$,PlotTitle$,Annotation$,xCoord$,yCoord$,zCoord$, _
                    XaxisLabel$,YaxisLabel$,ZaxisLabel$,zMin$,zMax$,GnuPlotEXE$,A$)

        LOCAL N%, ProcessID???

        N% = FREEFILE

        OPEN "cmd3d.gp" FOR OUTPUT AS #N%

            PRINT #N%, "set pm3d"
            PRINT #N%, "show pm3d"
            IF zMin$ <> "" AND zMax$ <> "" THEN  PRINT #N%, "set zrange ["+zMin$+":"+zMax$+"]"
            PRINT #N%, "set label "+Quote$+AnnoTation$+Quote$+" at graph "+xCoord$+","+yCoord$+","+zCoord$
            PRINT #N%, "show label"
            PRINT #N%, "set grid xtics ytics ztics"
```

```
        PRINT #N%, "show grid"
        PRINT #N%, "set title "+Quote$+PlotTitle$+Quote$
        PRINT #N%, "set xlabel "+Quote$+XaxisLabel$+Quote$
        PRINT #N%, "set ylabel "+Quote$+YaxisLabel$+Quote$
        PRINT #N%, "set zlabel "+Quote$+ZaxisLabel$+Quote$
        PRINT #N%, "splot "+Quote$+PlotFileName$+Quote$+A$+" notitle"' with lines"
     CLOSE #N%

     SHELL(GnuPlotEXE$+" cmd3d.gp -") : CALL Delay(0.5##)

  END SUB 'ThreeDplot()

'----

  SUB ThreeDplot2(PlotFileName$,PlotTitle$,Annotation$,xCoord$,yCoord$,zCoord$, _
                  XaxisLabel$,YaxisLabel$,ZaxisLabel$,zMin$,zMax$,GnuPlotEXE$,A$,xStart$,xStop$,yStart$,yStop$)

     LOCAL N%

     N% = FREEFILE

     OPEN "cmd3d.gp" FOR OUTPUT AS #N%

        PRINT #N%, "set pm3d"
        PRINT #N%, "show pm3d"
        PRINT #N%, "set hidden3d"
        PRINT #N%, "set view 45, 45, 1, 1"

        IF zMin$ <> "" AND zMax$ <> "" THEN  PRINT #N%, "set zrange ["+zMin$+":"+zMax$+"]"

        IF xStart$ <> "" THEN PRINT #N%, "set xrange [" + xStart$ + ":" + xStop$ + "]"
        IF yStart$ <> "" THEN PRINT #N%, "set yrange [" + yStart$ + ":" + yStop$ + "]"

        PRINT #N%, "set label "   + Quote$ + AnnoTation$ + Quote$+" at graph "+xCoord$+","+yCoord$+","+zCoord$
        PRINT #N%, "show label"
        PRINT #N%, "set grid xtics ytics ztics"
        PRINT #N%, "show grid"
        PRINT #N%, "set title " + Quote$+PlotTitle$    + Quote$
        PRINT #N%, "set xlabel " + Quote$+XaxisLabel$   + Quote$
        PRINT #N%, "set ylabel " + Quote$+YaxisLabel$   + Quote$
        PRINT #N%, "set zlabel " + Quote$+ZaxisLabel$   + Quote$
        PRINT #N%, "splot "      + Quote$+PlotFileName$ + Quote$ + A$ + " notitle with lines"
     CLOSE #N%

     SHELL(GnuPlotEXE$+" cmd3d.gp -")

  END SUB 'ThreeDplot2()

'----

  SUB TwoDplot2Curves(PlotFileName1$,PlotFileName2$,PlotTitle$,Annotation$,xCoord$,yCoord$,XaxisLabel$,YaxisLabel$, _
                      LogXaxis$,LogYaxis$,xMin$,xMax$,yMin$,yMax$,xTics$,yTics$,GnuPlotEXE$,LineSize)

     LOCAL N%, ProcessID???

     N% = FREEFILE

     OPEN "cmd2d.gp" FOR OUTPUT AS #N%
        'print #N%, "set output "+Quote$+"test.plt"+Quote$ 'tried this 3/11/06, didn't work...
        IF LogXaxis$ = "YES" AND LogYaxis$ = "NO"  THEN PRINT #N%, "set logscale x"
        IF LogXaxis$ = "NO"  AND LogYaxis$ = "YES" THEN PRINT #N%, "set logscale y"
        IF LogXaxis$ = "YES" AND LogYaxis$ = "YES" THEN PRINT #N%, "set logscale xy"

        IF xMin$ <> "" AND xMax$ <> "" THEN  PRINT #N%, "set xrange ["+xMin$+":"+xMax$+"]"

        IF yMin$ <> "" AND yMax$ <> "" THEN  PRINT #N%, "set yrange ["+yMin$+":"+yMax$+"]"

        PRINT #N%, "set label "+Quote$+Annotation$+Quote$+" at graph "+xCoord$+","+yCoord$
        PRINT #N%, "set grid xtics"
        PRINT #N%, "set grid ytics"
        PRINT #N%, "set xtics "+xTics$
        PRINT #N%, "set ytics "+yTics$
        PRINT #N%, "set grid mxtics"
        PRINT #N%, "set grid mytics"
        PRINT #N%, "set title "+Quote$+PlotTitle$+Quote$
        PRINT #N%, "set xlabel "+Quote$+XaxisLabel$+Quote$
        PRINT #N%, "set ylabel "+Quote$+YaxisLabel$+Quote$

        PRINT #N%, "plot "+Quote$+PlotFileName1$+Quote$+" with lines linewidth "+REMOVE$(STR$(LineSize),ANY" ")+","+ _
                          Quote$+PlotFileName2$+Quote$+" with points pointsize 0.05"+REMOVE$(STR$(LineSize),ANY" ")

     CLOSE #N%

     ProcessID??? = SHELL(GnuPlotEXE$+" cmd2d.gp -") : CALL Delay(0.5##)

  END SUB 'TwoDplot2Curves()

'----

  SUB Probe2Dplots(ProbePlotsFileList$,PlotTitle$,Annotation$,xCoord$,yCoord$,XaxisLabel$,YaxisLabel$, _
                   LogXaxis$,LogYaxis$,xMin$,xMax$,yMin$,yMax$,xTics$,yTics$,GnuPlotEXE$)

     LOCAL N%, ProcessID???

     N% = FREEFILE

     OPEN "cmd2d.gp" FOR OUTPUT AS #N%

        IF LogXaxis$ = "YES" AND LogYaxis$ = "NO"  THEN PRINT #N%, "set logscale x"
        IF LogXaxis$ = "NO"  AND LogYaxis$ = "YES" THEN PRINT #N%, "set logscale y"
        IF LogXaxis$ = "YES" AND LogYaxis$ = "YES" THEN PRINT #N%, "set logscale xy"

        IF xMin$ <> "" AND xMax$ <> "" THEN  PRINT #N%, "set xrange ["+xMin$+":"+xMax$+"]"

        IF yMin$ <> "" AND yMax$ <> "" THEN  PRINT #N%, "set yrange ["+yMin$+":"+yMax$+"]"

        PRINT #N%, "set label "+Quote$+Annotation$+Quote$+" at graph "+xCoord$+","+yCoord$
        PRINT #N%, "set grid xtics"
        PRINT #N%, "set grid ytics"
        PRINT #N%, "set xtics "+xTics$
        PRINT #N%, "set ytics "+yTics$
        PRINT #N%, "set grid mxtics"
        PRINT #N%, "set grid mytics"
        PRINT #N%, "set title "+Quote$+PlotTitle$+Quote$
        PRINT #N%, "set xlabel "+Quote$+XaxisLabel$+Quote$
        PRINT #N%, "set ylabel "+Quote$+YaxisLabel$+Quote$

        PRINT #N%, ProbePlotsFileList$

     CLOSE #N%

     ProcessID??? = SHELL(GnuPlotEXE$+" cmd2d.gp -") : CALL Delay(0.5##)

  END SUB 'Probe2Dplots()

'----
```

```
SUB Show2Dprobes(R(),Np%,Nt&,j&,Frep,BestFitness,BestProbeNumber%,BestTimeStep&,FunctionName$,RepositionFactor$,Gamma)

    LOCAL N%, p%

    LOCAL A$, PlotFileName$, PlotTitle$, Symbols$

    LOCAL xMin$, xMax$, yMin$, yMax$

    LOCAL s1, s2, s3, s4 AS EXT

    PlotFileName$ = "Probes("+REMOVE$(STR$(j&),ANY" ")+")"

    IF j& > 0 THEN 'PLOT PROBES AT THIS TIME STEP

        PlotTitle$ = "\nLOCATIONS OF "+REMOVE$(STR$(Np%),ANY" ") + " PROBES AT TIME STEP" + STR$(j&) + " / " + REMOVE$(STR$(Nt&),ANY" ") + "\n" + _
                    "Fitness = "+REMOVE$(STR$(ROUND(BestFitness,3)),ANY" ") + ", Probe #" + REMOVE$(STR$(BestProbeNumber%),ANY" ") + " at Step #" +
REMOVE$(STR$(BestTimeStep&),ANY" ") + _
                    " [Frep = "+REMOVE$(STR$(Frep,4),ANY" ") + " + " + RepositionFactor$ + "]\n"

    ELSE 'PLOT INITIAL PROBE DISTRIBUTION

        PlotTitle$ = "\nLOCATIONS OF "+REMOVE$(STR$(Np%),ANY" ") + " INITIAL PROBES FOR " + FunctionName$ + " FUNCTION\n[gamma = "+STR$(ROUND(Gamma,3))+"]\n"

    END IF

    N% = FREEFILE : OPEN PlotFileName$ FOR OUTPUT AS #N%

        FOR p% = 1 TO Np% : PRINT #N%, USING$("######.####    ######.####",R(p%,1,j&),R(p%,2,j&)) : NEXT p%

    CLOSE #N%

    s1 = 1.1## : s2 = 1.1## : s3 = 1.1## : s4 = 1.1## 'expand plots axes by 10%

    IF XiMin(1) > 0## THEN s1 = 0.9##
    IF XiMax(1) < 0## THEN s2 = 0.9##
    IF XiMin(2) > 0## THEN s3 = 0.9##
    IF XiMax(2) < 0## THEN s4 = 0.9##

    xMin$ = REMOVE$(STR$(s1*XiMin(1),2),ANY" ")
    xMax$ = REMOVE$(STR$(s2*XiMax(1),2),ANY" ")
    yMin$ = REMOVE$(STR$(s3*XiMin(2),2),ANY" ")
    yMax$ = REMOVE$(STR$(s4*XiMax(2),2),ANY" ")

    CALL TwoDplot(PlotFileName$,PlotTitle$,"0.6","0.7","x1\n\n","\nx2","NO","NO",xMin$,xMax$,yMin$,yMax$,"5","5","wgnuplot.exe"," pointsize 1 linewidth 2","")

    KILL PlotFileName$ 'erase plot data file after probes have been displayed

END SUB 'Show2Dprobes()

'---

SUB Show3Dprobes(R(),Np%,Nd%,Nt&,j&,Frep,BestFitness,BestProbeNumber%,BestTimeStep&,FunctionName$,RepositionFactor$,Gamma)

    LOCAL N%, p%, PlotWindowULC_X%, PlotWindowULC_Y%, PlotWindowWidth%, PlotWindowHeight%, PlotWindowOffset%

    LOCAL A$, PlotFileName$, PlotTitle$, Symbols$, Annotation$

    LOCAL xMin$, xMax$, yMin$, yMax$, zMin$, zMax$

    LOCAL s1, s2, s3, s4, s5, s6 AS EXT

    PlotFileName$ = "Probes("+REMOVE$(STR$(j&),ANY" ")+")"

    IF j& > 0 THEN 'PLOT PROBES AT THIS TIME STEP

        PlotTitle$ = "\nLOCATIONS OF "+REMOVE$(STR$(Np%),ANY" ") + " PROBES AT TIME STEP" + STR$(j&) + " / " + REMOVE$(STR$(Nt&),ANY" ") + "\n" + _
                    "Fitness = "+REMOVE$(STR$(ROUND(BestFitness,3)),ANY" ") + ", Probe #" + REMOVE$(STR$(BestProbeNumber%),ANY" ") + " at Step #" +
REMOVE$(STR$(BestTimeStep&),ANY" ") + _
                    " [Frep = "+REMOVE$(STR$(Frep,4),ANY" ") + " + " + RepositionFactor$ + "]\n"

    ELSE 'PLOT INITIAL PROBE DISTRIBUTION

        A$ = "" : IF Gamma > 0## AND Gamma < 1## THEN A$ = "0"

        PlotTitle$ = "\n"+REMOVE$(STR$(Np%),ANY" ") + "-PROBE IPD FOR FUNCTION " + FunctionName$ + ", GAMMA = "+A$+REMOVE$(STR$(ROUND(Gamma,3)),ANY" ")

    END IF

'   --------------- Probe Coordinates ----------------

    N% = FREEFILE : OPEN PlotFileName$ FOR OUTPUT AS #N%

        PRINT #N%, USING$("######.####    ######.####    ######.####",R(1,1,j&),R(1,2,j&),R(1,3,j&)) 'This line repeats Probe #1's coordinates.  It's necessary
                                                                                                    'to deal with a plotting artifact in Gnuplot!
        FOR p% = 1 TO Np% : PRINT #N%, USING$("######.####    ######.####    ######.####",R(p%,1,j&),R(p%,2,j&),R(p%,3,j&)) : NEXT p%

    CLOSE #N%

'   ----------- Principal Diagonal -------------

    N% = FREEFILE : OPEN "diag" FOR OUTPUT AS #N%

        PRINT #N%, USING$("######.####    ######.####    ######.####",XiMin(1),XiMin(2),XiMin(3))
        PRINT #N%, ""
        PRINT #N%, USING$("######.####    ######.####    ######.####",XiMax(1),XiMax(2),XiMax(3))

    CLOSE #N%

'   ----------------- Probe Line #1 -----------------

    N% = FREEFILE : OPEN "probeline1" FOR OUTPUT AS #N%

        PRINT #N%, USING$("######.####    ######.####    ######.####",R(1,1,j&),R(1,2,j&),R(1,3,j&))
        PRINT #N%, ""
        PRINT #N%, USING$("######.####    ######.####    ######.####",R(Np%/Nd%,1,j&),R(Np%/Nd%,2,j&),R(Np%/Nd%,3,j&))

    CLOSE #N%

'   ----------------- Probe Line #2 -----------------

    N% = FREEFILE : OPEN "probeline2" FOR OUTPUT AS #N%

        PRINT #N%, USING$("######.####    ######.####    ######.####",R(1+Np%/Nd%,1,j&),R(1+Np%/Nd%,2,j&),R(1+Np%/Nd%,3,j&))
        PRINT #N%, ""
        PRINT #N%, USING$("######.####    ######.####    ######.####",R(2*Np%/Nd%,1,j&),R(2*Np%/Nd%,2,j&),R(2*Np%/Nd%,3,j&))

    CLOSE #N%

'   ----------------- Probe Line #3 -----------------

    N% = FREEFILE : OPEN "probeline3" FOR OUTPUT AS #N%

        PRINT #N%, USING$("######.####    ######.####    ######.####",R(1+2*Np%/Nd%,1,j&),R(1+2*Np%/Nd%,2,j&),R(1+2*Np%/Nd%,3,j&))
        PRINT #N%, ""
        PRINT #N%, USING$("######.####    ######.####    ######.####",R(3*Np%/Nd%,1,j&),R(3*Np%/Nd%,2,j&),R(3*Np%/Nd%,3,j&))
```

```
        CLOSE #N%

'   -------- RE-PLOT PROBE #1 BECAUSE OF SOME ARTIFACT THAT DROPS IT FROM PROBE LINE #1 ?????? ------------------

    N% = FREEFILE : OPEN "probe1" FOR OUTPUT AS #N%

        PRINT #N%, USING$("######.####    ######.####   ######.####",R(1,1,j&),R(1,2,j&),R(1,3,j&))
        PRINT #N%, USING$("######.####    ######.####   ######.####",R(1,1,j&),R(1,2,j&),R(1,3,j&))

    CLOSE #N%

    s1 = 1.1## : s2 = s1 : s3 = s1 : s4 = s1 : s5 = s1 : s6 = s1 'expand plots axes by 10%

    IF XiMin(1) > 0## THEN s1 = 0.9##
    IF XiMax(1) < 0## THEN s2 = 0.9##
    IF XiMin(2) > 0## THEN s3 = 0.9##
    IF XiMax(2) < 0## THEN s4 = 0.9##
    IF XiMin(3) > 0## THEN s5 = 0.9##
    IF XiMax(3) < 0## THEN s6 = 0.9##

    xMin$ = REMOVE$(STR$(s1*XiMin(1),2),ANY" ")
    xMax$ = REMOVE$(STR$(s2*XiMax(1),2),ANY" ")
    yMin$ = REMOVE$(STR$(s3*XiMin(2),2),ANY" ")
    yMax$ = REMOVE$(STR$(s4*XiMax(2),2),ANY" ")
    zMin$ = REMOVE$(STR$(s5*XiMin(2),2),ANY" ")
    zMax$ = REMOVE$(STR$(s6*XiMax(2),2),ANY" ")

'USAGE: CALL ThreeDplot3(PlotFileName$,PlotTitle$,Annotation$,xCoord$,yCoord$,zCoord$, _
'                        XaxisLabel$,YaxisLabel$,ZaxisLabel$,zMin$,zMax$,GnuPlotEXE$,xStart$,xStop$,yStart$,yStop$)

    PlotWindowULC_X% = 50 : PlotWindowULC_Y% = 50 : PlotWindowWidth% = 1000 : PlotWindowHeight% = 800

    PlotWindowOffset% = 100*Gamma

    CALL CreateGNUplotINIfile(PlotWindowULC_X%+PlotWindowOffset%,PlotWindowULC_Y%+PlotWindowOffset%,PlotWindowWidth%,PlotWindowHeight%)

    CALL ThreeDplot3(PlotFileName$,PlotTitle$,Annotation$,"0.6","0.7","0.8", _
                     "x1","x2","x3",zMin$,zMax$,"wgnuplot.exe",xMin$,xMax$,yMin$,yMax$)

    'KILL PlotFileName$ 'erase plot data file after probes been displayed

END SUB 'Show3Dprobes()

'---

    SUB ThreeDplot3(PlotFileName$,PlotTitle$,Annotation$,xCoord$,yCoord$,zCoord$, _
                    XaxisLabel$,YaxisLabel$,ZaxisLabel$,zMin$,zMax$,GnuPlotEXE$,xStart$,xStop$,yStart$,yStop$)

        LOCAL N%, ProcID???

        N% = FREEFILE

        OPEN "cmd3d.gp" FOR OUTPUT AS #N%

            PRINT #N%, "set pm3d"
            PRINT #N%, "show pm3d"
            PRINT #N%, "set hidden3d"
'           PRINT #N%, "set view 45, 45, 1, 1"

            PRINT #N%, "set view 45, 60, 1, 1"

            IF zMin$ <> "" AND zMax$ <> "" THEN PRINT #N%, "set zrange ["+zMin$+":"+zMax$+"]"

            PRINT #N%, "set xrange [" + xStart$ + ":" + xStop$ + "]"
            PRINT #N%, "set yrange [" + yStart$ + ":" + yStop$ + "]"

            PRINT #N%, "set label "  + Quote$ + AnnoTation$ + Quote$+" at graph "+xCoord$+","+yCoord$+","+zCoord$
            PRINT #N%, "show label"
            PRINT #N%, "set grid xtics ytics ztics"
            PRINT #N%, "show grid"
            PRINT #N%, "set title "  + Quote$+PlotTitle$    + Quote$
            PRINT #N%, "set xlabel " + Quote$+XaxisLabel$   + Quote$
            PRINT #N%, "set ylabel " + Quote$+YaxisLabel$   + Quote$
            PRINT #N%, "set zlabel " + Quote$+ZaxisLabel$   + Quote$
            PRINT #N%, "unset colorbox"
'           print #N%, "set style fill"

            PRINT #N%, "splot "      + Quote$+PlotFileName$ + Quote$ + " notitle lw 1 pt 8," _
                                     + Quote$ + "diag"      + Quote$ + " notitle w l," _
                                     + Quote$ + "probeline1" + Quote$ + " notitle w l," _
                                     + Quote$ + "probeline2" + Quote$ + " notitle w l," _
                                     + Quote$ + "probeline3" + Quote$ + " notitle w l"
        CLOSE #N%

        ProcID??? = SHELL(GnuPlotEXE$+" cmd3d.gp -")

        CALL Delay(0.5##)

    END SUB 'ThreeDplot3()

'----

    SUB TwoDplot(PlotFileName$,PlotTitle$,xCoord$,yCoord$,XaxisLabel$,YaxisLabel$, _
                 LogXaxis$,LogYaxis$,xMin$,xMax$,yMin$,yMax$,xTics$,yTics$,GnuPlotEXE$,LineType$,Annotation$)

        LOCAL N%, ProcessID???

        N% = FREEFILE

        OPEN "cmd2d.gp" FOR OUTPUT AS #N%

            IF LogXaxis$ = "YES" AND LogYaxis$ = "NO"  THEN PRINT #N%, "set logscale x"
            IF LogXaxis$ = "NO"  AND LogYaxis$ = "YES" THEN PRINT #N%, "set logscale y"
            IF LogXaxis$ = "YES" AND LogYaxis$ = "YES" THEN PRINT #N%, "set logscale xy"

            IF xMin$ <> "" AND xMax$ <> "" THEN  PRINT #N%, "set xrange ["+xMin$+":"+xMax$+"]"
            IF yMin$ <> "" AND yMax$ <> "" THEN  PRINT #N%, "set yrange ["+yMin$+":"+yMax$+"]"

            PRINT #N%, "set label "   + Quote$ + Annotation$ + Quote$ + " at graph " + xCoord$ + "," + yCoord$
            PRINT #N%, "set grid xtics " + XTics$
            PRINT #N%, "set grid ytics " + yTics$
            PRINT #N%, "set grid mxtics"
            PRINT #N%, "set grid mytics"
            PRINT #N%, "show grid"
            PRINT #N%, "set title " + Quote$+PlotTitle$+Quote$
            PRINT #N%, "set xlabel " + Quote$+XaxisLabel$+Quote$
            PRINT #N%, "set ylabel " + Quote$+YaxisLabel$+Quote$

            PRINT #N%, "plot "+Quote$+PlotFileName$+Quote$+" notitle "+LineType$

        CLOSE #N%

        ProcessID??? = SHELL(GnuPlotEXE$+" cmd2d.gp -") : CALL Delay(0.5##)

    END SUB 'TwoDplot()

'-----
```

```
SUB CreateGNUplot INIfile(PlotWindowULC_X%,PlotWindowULC_Y%,PlotWindowWidth%,PlotWindowHeight%)

    LOCAL N%, WinPath$, A$, B$, WindowsDirectory$

    WinPath$ = UCASE$(ENVIRON$("Path"))'DIR$("C:\WINDOWS",23)

        DO
            B$ = A$

            A$ = EXTRACT$(WinPath$,";")

            WinPath$ = REMOVE$(WinPath$,A$+";")

            IF RIGHT$(A$,7) = "WINDOWS" OR A$ = B$ THEN EXIT LOOP

            IF RIGHT$(A$,5) = "WINNT"   OR A$ = B$ THEN EXIT LOOP

    LOOP

    WindowsDirectory$ = A$

    N% = FREEFILE

'   ----------- WGNUPLOT.INPUT FILE -----------
    OPEN WindowsDirectory$+"wgnuplot.ini" FOR OUTPUT AS #N%

        PRINT #N%,"[WGNUPLOT]"
        PRINT #N%,"TextOrigin=0 0"
        PRINT #N%,"TextSize=640 150"
        PRINT #N%,"TextFont=Terminal,9"
        PRINT #N%,"GraphOrigin="+REMOVE$(STR$(PlotWindowULC_X%),ANY" ")+" "+REMOVE$(STR$(PlotWindowULC_Y%),ANY" ")
        PRINT #N%,"GraphSize="  +REMOVE$(STR$(PlotWindowWidth%),ANY" ")+" "+REMOVE$(STR$(PlotWindowHeight%),ANY" ")
        PRINT #N%,"GraphFont=Arial,10"
        PRINT #N%,"GraphColor=1"
        PRINT #N%,"GraphToTop=1"
        PRINT #N%,"GraphBackground=255 255 255"
        PRINT #N%,"Border=0 0 0 0"
        PRINT #N%,"Axis=192 192 192 2 2"
        PRINT #N%,"Line1=0 0 255 0 0"
        PRINT #N%,"Line2=0 255 0 0 1"
        PRINT #N%,"Line3=255 0 0 0 2"
        PRINT #N%,"Line4=255 0 255 0 3"
        PRINT #N%,"Line5=0 0 128 0 4"

    CLOSE #N%

    END SUB 'CreateGNUplotINIfile()

'------

    SUB Delay(NumSecs)

        LOCAL StartTime, StopTime AS EXT

        StartTime = TIMER

        DO UNTIL (StopTime-StartTime) >= NumSecs

            StopTime = TIMER

        LOOP

    END SUB 'Delay()

'------

SUB MathematicalConstants
    EulerConst  = 0.57721566490153286060651288
    Pi          = 3.14159265358979323846264384##
    Pi2         = Pi/2##
    Pi4         = Pi/4##
    TwoPi       = 2##*Pi
    FourPi      = 4##*Pi
    e           = 2.71828182845904523536028784##
    Root2       = 1.41421356237309504884##
END SUB

'------

SUB AlphabetAndDigits
    Alphabet$   = "ABCDEFGHIJKLMNOPQRSTUVWXYZabcdefghijklmnopqrstuvwxyz"
    Digits$     = "0123456789"
    RunID$      = DATE$ + ", " + TIME$
END SUB

'------

SUB SpecialSymbols
    Quote$          = CHR$(34)  'Quotation mark "
    SpecialCharacters$ = "'(),#;:/_"
END SUB

'------

SUB EMconstants
    Mu0  = 4E-7##*Pi    'hy/meter
    Eps0 = 8.854##*1E-12 'fd/meter
    c    = 2.998E8##     'velocity of light, 1##/SQR(Mu0*Eps0) 'meters/sec
    eta0 = SQR(Mu0/Eps0) 'impedance of free space, ohms
END SUB

'------

SUB ConversionFactors
    Rad2Deg         = 180##/Pi
    Deg2Rad         = 1##/Rad2Deg
    Feet2Meters     = 0.3048##
    Meters2Feet     = 1##/Feet2Meters
    Inches2Meters   = 0.0254##
    Meters2Inches   = 1##/Inches2Meters
    Miles2Meters    = 1609.344##
    Meters2Miles    = 1##/Miles2Meters
    NautMi2Meters   = 1852##
    Meters2NautMi   = 1##/NautMi2Meters
END SUB

'------

SUB ShowConstants 'puts up msgbox showing all constants

LOCAL A$

A$ = _
"Mathematical Constants:"+CHR$(13)+_
"Euler const="+STR$(EulerConst)+CHR$(13)+_
"Pi="+STR$(Pi)+CHR$(13)+_
```

```
"Pi/2="+STR$(Pi2)+CHR$(13)+_
"Pi/4="+STR$(Pi4)+CHR$(13)+_
"2Pi="+STR$(TwoPi)+CHR$(13)+_
"4Pi="+STR$(FourPi)+CHR$(13)+_
"e="+STR$(e)+CHR$(13)+_
"Sqr2="+STR$(Root2)+CHR$(13)+CHR$(13)+_
"Alphabet, Digits & Special Characters:"+CHR$(13)+_
"Alphabet="+Alphabet$+CHR$(13)+_
"Digits="+Digits$+CHR$(13)+_
"quote="+Quote$+CHR$(13)+_
"Spec chars="+SpecialCharacters$+CHR$(13)+CHR$(13)+_
"E&M Constants:"+CHR$(13)+_
"Mu0="+STR$(Mu0)+CHR$(13)+_
"Eps0="+STR$(Eps0)+CHR$(13)+_
"c="+STR$(c)+CHR$(13)+_
"Eta0="+STR$(eta0)+CHR$(13)+CHR$(13)+_
"Conversion Factors:"+CHR$(13)+_
"Rad2Deg="+STR$(Rad2Deg)+CHR$(13)+_
"Deg2Rad="+STR$(Deg2Rad)+CHR$(13)+_
"Ft2meters="+STR$(Feet2Meters)+CHR$(13)+_
"Meters2Ft="+STR$(Meters2Feet)+CHR$(13)+_
"Inches2Meters="+STR$(Inches2Meters)+CHR$(13)+_
"Meters2Inches="+STR$(Meters2Inches)+CHR$(13)+_
"Miles2Meters="+STR$(Miles2Meters)+CHR$(13)+_
"Meters2Miles="+STR$(Meters2Miles)+CHR$(13)+_
"NautMi2Meters="+STR$(NautMi2Meters)+CHR$(13)+_
"Meters2NautMi="+STR$(Meters2NautMi)+CHR$(13)+CHR$(13)

MSGBOX(A$)

END SUB

'------

SUB DisplayRmatrix(Np%,Nd%,Nt&,R())

LOCAL p%, i%, j&, A$

    A$ = "Position Vector Matrix R()"+CHR$(13)

    FOR p% = 1 TO Np%

        FOR i% = 1 TO Nd%

            FOR j& = 0 TO Nt&

                A$ = A$ + "R("+STR$(p%)+", "+STR$(i%)+", "+STR$(j&)+" ) ="+STR$(R(p%,i%,j&)) + CHR$(13)

            NEXT j&

        NEXT i%

    NEXT p%

    MSGBOX(A$)

END SUB

'------

SUB DisplayRmatrixThisTimeStep(Np%,Nd%,j&,R(),Gamma)

LOCAL p%, i%, A$, B$

    A$ = "Position Vector Matrix R() at step "+STR$(j&)+", Gamma ="+STR$(Gamma)+"("+CHR$(13)+CHR$(13)

    FOR p% = 1 TO Np%

        A$ = A$ + "Probe#"+REMOVE$(STR$(p%),ANY" ")+": "

        B$ = ""

        FOR i% = 1 TO Nd%

            B$ = B$ + "  " + USING$("####.##",R(p%,i%,j&))

        NEXT i%

        A$ = A$ + B$ + CHR$(13)

    NEXT p%

    MSGBOX(A$)

END SUB

'------

SUB DisplayAmatrix(Np%,Nd%,Nt&,A())

LOCAL p%, i%, j&, A$

    A$ = "Acceleration Vector Matrix A()"+CHR$(13)

    FOR p% = 1 TO Np%

        FOR i% = 1 TO Nd%

            FOR j& = 0 TO Nt&

                A$ = A$ + "A("+STR$(p%)+", "+STR$(i%)+", "+STR$(j&)+" ) ="+STR$(A(p%,i%,j&)) + CHR$(13)

            NEXT j&

        NEXT i%

    NEXT p%

    MSGBOX(A$)

END SUB

'------

SUB DisplayAmatrixThisTimeStep(Np%,Nd%,j&,A())

LOCAL p%, i%, A$

    A$ = "Acceleration matrix A() at step "+STR$(j&)+":"+CHR$(13)

    FOR p% = 1 TO Np%

        FOR i% = 1 TO Nd%

            A$ = A$ + "A("+STR$(p%)+", "+STR$(i%)+", "+STR$(j&)+" ) ="+STR$(A(p%,i%,j&)) + CHR$(13)

        NEXT i%
```

```
        NEXT p%

        MSGBOX(A$)

END SUB

'------

SUB DisplayMmatrix(Np%,Nt&,M())

LOCAL p%, j&, A$

        A$ = "Fitness Matrix M()"+CHR$(13)

        FOR p% = 1 TO Np%

            FOR j& = 0 TO Nt&

                A$ = A$ + "M("+STR$(p%)+", "+STR$(j&)+" ) ="+STR$(M(p%,j&)) + CHR$(13)

            NEXT j&

        NEXT p%

        MSGBOX(A$)

END SUB

'------

SUB DisplayMbestMatrix(Np%,Nt&,Mbest())

LOCAL p%, j&, A$

        A$ = "Np= "+STR$(Np%)+"   Nt="+STR$(Nt&)+CHR$(13)+"Fitness Matrix Mbest()"+CHR$(13)

        FOR p% = 1 TO Np%

            FOR j& = 0 TO Nt&

                A$ = A$ + "Mbest("+STR$(p%)+", "+STR$(j&)+" ) ="+STR$(Mbest(p%,j&)) + CHR$(13)

            NEXT j&

        NEXT p%

        MSGBOX(A$)

END SUB

'------

SUB DisplayMmatrixThisTimeStep(Np%,j&,M())

LOCAL p%, A$

        A$ = "Fitness matrix M() at step "+STR$(j&)+":"+CHR$(13)

        FOR p% = 1 TO Np%

            A$ = A$ + "M("+STR$(p%)+", "+STR$(j&)+" ) ="+STR$(M(p%,j&)) + CHR$(13)

        NEXT p%

        MSGBOX(A$)

END SUB

'------

SUB DisplayXiMinMax(Nd%,XiMin(),XiMax())

LOCAL i%, A$

        A$ = ""

        FOR i% = 1 TO Nd%

            A$ = A$ + "XiMin("+STR$(i%)+" ) = "+STR$(XiMin(i%))+"    XiMax("+STR$(i%)+" ) = "+STR$(XiMax(i%)) + CHR$(13)

        NEXT i%

        MSGBOX(A$)

END SUB

'------

SUB DisplayRunParameters2(FunctionName$,Nd%,Np%,Nt&,G,DeltaT,Alpha,Beta,Frep,PlaceInitialProbes$,InitialAcceleration$,RepositionFactor$)

LOCAL A$

        A$ = "Function = "+ FunctionName$+CHR$(13)+_
             "Nd  = "+STR$(Nd%)+CHR$(13)+_
             "Np  = "+STR$(Np%)+CHR$(13)+_
             "Nt  = "+STR$(Nt&)+CHR$(13)+_
             "G   = "+STR$(G)+CHR$(13)+_
             "DeltaT  = "+STR$(DeltaT)+CHR$(13)+_
             "Alpha  = "+STR$(Alpha)+CHR$(13)+_
             "Beta   = "+STR$(Beta)+CHR$(13)+_
             "Frep  = "+STR$(Frep)+CHR$(13)+_
             "Init Probes: "+PlaceInitialProbes$+CHR$(13)+_
             "Init Accel:  "+InitialAcceleration$+CHR$(13)+_
             "Retrive Method: "+RepositionFactor$+CHR$(13)

        MSGBOX(A$)

END SUB

'------

SUB
Tabulate1DprobeCoordinates(MaxIDprobesPlotted%,Nd%,Np%,LastStep&,G,DeltaT,Alpha,Beta,Frep,R(),M(),PlaceInitialProbes$,InitialAcceleration$,RepositionFactor$,FunctionName$,Ga
mma)

LOCAL N%, ProbeNum%, FileHeader$, A$, B$, C$, D$, E$, F$, H$, StepNum&, FieldNumber%  'kludgy, yes, but it accomplishes its purpose...

        CALL GetPlotAnnotation(FileHeader$,Nd%,Np%,LastStep&,G,DeltaT,Alpha,Beta,Frep,M(),PlaceInitialProbes$,InitialAcceleration$,RepositionFactor$,FunctionName$,Gamma)

        REPLACE "\n" WITH "," IN FileHeader$

        FileHeader$ = LEFT$(FileHeader$,LEN(FileHeader$)-2)

        FileHeader$ = "PROBE COORDINATES" + CHR$(13) +_
                      "-----------------" + CHR$(13) + FileHeader$
```

```
        N% = FREEFILE : OPEN "ProbeCoordinates.DAT" FOR OUTPUT AS #N%

            A$ = "    Step #      " : B$ = "    ------     " : C$ = ""

            FOR ProbeNum% = 1 TO Np% 'create out data file header

                SELECT CASE ProbeNum%
                    CASE   1 TO   9 : E$ = ""   : F$ = "    "          : H$ = "        "
                    CASE  10 TO  99 : E$ = "-"  : F$ = "   "           : H$ = "       "
                    CASE 100 TO 999 : E$ = "--" : F$ = "  "            : H$ = "      "
                END SELECT

                A$ = A$ + "P" + NoSpaces$(ProbeNum%+0,0) + F$ 'note: adding zero to ProbeNum% necessary to convert to floating point...

                B$ = B$ + E$ + "--" + H$

                C$ = C$ + "#######.###    "

'               C$ = C$ + "##.#######"

            NEXT ProbeNum%

            PRINT #N%, FileHeader$ + CHR$(13) : PRINT #N%, A$ : PRINT #N%, B$

            FOR StepNum% = 0 TO LastStep&

                D$ = USING$("######   ",StepNum&)

                FOR ProbeNum% = 1 TO Np% : D$ = D$ + USING$(C$,R(ProbeNum%,1,StepNum&)) : NEXT ProbeNum%

                PRINT #N%, D$

            NEXT StepNum&

        CLOSE #N%

END SUB 'TabulatelDprobeCoordinates()

'------

SUB
PlotlDprobePositions(MaxlDprobesPlotted%,Nd%,Np%,LastStep&,G,DeltaT,Alpha,Beta,Prep,R(),M(),PlaceInitialProbes$,InitialAcceleration$,RepositionFactor$,FunctionName$,Gamma)
    'plots on-screen 1D function probe positions vs time step if Np =< 10

LOCAL ProcessID??, N%, n1%, n2&, n3%, n4%, n5%, n6%, n7%, n8%, n9%, n10%, n11%, n12%, n13%, n14%, n15%, ProbeNum%, StepNum&, A$

LOCAL PlotAnnotation$

    IF Np% > MaxlDprobesPlotted% THEN EXIT SUB

    CALL CLEANUP 'delete old "Px" plot files, if any

    ProbeNum% = 0

    DO 'create output data files, probe-by-probe
        INCR ProbeNum% : n1% = FREEFILE : OPEN "P"+REMOVE$(STR$(ProbeNum%),ANY" ") FOR OUTPUT AS #n1%  : IF ProbeNum% = Np% THEN EXIT LOOP
        INCR ProbeNum% : n2% = FREEFILE : OPEN "P"+REMOVE$(STR$(ProbeNum%),ANY" ") FOR OUTPUT AS #n2%  : IF ProbeNum% = Np% THEN EXIT LOOP
        INCR ProbeNum% : n3% = FREEFILE : OPEN "P"+REMOVE$(STR$(ProbeNum%),ANY" ") FOR OUTPUT AS #n3%  : IF ProbeNum% = Np% THEN EXIT LOOP
        INCR ProbeNum% : n4% = FREEFILE : OPEN "P"+REMOVE$(STR$(ProbeNum%),ANY" ") FOR OUTPUT AS #n4%  : IF ProbeNum% = Np% THEN EXIT LOOP
        INCR ProbeNum% : n5% = FREEFILE : OPEN "P"+REMOVE$(STR$(ProbeNum%),ANY" ") FOR OUTPUT AS #n5%  : IF ProbeNum% = Np% THEN EXIT LOOP
        INCR ProbeNum% : n6% = FREEFILE : OPEN "P"+REMOVE$(STR$(ProbeNum%),ANY" ") FOR OUTPUT AS #n6%  : IF ProbeNum% = Np% THEN EXIT LOOP
        INCR ProbeNum% : n7% = FREEFILE : OPEN "P"+REMOVE$(STR$(ProbeNum%),ANY" ") FOR OUTPUT AS #n7%  : IF ProbeNum% = Np% THEN EXIT LOOP
        INCR ProbeNum% : n8% = FREEFILE : OPEN "P"+REMOVE$(STR$(ProbeNum%),ANY" ") FOR OUTPUT AS #n8%  : IF ProbeNum% = Np% THEN EXIT LOOP
        INCR ProbeNum% : n9% = FREEFILE : OPEN "P"+REMOVE$(STR$(ProbeNum%),ANY" ") FOR OUTPUT AS #n9%  : IF ProbeNum% = Np% THEN EXIT LOOP
        INCR ProbeNum% : n10% = FREEFILE : OPEN "P"+REMOVE$(STR$(ProbeNum%),ANY" ") FOR OUTPUT AS #n10% : IF ProbeNum% = Np% THEN EXIT LOOP
        INCR ProbeNum% : n11% = FREEFILE : OPEN "P"+REMOVE$(STR$(ProbeNum%),ANY" ") FOR OUTPUT AS #n11% : IF ProbeNum% = Np% THEN EXIT LOOP
        INCR ProbeNum% : n12% = FREEFILE : OPEN "P"+REMOVE$(STR$(ProbeNum%),ANY" ") FOR OUTPUT AS #n12% : IF ProbeNum% = Np% THEN EXIT LOOP
        INCR ProbeNum% : n13% = FREEFILE : OPEN "P"+REMOVE$(STR$(ProbeNum%),ANY" ") FOR OUTPUT AS #n13% : IF ProbeNum% = Np% THEN EXIT LOOP
        INCR ProbeNum% : n14% = FREEFILE : OPEN "P"+REMOVE$(STR$(ProbeNum%),ANY" ") FOR OUTPUT AS #n14% : IF ProbeNum% = Np% THEN EXIT LOOP
        INCR ProbeNum% : n15% = FREEFILE : OPEN "P"+REMOVE$(STR$(ProbeNum%),ANY" ") FOR OUTPUT AS #n15% : IF ProbeNum% = Np% THEN EXIT LOOP
    LOOP

    ProbeNum% = 0

    DO 'output probe positions as a function of time step
        INCR ProbeNum% : FOR StepNum& = 0 TO LastStep& : PRINT #n1%,  USING$("######  #####.########",StepNum&,R(ProbeNum%,1,StepNum&)) : NEXT StepNum& : IF ProbeNum% = Np%
THEN EXIT LOOP
        INCR ProbeNum% : FOR StepNum& = 0 TO LastStep& : PRINT #n2%,  USING$("######  #####.########",StepNum&,R(ProbeNum%,1,StepNum&)) : NEXT StepNum& : IF ProbeNum% = Np%
THEN EXIT LOOP
        INCR ProbeNum% : FOR StepNum& = 0 TO LastStep& : PRINT #n3%,  USING$("######  #####.########",StepNum&,R(ProbeNum%,1,StepNum&)) : NEXT StepNum& : IF ProbeNum% = Np%
THEN EXIT LOOP
        INCR ProbeNum% : FOR StepNum& = 0 TO LastStep& : PRINT #n4%,  USING$("######  #####.########",StepNum&,R(ProbeNum%,1,StepNum&)) : NEXT StepNum& : IF ProbeNum% = Np%
THEN EXIT LOOP
        INCR ProbeNum% : FOR StepNum& = 0 TO LastStep& : PRINT #n5%,  USING$("######  #####.########",StepNum&,R(ProbeNum%,1,StepNum&)) : NEXT StepNum& : IF ProbeNum% = Np%
THEN EXIT LOOP
        INCR ProbeNum% : FOR StepNum& = 0 TO LastStep& : PRINT #n6%,  USING$("######  #####.########",StepNum&,R(ProbeNum%,1,StepNum&)) : NEXT StepNum& : IF ProbeNum% = Np%
THEN EXIT LOOP
        INCR ProbeNum% : FOR StepNum& = 0 TO LastStep& : PRINT #n7%,  USING$("######  #####.########",StepNum&,R(ProbeNum%,1,StepNum&)) : NEXT StepNum& : IF ProbeNum% = Np%
THEN EXIT LOOP
        INCR ProbeNum% : FOR StepNum& = 0 TO LastStep& : PRINT #n8%,  USING$("######  #####.########",StepNum&,R(ProbeNum%,1,StepNum&)) : NEXT StepNum& : IF ProbeNum% = Np%
THEN EXIT LOOP
        INCR ProbeNum% : FOR StepNum& = 0 TO LastStep& : PRINT #n9%,  USING$("######  #####.########",StepNum&,R(ProbeNum%,1,StepNum&)) : NEXT StepNum& : IF ProbeNum% = Np%
THEN EXIT LOOP
        INCR ProbeNum% : FOR StepNum& = 0 TO LastStep& : PRINT #n10%, USING$("######  #####.########",StepNum&,R(ProbeNum%,1,StepNum&)) : NEXT StepNum& : IF ProbeNum% = Np%
THEN EXIT LOOP
        INCR ProbeNum% : FOR StepNum& = 0 TO LastStep& : PRINT #n11%, USING$("######  #####.########",StepNum&,R(ProbeNum%,1,StepNum&)) : NEXT StepNum& : IF ProbeNum% = Np%
THEN EXIT LOOP
        INCR ProbeNum% : FOR StepNum& = 0 TO LastStep& : PRINT #n12%, USING$("######  #####.########",StepNum&,R(ProbeNum%,1,StepNum&)) : NEXT StepNum& : IF ProbeNum% = Np%
THEN EXIT LOOP
        INCR ProbeNum% : FOR StepNum& = 0 TO LastStep& : PRINT #n13%, USING$("######  #####.########",StepNum&,R(ProbeNum%,1,StepNum&)) : NEXT StepNum& : IF ProbeNum% = Np%
THEN EXIT LOOP
        INCR ProbeNum% : FOR StepNum& = 0 TO LastStep& : PRINT #n14%, USING$("######  #####.########",StepNum&,R(ProbeNum%,1,StepNum&)) : NEXT StepNum& : IF ProbeNum% = Np%
THEN EXIT LOOP
        INCR ProbeNum% : FOR StepNum& = 0 TO LastStep& : PRINT #n15%, USING$("######  #####.########",StepNum&,R(ProbeNum%,1,StepNum&)) : NEXT StepNum& : IF ProbeNum% = Np%
THEN EXIT LOOP
    LOOP

    ProbeNum% = 0

    DO 'close output data files
        INCR ProbeNum% : CLOSE #n1%  : IF ProbeNum% = Np% THEN EXIT LOOP
        INCR ProbeNum% : CLOSE #n2%  : IF ProbeNum% = Np% THEN EXIT LOOP
        INCR ProbeNum% : CLOSE #n3%  : IF ProbeNum% = Np% THEN EXIT LOOP
        INCR ProbeNum% : CLOSE #n4%  : IF ProbeNum% = Np% THEN EXIT LOOP
        INCR ProbeNum% : CLOSE #n5%  : IF ProbeNum% = Np% THEN EXIT LOOP
        INCR ProbeNum% : CLOSE #n6%  : IF ProbeNum% = Np% THEN EXIT LOOP
        INCR ProbeNum% : CLOSE #n7%  : IF ProbeNum% = Np% THEN EXIT LOOP
        INCR ProbeNum% : CLOSE #n8%  : IF ProbeNum% = Np% THEN EXIT LOOP
        INCR ProbeNum% : CLOSE #n9%  : IF ProbeNum% = Np% THEN EXIT LOOP
        INCR ProbeNum% : CLOSE #n10% : IF ProbeNum% = Np% THEN EXIT LOOP
        INCR ProbeNum% : CLOSE #n11% : IF ProbeNum% = Np% THEN EXIT LOOP
        INCR ProbeNum% : CLOSE #n12% : IF ProbeNum% = Np% THEN EXIT LOOP
        INCR ProbeNum% : CLOSE #n13% : IF ProbeNum% = Np% THEN EXIT LOOP
        INCR ProbeNum% : CLOSE #n14% : IF ProbeNum% = Np% THEN EXIT LOOP
        INCR ProbeNum% : CLOSE #n15% : IF ProbeNum% = Np% THEN EXIT LOOP
    LOOP
```

```
       ProbeNum% = 0 : A$ = ""

       DO 'create file string for plot command file
             INCR ProbeNum% : A$ = A$ + Quote$ + "P"+REMOVE$(STR$(ProbeNum%),ANY" ") + Quote$ + " w l lw 2, " : IF ProbeNum% = Np% THEN EXIT LOOP
             INCR ProbeNum% : A$ = A$ + Quote$ + "P"+REMOVE$(STR$(ProbeNum%),ANY" ") + Quote$ + " w l lw 2, " : IF ProbeNum% = Np% THEN EXIT LOOP
             INCR ProbeNum% : A$ = A$ + Quote$ + "P"+REMOVE$(STR$(ProbeNum%),ANY" ") + Quote$ + " w l lw 2, " : IF ProbeNum% = Np% THEN EXIT LOOP
             INCR ProbeNum% : A$ = A$ + Quote$ + "P"+REMOVE$(STR$(ProbeNum%),ANY" ") + Quote$ + " w l lw 2, " : IF ProbeNum% = Np% THEN EXIT LOOP
             INCR ProbeNum% : A$ = A$ + Quote$ + "P"+REMOVE$(STR$(ProbeNum%),ANY" ") + Quote$ + " w l lw 2, " : IF ProbeNum% = Np% THEN EXIT LOOP
             INCR ProbeNum% : A$ = A$ + Quote$ + "P"+REMOVE$(STR$(ProbeNum%),ANY" ") + Quote$ + " w l lw 2, " : IF ProbeNum% = Np% THEN EXIT LOOP
             INCR ProbeNum% : A$ = A$ + Quote$ + "P"+REMOVE$(STR$(ProbeNum%),ANY" ") + Quote$ + " w l lw 2, " : IF ProbeNum% = Np% THEN EXIT LOOP
             INCR ProbeNum% : A$ = A$ + Quote$ + "P"+REMOVE$(STR$(ProbeNum%),ANY" ") + Quote$ + " w l lw 2, " : IF ProbeNum% = Np% THEN EXIT LOOP
             INCR ProbeNum% : A$ = A$ + Quote$ + "P"+REMOVE$(STR$(ProbeNum%),ANY" ") + Quote$ + " w l lw 2, " : IF ProbeNum% = Np% THEN EXIT LOOP
             INCR ProbeNum% : A$ = A$ + Quote$ + "P"+REMOVE$(STR$(ProbeNum%),ANY" ") + Quote$ + " w l lw 2, " : IF ProbeNum% = Np% THEN EXIT LOOP
             INCR ProbeNum% : A$ = A$ + Quote$ + "P"+REMOVE$(STR$(ProbeNum%),ANY" ") + Quote$ + " w l lw 2, " : IF ProbeNum% = Np% THEN EXIT LOOP
             INCR ProbeNum% : A$ = A$ + Quote$ + "P"+REMOVE$(STR$(ProbeNum%),ANY" ") + Quote$ + " w l lw 2, " : IF ProbeNum% = Np% THEN EXIT LOOP
             INCR ProbeNum% : A$ = A$ + Quote$ + "P"+REMOVE$(STR$(ProbeNum%),ANY" ") + Quote$ + " w l lw 2, " : IF ProbeNum% = Np% THEN EXIT LOOP
             INCR ProbeNum% : A$ = A$ + Quote$ + "P"+REMOVE$(STR$(ProbeNum%),ANY" ") + Quote$ + " w l lw 2, " : IF ProbeNum% = Np% THEN EXIT LOOP
       LOOP

       A$ = LEFT$(A$,LEN(A$)-2)

       CALL GetPlotAnnotation(PlotAnnotation$,Nd%,Np%,LastStep&,G,DeltaT,Alpha,Beta,Frep,M(),PlaceInitialProbes$,InitialAcceleration$,RepositionFactor$,FunctionName$,Gamma)

       N% = FREEFILE

       OPEN "cmd2d.gp" FOR OUTPUT AS #N%

             PRINT #N%, "set label "       + Quote$ + PlotAnnotation$ + Quote$ + " at graph 0.5,0.95"
             PRINT #N%, "set grid xtics"
             PRINT #N%, "set grid ytics"
             PRINT #N%, "set title "   + Quote$ + "Evolution of "    + FunctionName$ + " Probe Positions"+ "\n" + RunID$ + Quote$
             PRINT #N%, "set xlabel "  + Quote$ + "Time Step"       + Quote$
             PRINT #N%, "set ylabel "  + Quote$ + "Probe Coordinate" + Quote$
             PRINT #N%, "plot "        + A$

       CLOSE #N%

       CALL CreateGNUplotINIfile(0.2##*ScreenWidth&,0.2##*ScreenHeight&,0.6##*ScreenWidth&,0.6##*ScreenHeight&) 'USAGE: CALL CreateGNUplotINIfile(PlotWindowGLC_X%,PlotWindowGLC_Y%,PlotWindowWidth%,PlotWindowWeight%)

       ProcessID??? = SHELL("wgnuplot.exe"+" cmd2d.gp -") : CALL Delay(5##) 'before SUB Cleanup is called

END SUB

'------

SUB
DisplayRunParameters(FunctionName$,Nd%,Np%,Nt&,G,DeltaT,Alpha,Beta,Frep,R(),A(),M(),PlaceInitialProbes$,InitialAcceleration$,RepositionFactor$,RunCFO$,ShrinkDS$,CheckForEarl
yTermination$)

LOCAL A$, B$, YN&

       B$ = "" : IF PlaceInitialProbes$ = "UNIFORM ON-AXIS" AND Nd% > 1 THEN B$ = "  ["+REMOVE$(STR$(Np%/Nd%),ANY" ") + "/axis]"

       RunCFO$ = "NO"

       A$ = "RUN CFO WITH THE" + CHR$(13) +_
             "FOLLOWING PARAMETERS?"                               + CHR$(13) + CHR$(13) +_
             "Function "       + FunctionName$                   + " (" + REMOVE$(STR$(Nd%),ANY" ") + "-D)" + CHR$(13) +_
             "# time steps = " + REMOVE$(STR$(Nt&),ANY" ")       + CHR$(13) + _
             "Grav Const G = " + REMOVE$(STR$(G,2),ANY" ")       + CHR$(13) + _
             "Delta-T = "      + REMOVE$(STR$(DeltaT,3),ANY" ")  + CHR$(13) + _
             "Exp Alpha = "    + REMOVE$(STR$(Alpha,3),ANY" ")   + CHR$(13) + _
             "Exp Beta = "     + REMOVE$(STR$(Beta,3),ANY" ")    + CHR$(13) + _
             "Frep = "         + REMOVE$(STR$(Frep,4),ANY" ")    + " ["+RepositionFactor$ + "]" + CHR$(13) + _
             "Initial Probes: " + PlaceInitialProbes$            + CHR$(13) + _
             "Initial Accel: " + InitialAcceleration$            + CHR$(13) + _
             "Check for Early Termination? " + CheckForEarlyTermination$ + CHR$(13) + _
             "Shrink Decision Space? "       + ShrinkDS$ + CHR$(13) +CHR$(13)
'         lResult& = MSGBOX(txt$ [, [style&], title$])

       A$ = "RUN CFO ON FUNCTION " + FunctionName$ + "?"

       YN& = MSGBOX(A$,%MB_YESNO,"CONFIRM RUN")

       IF YN& = %IDYES THEN RunCFO$ = "YES"

END SUB

'------

SUB StatusWindow(FunctionName$,StatusWindowHandle???)

       GRAPHIC WINDOW "Run Progress, "+FunctionName$,0.08##*ScreenWidth&,0.08##*ScreenHeight&,0.25##*ScreenWidth&,0.17##*ScreenHeight& TO StatusWindowHandle???

       GRAPHIC ATTACH StatusWindowHandle???,0,REDRAW

       GRAPHIC FONT "Lucida Console",8,0 '"Courier New",8,0 'Fixed width fonts

       GRAPHIC SET PIXEL (35,15) : GRAPHIC PRINT " Initializing...     " : GRAPHIC REDRAW

END SUB

'------

SUB GetTestFunctionNumber(FunctionName$)

  LOCAL hDlg AS DWORD

  LOCAL N%, M%

  LOCAL FrameWidth&, FrameHeight&, BoxWidth&, BoxHeight&

  BoxWidth& = 276 : BoxHeight& = 300 : FrameWidth& = 82 : FrameHeight& = BoxHeight&-5

  DIALOG NEW 0, "CENTRAL FORCE OPTIMIZATION TEST FUNCTIONS",,, BoxWidth&, BoxHeight&,%WS_CAPTION OR %WS_SYSMENU, 0 TO hDlg
'-------------------------------------------------------------------------------------
       CONTROL ADD FRAME,  hDlg, %IDC_FRAME1, "Test Functions",       5,  2, FrameWidth&, FrameHeight&
       CONTROL ADD FRAME,  hDlg, %IDC_FRAME2, "GSO Test Functions",  95,  2, FrameWidth&, 255

       CONTROL ADD OPTION, hDlg, %IDC_Function_Number1,  "Parrott F4",    10,  14, 62, 10, %WS_GROUP OR %WS_TABSTOP
       CONTROL ADD OPTION, hDlg, %IDC_Function_Number2,  "SGO",           10,  24, 62, 10
       CONTROL ADD OPTION, hDlg, %IDC_Function_Number3,  "Goldstein-Price",10,  34, 62, 10
       CONTROL ADD OPTION, hDlg, %IDC_Function_Number4,  "Step",          10,  44, 62, 10
       CONTROL ADD OPTION, hDlg, %IDC_Function_Number5,  "Schwefel 2.26", 10,  54, 62, 10
       CONTROL ADD OPTION, hDlg, %IDC_Function_Number6,  "Colville",      10,  64, 62, 10
       CONTROL ADD OPTION, hDlg, %IDC_Function_Number7,  "Griewank",      10,  74, 62, 10

       CONTROL ADD OPTION, hDlg, %IDC_Function_Number31, "PBM #1",        10,  84, 62, 10
       CONTROL ADD OPTION, hDlg, %IDC_Function_Number32, "PBM #2",        10,  94, 62, 10
       CONTROL ADD OPTION, hDlg, %IDC_Function_Number33, "PBM #3",        10, 104, 62, 10
       CONTROL ADD OPTION, hDlg, %IDC_Function_Number34, "PBM #4",        10, 114, 62, 10
       CONTROL ADD OPTION, hDlg, %IDC_Function_Number35, "PBM #5",        10, 124, 62, 10
```

```
CONTROL ADD OPTION, hDlg, %IDC_Function_Number36, "Himmelblau",    10, 134, 62, 10
CONTROL ADD OPTION, hDlg, %IDC_Function_Number37, "Rosenbrock",    10, 144, 62, 10
CONTROL ADD OPTION, hDlg, %IDC_Function_Number38, "Sphere",        10, 154, 62, 10
CONTROL ADD OPTION, hDlg, %IDC_Function_Number39, "HimmelblauNLO", 10, 164, 62, 10
CONTROL ADD OPTION, hDlg, %IDC_Function_Number40, "Reserved",      10, 174, 62, 10
CONTROL ADD OPTION, hDlg, %IDC_Function_Number41, "Reserved",      10, 184, 62, 10
CONTROL ADD OPTION, hDlg, %IDC_Function_Number42, "Reserved",      10, 194, 62, 10
CONTROL ADD OPTION, hDlg, %IDC_Function_Number43, "Reserved",      10, 204, 62, 10
CONTROL ADD OPTION, hDlg, %IDC_Function_Number44, "Reserved",      10, 214, 62, 10
CONTROL ADD OPTION, hDlg, %IDC_Function_Number45, "Reserved",      10, 224, 62, 10
CONTROL ADD OPTION, hDlg, %IDC_Function_Number46, "Reserved",      10, 234, 62, 10
CONTROL ADD OPTION, hDlg, %IDC_Function_Number47, "Reserved",      10, 244, 62, 10
CONTROL ADD OPTION, hDlg, %IDC_Function_Number48, "Reserved",      10, 254, 62, 10
CONTROL ADD OPTION, hDlg, %IDC_Function_Number49, "Reserved",      10, 264, 62, 10
CONTROL ADD OPTION, hDlg, %IDC_Function_Number50, "Reserved",      10, 274, 62, 10

' ---------------------- Test Functions from GSO Paper ----------------------
CONTROL ADD OPTION, hDlg, %IDC_Function_Number8,  "f1" , 120,  14, 40, 10
CONTROL ADD OPTION, hDlg, %IDC_Function_Number9,  "f2" , 120,  24, 40, 10
CONTROL ADD OPTION, hDlg, %IDC_Function_Number10, "f3" , 120,  34, 40, 10
CONTROL ADD OPTION, hDlg, %IDC_Function_Number11, "f4" , 120,  44, 40, 10
CONTROL ADD OPTION, hDlg, %IDC_Function_Number12, "f5" , 120,  54, 40, 10
CONTROL ADD OPTION, hDlg, %IDC_Function_Number13, "f6" , 120,  64, 40, 10
CONTROL ADD OPTION, hDlg, %IDC_Function_Number14, "f7" , 120,  74, 40, 10
CONTROL ADD OPTION, hDlg, %IDC_Function_Number15, "f8" , 120,  84, 40, 10
CONTROL ADD OPTION, hDlg, %IDC_Function_Number16, "f9" , 120,  94, 40, 10
CONTROL ADD OPTION, hDlg, %IDC_Function_Number17, "f10", 120, 104, 40, 10
CONTROL ADD OPTION, hDlg, %IDC_Function_Number18, "f11", 120, 114, 40, 10
CONTROL ADD OPTION, hDlg, %IDC_Function_Number19, "f12", 120, 124, 40, 10
CONTROL ADD OPTION, hDlg, %IDC_Function_Number20, "f13", 120, 134, 40, 10
CONTROL ADD OPTION, hDlg, %IDC_Function_Number21, "f14", 120, 144, 40, 10
CONTROL ADD OPTION, hDlg, %IDC_Function_Number22, "f15", 120, 154, 40, 10
CONTROL ADD OPTION, hDlg, %IDC_Function_Number23, "f16", 120, 164, 40, 10
CONTROL ADD OPTION, hDlg, %IDC_Function_Number24, "f17", 120, 174, 40, 10
CONTROL ADD OPTION, hDlg, %IDC_Function_Number25, "f18", 120, 184, 40, 10
CONTROL ADD OPTION, hDlg, %IDC_Function_Number26, "f19", 120, 194, 40, 10
CONTROL ADD OPTION, hDlg, %IDC_Function_Number27, "f20", 120, 204, 40, 10
CONTROL ADD OPTION, hDlg, %IDC_Function_Number28, "f21", 120, 214, 40, 10
CONTROL ADD OPTION, hDlg, %IDC_Function_Number29, "f22", 120, 224, 40, 10
CONTROL ADD OPTION, hDlg, %IDC_Function_Number30, "f23", 120, 234, 40, 10

CONTROL SET OPTION  hDlg, %IDC_Function_Number1, %IDC_Function_Number1, %IDC_Function_Number3 'default to Parrott F4

'-----------------------------------------------------------------------

CONTROL ADD BUTTON, hDlg, %IDOK, "&OK", 200, 0.45##*BoxHeight&, 50, 14

'-----------------------------------------------------------------------

DIALOG SHOW MODAL hDlg CALL DlgProc

CALL Delay(0.5##)

IF FunctionNumber% < 1 OR FunctionNumber% > 39 THEN

  FunctionNumber% = 1 : MSGBOX("Error in function number...")

END IF

  SELECT CASE FunctionNumber%

      CASE 1 : FunctionName$ = "ParrottF4"
      CASE 2 : FunctionName$ = "SGO"
      CASE 3 : FunctionName$ = "GP"
      CASE 4 : FunctionName$ = "STEP"
      CASE 5 : FunctionName$ = "SCHWEFEL_226"
      CASE 6 : FunctionName$ = "COLVILLE"
      CASE 7 : FunctionName$ = "GRIEWANK"
      CASE 8 : FunctionName$ = "F1"
      CASE 9 : FunctionName$ = "F2"
      CASE 10: FunctionName$ = "F3"
      CASE 11: FunctionName$ = "F4"
      CASE 12: FunctionName$ = "F5"
      CASE 13: FunctionName$ = "F6"
      CASE 14: FunctionName$ = "F7"
      CASE 15: FunctionName$ = "F8"
      CASE 16: FunctionName$ = "F9"
      CASE 17: FunctionName$ = "F10"
      CASE 18: FunctionName$ = "F11"
      CASE 19: FunctionName$ = "F12"
      CASE 20: FunctionName$ = "F13"
      CASE 21: FunctionName$ = "F14"
      CASE 22: FunctionName$ = "F15"
      CASE 23: FunctionName$ = "F16"
      CASE 24: FunctionName$ = "F17"
      CASE 25: FunctionName$ = "F18"
      CASE 26: FunctionName$ = "F19"
      CASE 27: FunctionName$ = "F20"
      CASE 28: FunctionName$ = "F21"
      CASE 29: FunctionName$ = "F22"
      CASE 30: FunctionName$ = "F23"
      CASE 31: FunctionName$ = "PBM_1"
      CASE 32: FunctionName$ = "PBM_2"
      CASE 33: FunctionName$ = "PBM_3"
      CASE 34: FunctionName$ = "PBM_4"
      CASE 35: FunctionName$ = "PBM_5"
      CASE 36: FunctionName$ = "HIMMELBLAU"
      CASE 37: FunctionName$ = "ROSENBROCK"
      CASE 38: FunctionName$ = "SPHERE"
      CASE 39: FunctionName$ = "HIMMELBLAUNLO"

  END SELECT

END SUB

'-----------

CALLBACK FUNCTION DlgProc() AS LONG

  '-----------------------------------------------------------------
  ' Callback procedure for the main dialog
  '-----------------------------------------------------------------
  LOCAL c, lRes AS LONG, sText AS STRING

  SELECT CASE AS LONG CBMSG

  CASE %WM_INITDIALOG' %WM_INITDIALOG is sent right before the dialog is shown.

  CASE %WM_COMMAND          ' <- a control is calling

    SELECT CASE AS LONG CBCTL  ' <- look at control's id

    CASE %IDOK                ' OK button or Enter key was pressed

        IF CBCTLMSG = %BN_CLICKED THEN
          '-----------------------------------------
          ' Loop through the Function_Number controls
```

```
            ' to see which one is selected
            '----------------------------------------
            FOR c = %IDC_Function_Number1 TO %IDC_Function_Number50

                CONTROL GET CHECK CBHNDL, c TO lRes

                IF lRes THEN EXIT FOR

            NEXT 'c holds the id for selected test function.

            FunctionNumber% = c-120

            DIALOG END CBHNDL

        END IF

    END SELECT

  END SELECT

END FUNCTION

'------------------------------ PBM ANTENNA BENCHMARK FUNCTIONS ----------------------------

'Reference for benchmarks PBM_1 through PBM_5:

'Pantoja, M F., Bretones, A. R., Martin, R. G., "Benchmark Antenna Problems for Evolutionary
'Optimization Algorithms," IEEE Trans. Antennas & Propagation, vol. 55, no. 4, April 2007,
'pp. 1111-1121

FUNCTION PBM_1(R(),Nd%,p%,j&) 'PBM Benchmark #1: Max D for Variable-Length CF Dipole

    LOCAL Z, LengthWaves, ThetaRadians AS EXT

    LOCAL N%, Nsegs%, FeedSegNum%

    LOCAL NumSegs$, FeedSeg$, HalfLength$, Radius$, ThetaDeg$, Lyne$, GainDB$

    LengthWaves  = R(p%,1,j&)

    ThetaRadians = R(p%,2,j&)

    ThetaDeg$ = REMOVE$(STR$(ROUND(ThetaRadians*Rad2Deg,2)),ANY" ")

    IF TALLY(ThetaDeg$,".") = 0 THEN ThetaDeg$ = ThetaDeg$+"."

    Nsegs% = 2*(INT(100*LengthWaves)\2)+1 '100 segs per wavelength, must be an odd #, VOLTAGE SOURCE

    FeedSegNum% = Nsegs%\2 + 1 'center segment number, VOLTAGE SOURCE

    NumSegs$    = REMOVE$(STR$(Nsegs%),ANY" ")

    FeedSeg$    = REMOVE$(STR$(FeedSegNum%),ANY" ")

    HalfLength$ = REMOVE$(STR$(ROUND(LengthWaves/2##,6)),ANY" ")

    IF TALLY(HalfLength$,".") = 0 THEN HalfLength$ = HalfLength$+"."

    Radius$     = "0.00001" +REMOVE$(STR$(ROUND(LengthWaves/1000##,6)),ANY" ")

    N% = FREEFILE

    OPEN "PBM1.NEC" FOR OUTPUT AS #N%

        PRINT #N%,"CM File: PBM1.NEC"
        PRINT #N%,"CM Run ID "+DATE$+" "+TIME$
        PRINT #N%,"CM Nd="+STR$(Nd%)+", p="+STR$(p%)+", j="+STR$(j&)
        PRINT #N%,"CM R(p,1,j)="+STR$(R(p%,1,j&))+", R(p,2,j)="+STR$(R(p%,2,j&))
        PRINT #N%,"CE"
        PRINT #N%,"GW 1,"+NumSegs$+",0.,0.,-"+HalfLength$+",0.,0.,"+HalfLength$+","+Radius$
        PRINT #N%,"GE"
        PRINT #N%,"EX 0,1,"+FeedSeg$+",0,1.,0." 'VOLTAGE SOURCE
        PRINT #N%,"FR 0,1,0,0,299.79564,0."
        PRINT #N%,"RP 0,1,1,1001,"+ThetaDeg$+",0.,0.,0.,1000." 'gain at 1000 wavelengths range
        PRINT #N%,"XQ"
        PRINT #N%,"EN"

    CLOSE #N%

'      - - ANGLES - - -            - POWER GAINS -        - - - POLARIZATION - - -   - - E(THETA) - - -   - - - E(PHI) - - -
'  THETA    PHI        VERT.    HOR.    TOTAL      AXIAL    TILT  SENSE    MAGNITUDE   PHASE     MAGNITUDE   PHASE
' DEGREES  DEGREES      DB       DB      DB        RATIO    DEG.            VOLTS/M    DEGREES    VOLTS/M    DEGREES
'  90.00    0.00      3.91  -999.99   3.91     0.00000     0.00  LINEAR   1.29504E-04   5.37   0.00000E+00   -5.24
'123456789x123456789x123456789x123456789x123456789x123456789x123456789x123456789x123456789x123456789x123456789x123456789x
'     10        20        30        40        50        60        70        80        90       100       110       120

    SHELL "n41_2k1.exe",0

    N% = FREEFILE

    OPEN "PBM1.OUT" FOR INPUT AS #N%

        WHILE NOT EOF(N%)

            LINE INPUT #N%, Lyne$

            IF INSTR(Lyne$,"DEGREES  DEGREES") > 0 THEN EXIT LOOP

        WEND 'position at next data line

        LINE INPUT #N%, Lyne$

    CLOSE #N%

    GainDB$ = REMOVE$(MID$(Lyne$,37,8),ANY" ")

    PBM_1 = 10^(VAL(GainDB$)/10##) 'Directivity
END FUNCTION 'PBM_1()

'----

FUNCTION PBM_2(R(),Nd%,p%,j&) 'PBM Benchmark #2: Max D for Variable-Separation Array of CF Dipoles

    LOCAL Z, DipoleSeparationWaves, ThetaRadians AS EXT

    LOCAL N%, i%

    LOCAL NumSegs$, FeedSeg$, Radius$, ThetaDeg$, Lyne$, GainDB$, Xcoord$, WireNum$

    DipoleSeparationWaves = R(p%,1,j&)

    ThetaRadians          = R(p%,2,j&)

    ThetaDeg$ = REMOVE$(STR$(ROUND(ThetaRadians*Rad2Deg,2)),ANY" ")

    IF TALLY(ThetaDeg$,".") = 0 THEN ThetaDeg$ = ThetaDeg$+"."
```

```
    NumSegs$ = "49"

    FeedSeg$ = "25"

    Radius$  = "0.00001"

    N% = FREEFILE

    OPEN "PBM2.NEC" FOR OUTPUT AS #N%

        PRINT #N%,"CM File: PBM2.NEC"
        PRINT #N%,"CM Run ID "+DATE$+" "+TIME$
        PRINT #N%,"CM Nd="+STR$(Nd%)+", p="+STR$(p%)+", j="+STR$(j&)
        PRINT #N%,"CM R(p,1,j)="+STR$(R(p%,1,j&))+", R(p,2,j)="+STR$(R(p%,2,j&))
        PRINT #N%,"CE"

        FOR i% = -9 TO 9 STEP 2
            WireNum$ = REMOVE$(STR$((i%+11)\2),ANY" ")
            Xcoord$  = REMOVE$(STR$(i%*DipoleSeparationWaves/2##),ANY" ")
            PRINT #N%,"GW "+WireNum$+","+NumSegs$+","+Xcoord$+",0,-0.25,"+Xcoord$+",0,0.25,"+Radius$
        NEXT i%

        PRINT #N%,"GE"

        FOR i% = 1 TO 10
            PRINT #N%,"EX 0,"+REMOVE$(STR$(i%),ANY" ")+","+FeedSeg$+",0,1,,0." 'VOLTAGE SOURCE
        NEXT i%
        PRINT #N%,"FR 0,1,0,0,299.79564,0."
        PRINT #N%,"RP 0,1,1,1001,"+ThetaDeg$+",90.,0.,0.,1000." 'gain at 1000 wavelengths range
        PRINT #N%,"XQ"
        PRINT #N%,"EN"

    CLOSE #N%
```

```
'     - - ANGLES - -         - POWER GAINS -        - - - POLARIZATION - - -    - - - E(THETA) - - -   - - - E(PHI) - - -
'    THETA      PHI     VERT.  HOR.    TOTAL    AXIAL    TILT   SENSE    MAGNITUDE   PHASE    MAGNITUDE     PHASE
'   DEGREES  DEGREES     DB     DB       DB      RATIO    DEG.            VOLTS/M    DEGREES    VOLTS/M     DEGREES
'    90.00     0.00     3.91  -999.99   3.91    0.00000   0.00  LINEAR   1.29504E-04   5.37   0.00000E+00   -5.24
'123456789x123456789x123456789x123456789x123456789x123456789x123456789x123456789x123456789x123456789x123456789x
'     10      20       30      40       50       60        70      80        90       100       110       120
```

```
    SHELL "n41_2k1.exe",0

    N% = FREEFILE

    OPEN "PBM2.OUT" FOR INPUT AS #N%

        WHILE NOT EOF(N%)

            LINE INPUT #N%, Lyne$

            IF INSTR(Lyne$,"DEGREES  DEGREES") > 0 THEN EXIT LOOP

        WEND 'position at next data line

        LINE INPUT #N%, Lyne$

    CLOSE #N%

    GainDB$ = REMOVE$(MID$(Lyne$,37,8),ANY" ")

    IF AddNoiseToPBM2 = "YES" THEN

        Z = 10^(VAL(GainDB$)/10##) + GaussianDeviate(0##,0.4472##) 'Directivity with Gaussian noise (zero mean, 0.2 variance)

    ELSE

        Z = 10^(VAL(GainDB$)/10##) 'Directivity without noise

    END IF

    PBM_2 = Z

END FUNCTION 'PBM_2()

'----

FUNCTION PBM_3(R(),Nd%,j&) 'PBM Benchmark #3: Max D for Circular Dipole Array

    LOCAL Beta, ThetaRadians, Alpha, ReV, ImV AS EXT

    LOCAL N%, i%

    LOCAL NumSegs$, FeedSeg$, Radius$, ThetaDeg$, Lyne$, GainDB$, Xcoord$, Ycoord$, WireNum$, ReEX$, ImEX$

    Beta         = R(p%,1,j&)

    ThetaRadians = R(p%,2,j&)

    ThetaDeg$ = REMOVE$(STR$(ROUND(ThetaRadians*Rad2Deg,2)),ANY" ")

    IF TALLY(ThetaDeg$,".") = 0 THEN ThetaDeg$ = ThetaDeg$+"."

    NumSegs$ = "49"

    FeedSeg$ = "25"

    Radius$  = "0.00001"

    N% = FREEFILE

    OPEN "PBM3.NEC" FOR OUTPUT AS #N%
        PRINT #N%,"CM File: PBM3.NEC"
        PRINT #N%,"CM Run ID "+DATE$+" "+TIME$
        PRINT #N%,"CM Nd="+STR$(Nd%)+", p="+STR$(p%)+", j="+STR$(j&)
        PRINT #N%,"CM R(p,1,j)="+STR$(R(p%,1,j&))+", R(p,2,j)="+STR$(R(p%,2,j&))
        PRINT #N%,"CE"

        FOR i% = 1 TO 8
            WireNum$ = REMOVE$(STR$(i%),ANY" ")

            SELECT CASE i%
                CASE 1 : Xcoord$ = "1"        : Ycoord$ = "0"
                CASE 2 : Xcoord$ = "0.70711"  : Ycoord$ = "0.70711"
                CASE 3 : Xcoord$ = "0"        : Ycoord$ = "1"
                CASE 4 : Xcoord$ = "-0.70711" : Ycoord$ = "0.70711"
                CASE 5 : Xcoord$ = "-1"       : Ycoord$ = "0"
                CASE 6 : Xcoord$ = "-0.70711" : Ycoord$ = "-0.70711"
                CASE 7 : Xcoord$ = "0"        : Ycoord$ = "-1"
                CASE 8 : Xcoord$ = "0.70711"  : Ycoord$ = "-0.70711"
            END SELECT

            PRINT #N%,"GW "+WireNum$+","+NumSegs$+","+Xcoord$+","+Ycoord$+",-0.25,"+Xcoord$+","+Ycoord$+",0.25,"+Radius$
        NEXT i%
```

```
        PRINT #N%,"GE"

        FOR i% = 1 TO 8
            Alpha = -COS(TwoPi*Beta*(i%-1))

            ReV  = COS(Alpha)
            ImV  = SIN(Alpha)

            ReEX$ = REMOVE$(STR$(ROUND(ReV,6)),ANY" ")
            ImEX$ = REMOVE$(STR$(ROUND(ImV,6)),ANY" ")

            IF TALLY(ReEX$,".") = 0 THEN ReEX$ = ReEX$+"."
            IF TALLY(ImEX$,".") = 0 THEN ImEX$ = ImEX$+"."

            PRINT #N%,"EX 0,"+REMOVE$(STR$(i%),ANY" ")+","+FeedSeg$+",0,"+ReEX$+","+ImEX$ 'VOLTAGE SOURCE
        NEXT i%

        PRINT #N%,"FR 0,1,0,0,299.79564,0."
        PRINT #N%,"RP 0,1,1,1001,"+ThetaDegS$+",0.,0.,0.,1000." 'gain at 1000 wavelengths range
        PRINT #N%,"XQ"
        PRINT #N%,"EN"

    CLOSE #N%
```

```
'       - - ANGLES - -      - POWER GAINS -      - - - POLARIZATION - - -     - - E(THETA) - - -   - - E(PHI) - - -
'  THETA    PHI       VERT.   HOR.   TOTAL      AXIAL    TILT  SENSE    MAGNITUDE   PHASE     MAGNITUDE    PHASE
' DEGREES DEGREES      DB     DB      DB        RATIO    DEG.          VOLTS/M    DEGREES    VOLTS/M    DEGREES
'  90.00    0.00      3.91  -999.99   3.91    0.00000    0.00  LINEAR  1.29504E-04   5.37   0.00000E+00   -5.24
'123456789x123456789x123456789x123456789x123456789x123456789x123456789x123456789x123456789x123456789x123456789x123456789x
'     10       20       30      40      50      60      70      80      90      100      110      120
```

```
    SHELL "n41_2k1.exe",0

    N% = FREEFILE

    OPEN "PBM3.OUT" FOR INPUT AS #N%

        WHILE NOT EOF(N%)

            LINE INPUT #N%, Lyne$

            IF INSTR(Lyne$,"DEGREES  DEGREES") > 0 THEN EXIT LOOP

        WEND 'position at next data line

        LINE INPUT #N%, Lyne$

    CLOSE #N%

    GainDB$ = REMOVE$(MID$(Lyne$,37,8),ANY" ")

    PBM_3 = 10^(VAL(GainDB$)/10##) 'Directivity

END FUNCTION 'PBM_3()

'----
FUNCTION PBM_4(R(),Nd%,p%,j&) 'PBM Benchmark #4: Max D for Vee Dipole

    LOCAL TotalLengthWaves, AlphaRadians, ArmLength, Xlength, Zlength, Lfeed AS EXT

    LOCAL N%, i%, Nsegs%, FeedZcoord$

    LOCAL NumSegs%, Lyne$, GainDB$, Xcoord$, Zcoord$

    TotalLengthWaves = 2##*R(p%,1,j&)

    AlphaRadians     = R(p%,2,j&)

    Lfeed            = 0.01##

    FeedZcoord$      = REMOVE$(STR$(Lfeed),ANY" ")

    ArmLength = (TotalLengthWaves-2##*Lfeed)/2##

    Xlength  = ROUND(ArmLength*COS(AlphaRadians),6)

    Xcoord$ = REMOVE$(STR$(Xlength),ANY" ") : IF TALLY(Xcoord$,".") = 0 THEN Xcoord$ = Xcoord$+"."

    Zlength  = ROUND(ArmLength*SIN(AlphaRadians),6)

    Zcoord$ = REMOVE$(STR$(Zlength+Lfeed),ANY" ") : IF TALLY(Zcoord$,".") = 0 THEN Zcoord$ = Zcoord$+"."

    Nsegs%  = 2*(INT(TotalLengthWaves*100)\2) 'even number, total # segs

    NumSegs$ = REMOVE$(STR$(Nsegs%\2),ANY" ") '# segs per arm

    N% = FREEFILE

    OPEN "PBM4.NEC" FOR OUTPUT AS #N%

        PRINT #N%,"CM File: PBM4.NEC"
        PRINT #N%,"CM Run ID "+DATE$+" "+TIME$
        PRINT #N%,"CM Nd="+STR$(Nd%)+", p="+STR$(p%)+", j="+STR$(j&)
        PRINT #N%,"CM R(p,1,j)="+STR$(R(p%,1,j&))+", R(p,2,j)="+STR$(R(p%,2,j&))
        PRINT #N%,"CE"

        PRINT #N%,"GW 1,5,0.,0.,-"+FeedZcoord$+",0.,0.,-"+FeedZcoord$+",0.00001 'feed wire, 1 segment, 0.01 wvln

        PRINT #N%,"GW 2,"+NumSegs$+",0.,0.,-"+FeedZcoord$+","+Xcoord$+",0.,-"+Zcoord$+",0.00001 'upper arm

        PRINT #N%,"GW 3,"+NumSegs$+",0.,0.,-"+FeedZcoord$+","+Xcoord$+",0.,-"+Zcoord$+",0.00001 'lower arm

        PRINT #N%,"GE"

        PRINT #N%,"EX 0,1,3,0,1.,0." 'VOLTAGE SOURCE

        PRINT #N%,"FR 0,1,0,0,299.79564,0."
        PRINT #N%,"RP 0,1,1,1001,90.,0.,0.,0.,1000." 'ENDFIRE gain at 1000 wavelengths range
        PRINT #N%,"XQ"
        PRINT #N%,"EN"

    CLOSE #N%
```

```
'       - - ANGLES - -      - POWER GAINS -      - - - POLARIZATION - - -     - - E(THETA) - - -   - - E(PHI) - - -
'  THETA    PHI       VERT.   HOR.   TOTAL      AXIAL    TILT  SENSE    MAGNITUDE   PHASE     MAGNITUDE    PHASE
' DEGREES DEGREES      DB     DB      DB        RATIO    DEG.          VOLTS/M    DEGREES    VOLTS/M    DEGREES
'  90.00    0.00      3.91  -999.99   3.91    0.00000    0.00  LINEAR  1.29504E-04   5.37   0.00000E+00   -5.24
'123456789x123456789x123456789x123456789x123456789x123456789x123456789x123456789x123456789x123456789x123456789x123456789x
'     10       20       30      40      50      60      70      80      90      100      110      120
```

```
    SHELL "n41_2k1.exe",0

    N% = FREEFILE

    OPEN "PBM4.OUT" FOR INPUT AS #N%
```

```
        WHILE NOT EOF(N%)

            LINE INPUT #N%, Lyne$

            IF INSTR(Lyne$,"DEGREES  DEGREES") > 0 THEN EXIT LOOP

        WEND 'position at next data line

        LINE INPUT #N%, Lyne$

    CLOSE #N%

    GainDB$ = REMOVE$(MID$(Lyne$,37,8),ANY" ")

    PBM_4 = 10^(VAL(GainDB$)/10##) 'Directivity

END FUNCTION 'PBM_4()

'----

FUNCTION PBM_5(R(),Nd%,p%,j&) 'PBM Benchmark #5: N-element collinear array (Nd=N-1)

    LOCAL TotalLengthWaves, Di(), Ystart, Y1, Y2, SumDi AS EXT

    LOCAL N%, i%, q%

    LOCAL Lyne$, GainDB$

    REDIM Di(1 TO Nd%)

    FOR i% = 1 TO Nd%

        Di(i%) = R(p%,i%,j&) 'dipole separation, wavelengths

    NEXT i%

    TotalLengthWaves = 0##

    FOR i% = 1 TO Nd%

        TotalLengthwaves = TotalLengthWaves + Di(i%)

    NEXT i%

    TotalLengthWaves = TotalLengthWaves + 0.5## 'add half-wavelength of 1 meter at 299.8 MHz

    Ystart = -TotalLengthWaves/2##

    N% = FREEFILE

    OPEN "PBM5.NEC" FOR OUTPUT AS #N%

        PRINT #N%,"CM File: PBM5.NEC"
        PRINT #N%,"CM Run ID "+DATE$+" "+TIME$
        PRINT #N%,"CM Nd="+STR$(Nd%)+", p="+STR$(p%)+", j="+STR$(j&)
        PRINT #N%,"CM R(p,1,j)="+STR$(R(p%,1,j&))+", R(p,2,j)="+STR$(R(p%,2,j&))
        PRINT #N%,"CE"

        FOR i% = 1 TO Nd%+1

            SumDi = 0##

            FOR q% = 1 TO i%-1

                SumDi = SumDi + Di(q%)

            NEXT q%

            Y1 = ROUND(Ystart + SumDi,6)

            Y2 = ROUND(Y1+0.5##,6) 'add one-half wavelength for other end of dipole

            PRINT #N%,"GW "+REMOVE$(STR$(i%),ANY" ")+",49,0.,"+REMOVE$(STR$(Y1),ANY" ")+",0.,0.,"+REMOVE$(STR$(Y2),ANY" ")+",0.,0.00001"

        NEXT i%

        PRINT #N%,"GE"

        FOR i% = 1 TO Nd%+1
            PRINT #N%,"EX 0,"+REMOVE$(STR$(i%),ANY" ")+",25,0,1.,0." 'VOLTAGE SOURCES
        NEXT i%

        PRINT #N%,"FR 0,1,0,0,299.79564,0."
        PRINT #N%,"RP 0,1,1,1001,90.,0.,0.,0.,1000." 'gain at 1000 wavelengths range
        PRINT #N%,"XQ"
        PRINT #N%,"EN"

    CLOSE #N%

'          - - ANGLES - -        - POWER GAINS -      - - - POLARIZATION - - -   - - E(THETA) - - -   - - E(PHI) - - -
'   THETA     PHI      VERT.   HOR.   TOTAL    AXIAL    TILT  SENSE   MAGNITUDE    PHASE   MAGNITUDE    PHASE
' DEGREES  DEGREES      DB      DB      DB     RATIO    DEG.          VOLTS/M    DEGREES    VOLTS/M    DEGREES
'   90.00    0.00       3.91 -999.99   3.91   0.00000   0.00 LINEAR  1.29504E-04    5.37  0.00000E+00   -5.24
'123456789x123456789x123456789x123456789x123456789x123456789x123456789x123456789x123456789x123456789x123456789x
'     10      20      30      40      50      60      70      80      90      100     110     120

    SHELL "n41_2k1.exe",0

    N% = FREEFILE

    OPEN "PBM5.OUT" FOR INPUT AS #N%

        WHILE NOT EOF(N%)

            LINE INPUT #N%, Lyne$

            IF INSTR(Lyne$,"DEGREES  DEGREES") > 0 THEN EXIT LOOP

        WEND 'position at next data line

        LINE INPUT #N%, Lyne$

    CLOSE #N%

    GainDB$ = REMOVE$(MID$(Lyne$,37,8),ANY" ")

    PBM_5 = 10^(VAL(GainDB$)/10##) 'Directivity

END FUNCTION 'PBM_5()

'************************************************ END PROGRAM 'CFO_02-27-2010(arXiv#5).BAS' ************************************************
```